\pdfoutput=1

\documentclass[11pt,twoside,a4paper,cmspaper,final,collab]{cms-tdr}
\def\svnVersion{199448}\def\cmsCernNo{2013-073}\def\cmsCernDate{\today}\def\cmsMessage{Published in the Journal of High Energy Physics as \href{http://dx.doi.org/10.1007/JHEP07(2013)122}{\doi{10.1007/JHEP07(2013)122.}}}

\begin{document}\cmsNoteHeader{EXO-12-026}

\hyphenation{had-ron-i-za-tion}
\hyphenation{cal-or-i-me-ter}
\hyphenation{de-vices}

\RCS$Revision: 198699 $
\RCS$HeadURL: svn+ssh://svn.cern.ch/reps/tdr2/papers/EXO-12-026/trunk/EXO-12-026.tex $
\RCS$Id: EXO-12-026.tex 198699 2013-07-26 07:56:39Z querten $
\newlength\cmsFigWidth
\ifthenelse{\boolean{cms@external}}{\setlength\cmsFigWidth{0.85\columnwidth}}{\setlength\cmsFigWidth{0.4\textwidth}}
\ifthenelse{\boolean{cms@external}}{\providecommand{\cmsLeft}{top}}{\providecommand{\cmsLeft}{left}}
\ifthenelse{\boolean{cms@external}}{\providecommand{\cmsRight}{bottom}}{\providecommand{\cmsRight}{right}}
\cmsNoteHeader{EXO-12-026} % This is over-written in the CMS environment: useful as preprint no. for export versions
\title{Searches for long-lived charged particles in pp collisions at $\sqrt{s}=$7 and 8\TeV}

\date{\today}
\newcommand{\stau}{\ensuremath{\sTau_1}\xspace}
\newcommand{\smu}{\ensuremath{\PSgm_1}\xspace}
\newcommand{\sel}{\ensuremath{\PSe_1}\xspace}
\newcommand{\sto}{\ensuremath{\sTop_1}\xspace}
\newcommand{\ias}{\ensuremath{I_{as}}}
\newcommand{\iasp}{\ensuremath{I_{as}^\prime}}
\newcommand{\id}{\ensuremath{I_{d}}}
\newcommand{\ih}{\ensuremath{I_{h}}}
\providecommand{\dedx}{\ensuremath{\cmsSymbolFace{d}E\kern-0.12em/\kern-0.2em\cmsSymbolFace{d}x}\xspace}
\providecommand{\MeVcm}{\ensuremath{{\,\text{Me\hspace{-.08em}V\hspace{-0.16em}/\hspace{-0.08em}}\text{cm}}}\xspace}
\newcommand{\tof}{TOF}
\newcommand{\invbeta}{\ensuremath{1/\beta}\xspace}
\newcommand{\tkonly}{{tracker-only}}
\newcommand{\tktof}{{tracker+\tof}}
\newcommand{\multicharge}{{multiply charged}}
\newcommand{\muononly}{{muon-only}}
\newcommand{\fractionalcharge}{{fractionally charged}}
\renewcommand{\labelitemi}{$-$}
\renewcommand{\labelitemii}{$\ast$}

\abstract{
Results of
searches for heavy stable
charged particles produced in pp collisions at $\sqrt{s} = 7$
and 8\TeV are presented corresponding to an integrated luminosity of 5.0\fbinv and 18.8\fbinv, respectively.  Data collected with the CMS
detector are used to study the momentum, energy deposition,
and time-of-flight of signal candidates.
Leptons with an electric charge between $e$/3 and 8$e$, as well as bound states that can undergo charge exchange with the detector material, are studied.
Analysis results are presented
for various combinations of signatures in the inner
tracker only, inner tracker and muon detector, and muon detector
only.  Detector signatures utilized are long time-of-flight
to the outer muon system and anomalously high (or low) energy
deposition in the inner tracker.
The data are consistent with the expected
background, and upper limits are set on the production
cross section of long-lived
gluinos, scalar top quarks, and scalar $\tau$ leptons, as well as
pair produced long-lived leptons.
Corresponding lower mass limits, ranging up to 1322\GeVcc
for gluinos, are the most stringent to date.
}

\hypersetup{%
pdfauthor={CMS Collaboration},%
pdftitle={Searches for long-lived charged particles in pp collisions at sqrt(s)=7 and 8 TeV},%
pdfsubject={CMS},%
pdfkeywords={CMS, physics}}

\maketitle %maketitle comes after all the front information has been supplied
\section{Introduction}

Many extensions of the standard model (SM) include
heavy, long-lived, charged particles that have speed $v$
significantly less than the speed of light $c$~\cite{Drees:1990yw, Fairbairn:2006gg, Bauer:2009cc}
or charge $Q$ not equal to the elementary positive or negative charge ${\pm}1e$~\cite{Schwinger:1966nj, Kusenko:1997si,  Fargion:2005ep, Koch:2007um, Kang:2007ib}, or both.
With lifetimes greater than a few nanoseconds, these particles can
travel distances comparable to the size of modern detectors and
thus appear to be stable.
These particles, generically referred to as heavy stable
charged particles (HSCP), can be singly charged ($|Q|=1e$),
fractionally charged ($|Q|<1e$), or multiply charged ($|Q|>1e$).
Without dedicated searches, HSCPs may be misidentified or even completely missed, as particle identification algorithms at hadron collider experiments
generally assume signatures appropriate for SM particles,
\eg, $v \approx c$ and $Q = 0$ or ${\pm}1e$.
Additionally, some HSCPs may combine with SM particles to form
composite objects.  Interactions of these composite objects
with the detector may change their constituents and possibly their electric charge,
further limiting the ability of standard algorithms to identify them.

\begin{figure}[ht]
 \begin{center}
  \includegraphics[clip=true, trim=0.0cm 0cm 0.0cm 0cm, width=0.48\linewidth]{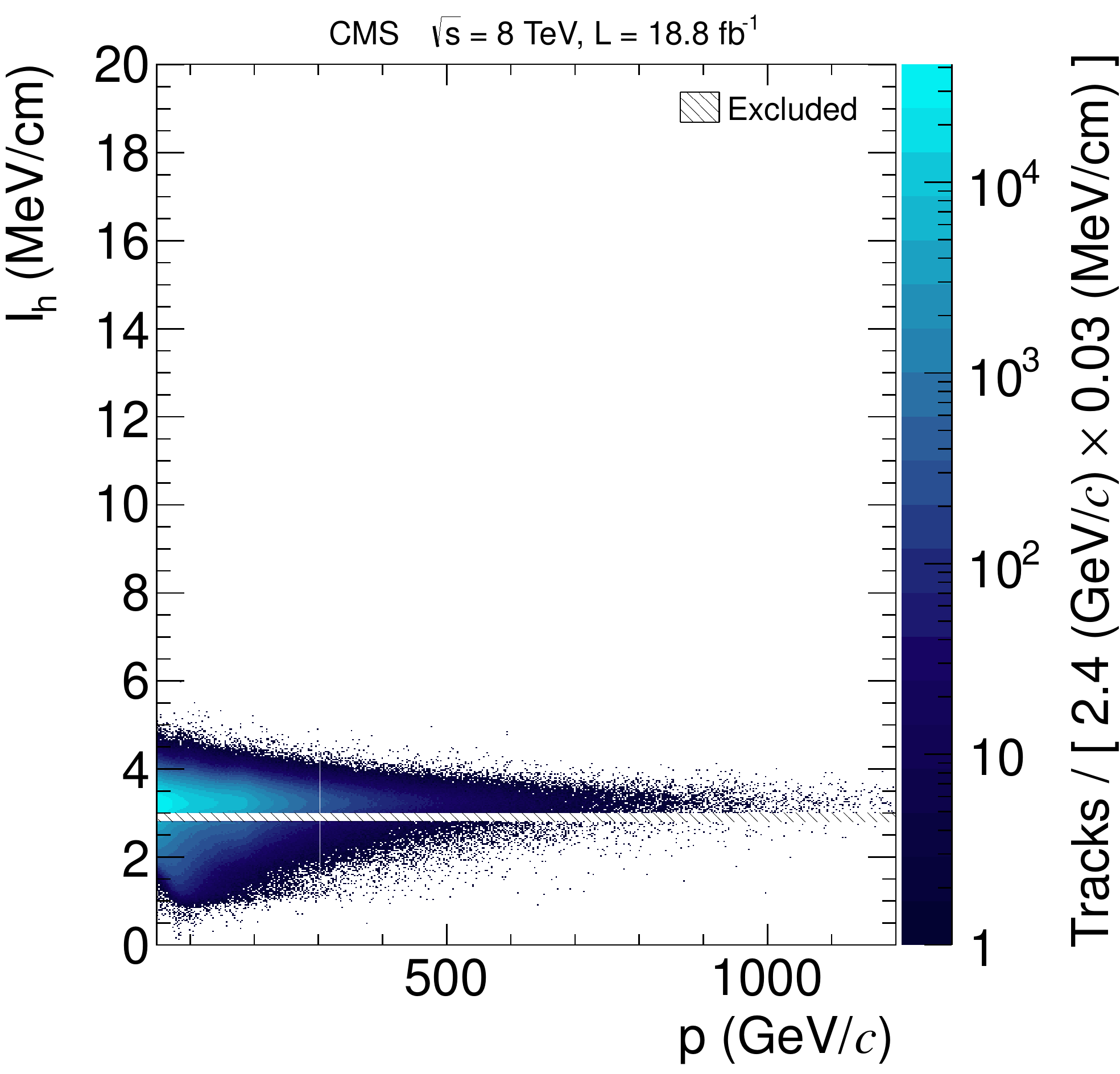}~~~
  \includegraphics[clip=true, trim=0.0cm 0cm 0.0cm 0cm, width=0.48\linewidth]{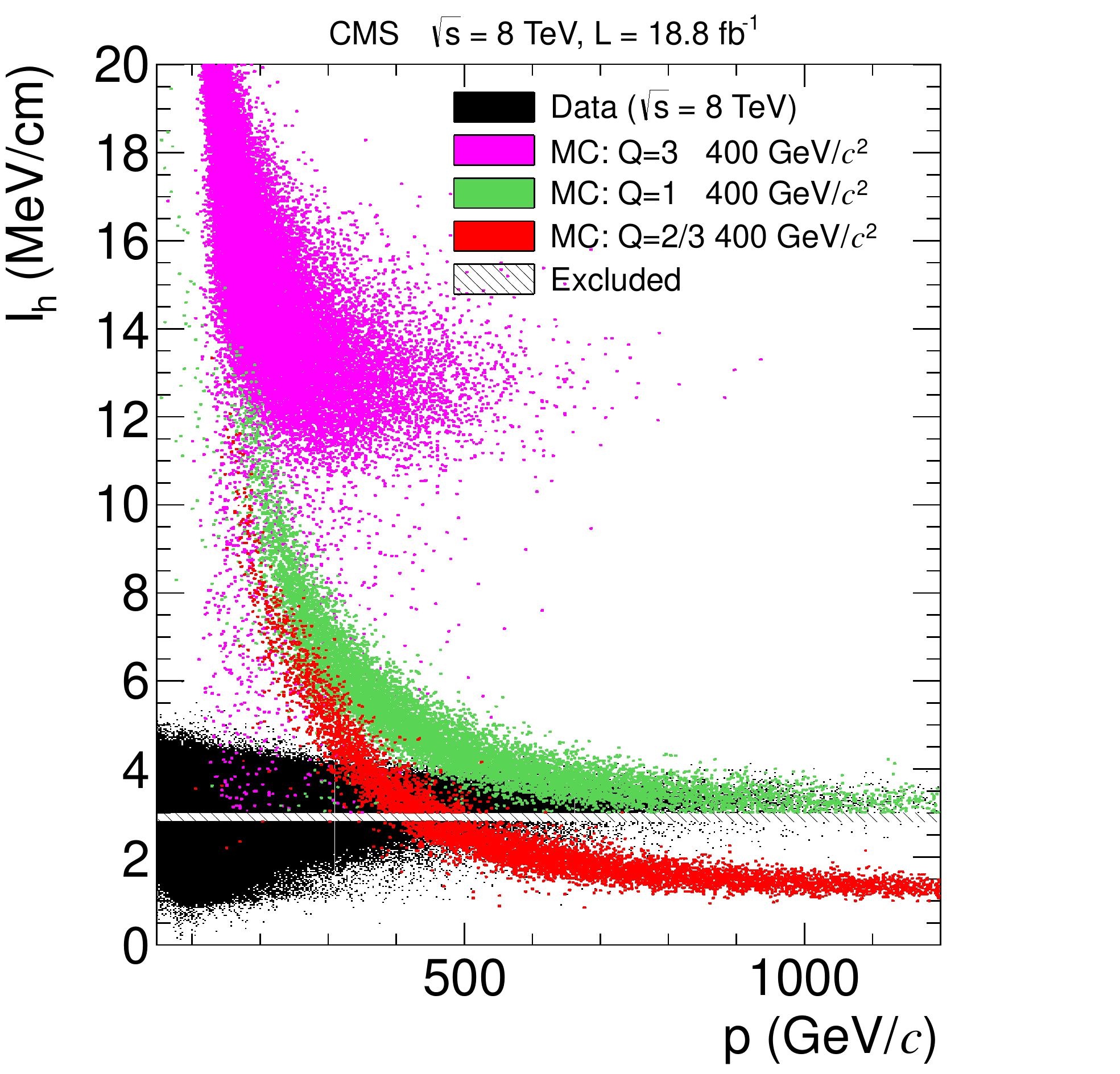}
 \end{center}
 \caption{Distribution of $I_h$, a \dedx\ estimator that is defined in Section~\ref{sec:dEdx}, versus particle momentum for
   $\sqrt{s} = 8$\TeV data (left) and also including MC simulated HSCP candidates of different charges (right).
   Tracks with $2.8\leq I_h \leq 3.0\MeVcm$ are excluded by preselection requirements, as discussed in Section~\ref{sec:presel}.
   \label{fig:dedxdists}}
\end{figure}

For HSCP masses greater than 100\GeVcc, a significant
fraction of particles produced at the Large Hadron Collider (LHC) have
$\beta$ ($\equiv{v/c}$) values less than 0.9.
These HSCPs can be identified by their longer time-of-flight (\tof) to outer detectors
or their anomalous energy loss (\dedx).
The \dedx\ of a particle depends on both its electric charge (varying as $Q^2$) and its $\beta$.
The dependence of \dedx\ on these variables is described by the Bethe-Bloch
formula~\cite{Beringer:1900zz}. This dependence can be seen in Fig.~\ref{fig:dedxdists},
which shows a \dedx\ estimate versus momentum for tracks from data and Monte Carlo (MC) simulations of HSCP signals with various charges.  In
the momentum range of interest (10--1000\GeVc), SM charged particles
have a relatively flat ionization energy loss and $\beta$ values very close to one.
Searching for candidates with long time-of-flight or large \dedx\ gives sensitivity
to massive particles with $|Q| = 1e$, particles with $|Q| > 1e$,
and low-momentum particles with $|Q| < 1e$.  On the other hand, searching
for candidates with lower \dedx\ yields sensitivity to high-momentum
particles with $|Q| < 1e$.

Previous collider searches for HSCPs have been performed at
LEP~\cite{Barate:1997dr, Abreu:2000tn,Achard:2001qw,Abbiendi:2003yd},
HERA~\cite{Aktas:2004pq}, the
Tevatron~\cite{Abazov:2008qu, Aaltonen:2009kea, Abazov:2011pf,Abazov:2012ab},
and the LHC~\cite{Khachatryan:2011ts, Aad:2011mb,  Aad:2011yf, Aad:2011hz, Chatrchyan::2012dr, Chatrchyan:2012sp, Aad:2012vd, ATLASmCHAMPs}.
The results from such searches have placed important bounds on
beyond the standard model (BSM)
theories~\cite{Berger:2008cq,CahillRowley:2012kx}, such as lower limits
on the mass of gluinos, scalar top quarks (stops), and pair-produced scalar $\tau$ leptons (staus) at 1098, 737, and 223\GeVcc, respectively.
Presented here are several searches for singly, fractionally,
and multiply charged HSCPs using data collected with
the Compact Muon Solenoid (CMS) detector during the 2011 ($\sqrt{s} = 7$\TeV, 5.0\fbinv)
and 2012 ($\sqrt{s} = 8$\TeV, 18.8\fbinv) data taking periods.

\section{Signal benchmarks \label{sec:signals}}

The searches presented here are sensitive to a wide variety
of signals of new charged massive particles.  Several
BSM models are used to benchmark the sensitivity.
The HSCPs can be classified as either {\em lepton-like} or
{\em hadron-like}.  Lepton-like HSCPs interact primarily
through the electromagnetic force, while hadron-like
HSCPs additionally interact through the strong force and
form bound states with SM quarks (or gluons) called
$R$-hadrons~\cite{Farrar:1978xj}.
The $R$-hadrons can be charged or neutral.  Strong interactions between
the SM quarks and detector material increase energy
loss and can lead to charge exchange, \eg, conversion of charged
$R$-hadrons into neutral ones (and vice-versa).  There is some
uncertainty in the modeling of $R$-hadrons' strong interactions with
detector material.  For this analysis, two separate models are
considered: (1) the model described in
Refs.~\cite{Kraan:2004tz,Mackeprang:2006gx}, referred to as
the cloud model, and (2) a model in which
any strong interaction results in a neutral
$R$-hadron~\cite{Mackeprang:2009ad}, referred to as the charge-suppressed model.
The cloud model envisions the $R$-hadron as composed of a
spectator HSCP surrounded by a cloud of colored, light
constituents.
The charge-suppressed model
results in essentially all $R$-hadrons being neutral by the
time they enter the muon system.
For each of the models considered, particle interactions with
the CMS apparatus and detector response are simulated using
\GEANTfour v9.2~\cite{Agostinelli2003250, Allison:2006ve}.
To produce the effect of multiple interactions per bunch
crossing (pileup), simulated minimum bias events are
overlaid with the primary collision.

The minimal gauge-mediated supersymmetry breaking (GMSB) model~\cite{Giudice:1998bp}  predicts the gravitino to be the lightest supersymmetric particle (LSP)
and allows for the next-to-lightest supersymmetric particle (NLSP) to be long-lived because of the weakness of the gravitational coupling,
which governs the decay of the NLSP to the LSP.
For this analysis the NLSP is taken to be a lepton-like stau (\stau) with an assumed lifetime that exceeds the time-of-flight through the CMS detector.
For $\sqrt{s}=7$ (8)\TeV simulation, \PYTHIA\ v6.422 (v6.426)~\cite{Sjostrand:2006za}
is used to model both
Drell--Yan production of a \stau pair (direct pair-production) and production of heavier supersymmetric particles
whose decay chains lead to indirect stau production.
Events with \stau\ masses in the range 100--557\GeVcc
are generated using line 7 of the "Snowmass Points and Slopes" benchmarks~\cite{Allanach:2002nj}.
They correspond to $N=3$ chiral SU(5) multiplets added to the theory at a
scale $F$ from $60$ to $360$\TeV~\cite{Giudice:1998bp} depending on the \stau mass, and an
effective supersymmetry-breaking scale of $\Lambda = F/2$.
All points have a value of 10 for the ratio of neutral Higgs field vacuum expectation values ($\tan{\beta}$),
a positive sign for the Higgs-Higgsino mass parameter ($\sgn(\mu$)),
and a value of $10^4$ for the ratio of the gravitino mass to the value it would have if the only supersymmetry-breaking scale were that in the messenger sector ($c_\text{grav}$).
The particle mass spectrum and the decay table were produced
with the program \ISASUGRA version 7.69~\cite{Paige:2003mg}.
Theoretical production cross sections ($\sigma_{\text{th}}$) for staus are calculated
at next-to-leading order (NLO) with
\PROSPINO\ v2.1~\cite{Beenakker:1999xh}.
Compared to the previous publication~\cite{Chatrchyan:2012sp}, the theoretical NLO cross section used for the indirect production of staus also includes processes involving pairs of neutralinos/charginos.

$R$-hadron signals from gluino (\PSg) and scalar top (\sto) pair production are studied
using \PYTHIA\ v6.442 (v8.153) \cite{Sjostrand:2006za,Sjostrand:2007gs} for $\sqrt{s}=7$ (8)\TeV generation.
Stop pair production is modeled for
masses in the range 100--1000\GeVcc.
For \PSg\ production,
split supersymmetry~\cite{Giudice:2004tc, ArkaniHamed:2004fb}
is modeled by setting the squark masses to greater than 10 \TeVcc
and generating \PSg\ masses of 300--1500\GeVcc.
The fraction $f$ of gluinos hadronizing into $\tilde g$-gluon
bound states is unknown.  These neutral states would not leave a
track in the inner detectors.
Therefore, several scenarios
are considered for the singly charged analysis:
$f=0.1$, 0.5, and 1.0.
In the extreme case where $f=1.0$,  $R$-hadrons are always neutral in the inner tracker, but a fraction of them may interact with the detector material and
be electrically charged during their passage through the muon system.
Gluino and scalar top pair production cross sections are calculated
at NLO plus next-to-leading logarithmic (NLL)
accuracy with \PROSPINO\ v2.1~\cite{Beenakker:1996ch, Beenakker:1997ut, Kulesza:2008jb,
Beenakker:2009ha, Kulesza:2009kq,Beenakker:2010nq,Beenakker:2011fu,  Kramer:2012bx}.

The last of the signal samples studied is the modified Drell--Yan production of long-lived leptons.
In this scenario, new massive spin-1/2
particles may have an electric charge different than $|Q|=1e$ and are neutral under $SU(3)_C$
and $SU(2)_L$; therefore they
couple only to the photon and the Z boson via $U(1)$ couplings~\cite{Langacker:2011db}.
Signal samples are simulated
using \PYTHIA v6.422 (v6.426)~\cite{Sjostrand:2006za} for $\sqrt{s}=7$ (8)\TeV.
The analysis uses simulations of $|Q|$ = $e/3$ and $2e/3$ for masses of 100--600\GeVcc, of $|Q|$ ranging from $1e$ to $6e$ for masses of 100--1000\GeVcc,
and of $|Q|$ = $7e$ and $8e$ for masses of 200--1000\GeVcc.

The CTEQ6L1 parton distribution functions (PDF)~\cite{Pumplin:2002vw} are
used for the sample generation.

\section{The CMS detector}

The CMS  experiment uses a right-handed coordinate system, with the origin at the
nominal interaction point, the $x$ axis pointing to the center of
the LHC ring, the $y$ axis pointing up (perpendicular to the plane of the LHC ring),
and the $z$ axis along the counterclockwise-beam direction. The polar
angle $\theta$ is measured from the positive $z$ axis and the
azimuthal angle $\phi$ in the $x$-$y$ plane.
The pseudorapidity is given by $\eta = -\ln[\tan(\theta/2)]$.

The central feature of the CMS apparatus
is a superconducting solenoid of 6\unit{m} internal diameter.
Within the field volume are
a silicon pixel and strip tracker, a lead tungstate crystal
electromagnetic calorimeter, and a brass and scintillator
hadron calorimeter. Muons are measured in gas-ionization
detectors embedded in the steel flux-return yoke of the magnet. Extensive forward
calorimetry complements the coverage provided by the barrel
and endcap detectors.
The inner tracker measures charged particles within the pseudorapidity
range $|\eta|< 2.5$. It consists of 1440 silicon pixel and
15\,148 silicon strip detector modules and is located in the
3.8\unit{T} field of the solenoid. The inner tracker provides
a transverse momentum (\pt) resolution of about 1.5\% for 100\GeVc particles.
Muons are measured in the pseudorapidity range $|\eta|< 2.4$, with
detection planes made using three technologies: drift tubes (DT),
cathode strip chambers (CSC), and resistive plate chambers (RPC).
The muon system extends out to eleven meters from the interaction point in the $z$ direction and
seven meters radially.
Matching tracks in the muon system to tracks measured in the silicon tracker results in a transverse
momentum resolution between 1 and 5\%, for \pt values up to 1\TeVc.
The first level (L1) of the CMS trigger system, composed of custom
hardware processors, uses information from the calorimeters and
muon detectors to select events of interest.
The high level trigger (HLT) processor
farm further decreases the event rate from around 100\unit{kHz} to
around 300\unit{Hz} for data storage.
A more detailed description of the CMS detector can be found in Ref.~\cite{Chatrchyan:2008zzk}.

\subsection{The dE/dx measurements \label{sec:dEdx}}
As in Ref.~\cite{Khachatryan:2011ts}, \dedx for a track is estimated as:
\begin{equation}
 I_h= \biggl( \cfrac{1}{N} \sum_i c_{i}^{-2} \biggr)^{-1/2},
 \label{eq:HarmonicEstimator}
\end{equation}
where $N$ is the number of measurements in the silicon-strip detectors and $c_{i}$ is the energy loss per unit path length in the sensitive part
of the silicon detector of the $i$th measurement; \ih\
has units \MeVcm.
In addition, two modified versions of the Smirnov--Cramer--von Mises~\cite{Eadie, James} discriminator, \ias\ (\iasp), are used to separate SM particles from candidates with
large (small) \dedx.  The discriminator is given by:
\begin{equation}
 I_{as}^{(\prime)} = \frac{3}{N} \times \left(
   \frac{1}{12N} + \sum_{i=1}^N
   \left[
   P_i^{(\prime)} \times \left( P_i^{(\prime)} - \frac{2i-1}{2N} \right)^2 \right] \right),
\end{equation}
where $P_i$ ($P_i^{\prime}$) is the probability for a minimum ionizing particle (MIP) to
produce a charge smaller (larger) or equal to that of the $i$th measurement
for the observed path length in the detector, and the sum is over the
measurements ordered in terms of increasing $P_i^{(\prime)}$.

As in Ref.~\cite{Khachatryan:2011ts}, the mass of a $|Q|=1e$ candidate particle
is calculated based on the relationship:
\begin{equation}
 I_h= K\cfrac{m^2}{p^2}+C,
 \label{eq:MassFromHarmonicEstimator}
\end{equation}
where the empirical parameters $K=2.559 \pm 0.001\MeV\cdot c^2/\text{cm}$
and $C=2.772 \pm 0.001\MeVcm$ are determined from
data using a sample of low-momentum protons in a minimum-bias dataset.

The number of silicon-strip measurements associated with a track, 15 on average, is sufficient to ensure good \dedx and mass resolutions.

\subsection{Time-of-flight measurements}

As in Ref.~\cite{Chatrchyan:2012sp}, the time-of-flight to the muon system can be used to discriminate between
$\beta \approx 1$ particles and slower candidates.
The measured time difference ($\delta_t$) of a hit relative to that
expected for a $\beta = 1$ particle can be used to determine the particle
\invbeta via the equation:
\begin{equation}
 1/\beta= 1+ \frac{c \delta_t}{L},
 \label{betatotof}
\end{equation}
where $L$ is the flight distance from the interaction point.
The track \invbeta value is calculated as the weighted
average of the \invbeta measurements from the DT and CSC hits associated with the track.
A description of how the DT and CSC systems measure the time of hits is given below.

As tubes in consecutive layers of DT chambers are staggered by half a tube, a typical track passes alternatively to the left and to the right of the
sensitive wires in consecutive layers.
The position of hits is inferred from the drift time of the ionization electrons assuming the hits come from a prompt muon.
For a late arriving HSCP, the delay will result in a longer drift time being attributed, so hits drifting
left will be to the right of their true position while hits drifting right will be to the left.
The DT measurement of $\delta_t$ then comes from the residuals of a straight line fit to the track hits in the chamber.  Only phi-projections from the DT chambers are used for this purpose.
The weight for the $i${th} DT measurement is given by:
\begin{equation}
 w_{i} = \frac{(n-2)}{n}\frac{L_{i}^{2}}{c^2 \sigma_{DT}^{2}},
\end{equation}
where $n$ is the number of $\phi$ projection measurements found in the
chamber from which the measurement comes
and $\sigma_{DT}= 3$\unit{ns} is the time resolution of the DT
measurements.
The factor $(n-2)/n$ accounts for the fact that residuals are computed
using two parameters of a straight line determined from the same
$n$ measurements (the minimum number of hits in a DT chamber
needed for a residual calculation is $n=3$).
Particles passing through the DTs have on average 16 time measurements.

The CSC measurement of $\delta_t$ is found by measuring the arrival time of the signals from both the cathode strips
and anode wires with respect to the time expected for prompt muons. The weight for the $i$th CSC measurement is given by:
\begin{equation}
 w_{i} =\frac{L_{i}^{2}}{c^2 \sigma_{i}^{2}},
\end{equation}
where $\sigma_{i}$, the measured time resolution, is 7.0\unit{ns}
for cathode strip measurements and 8.6\unit{ns} for anode wire measurements.
Particles passing through the CSCs have on average 30 time measurements, where cathode strip and anode wire measurements are counted separately.

The uncertainty ($\sigma_{1/\beta}$) on \invbeta for the track is found via the equation:
\begin{equation}
 \sigma_{1/\beta} = \sqrt{\sum_{i=1}^N \frac{(1/\beta_i - \overline{1/\beta})^2 \times w_{i}}{N-1}},
 \label{betaerr}
\end{equation}
where $\overline{1/\beta}$ is the average \invbeta of the track and N is the number of measurements associated with the track.

Several factors including the intrinsic time resolution of the subsystems,
the typical number of measurements per track, and the distance from the
interaction point lead to
a resolution of about $0.065$ for \invbeta in both the DT and CSC subsystems
over the full $\eta$ range.

\section{Data selection \label{sec:presel}}

Multiple search strategies are used to separate signal from background
depending on the nature of the HSCP under investigation.
\begin{itemize}
 \item For singly charged HSCPs,
  \begin{itemize}{}
   \item [$\circ$] the ``\tktof'' analysis requires tracks to be reconstructed in
      the inner tracker and the muon system,
   \item [$\circ$] the ``\tkonly'' analysis only requires tracks to be reconstructed
      in the inner tracker, and
   \item [$\circ$] the ``\muononly'' analysis only requires tracks to be reconstructed
      in the muon system.
  \end{itemize}
 \item For fractionally charged HSCPs, the ``\fractionalcharge'' ($|Q|<1e$) analysis only requires tracks to be reconstructed
      in the inner tracker and to have a \dedx smaller than SM particles.
 \item For multiply charged HSCPs, the ``\multicharge'' ($|Q|>1e$) analysis requires tracks to be reconstructed in
      the inner tracker and the muon system.  The analysis is optimized for
      much larger ionization in the detector compared to the \tktof\ analysis.
\end{itemize}

HSCP signal events have unique characteristics.
For each analysis, the primary background arises from SM particles
with random fluctuations in energy deposition/timing or
mis-measurement of the energy, timing, or momentum.

The \tkonly\ and \muononly\ cases allow for the possibility
of charge flipping (charged to neutral or vice versa) within
the calorimeter or tracker material.
The \muononly\ analysis is the first CMS search
that does not require an HSCP to be charged in the inner tracker.
The singly, multiply, and fractionally charged analyses feature
different selections, background estimates, and systematic uncertainties.
The preselection requirements for the analyses are described below.

All events must pass
a trigger requiring either the reconstruction of (i) a muon with high
\pt  or (ii) large missing transverse
energy (\MET) defined as the magnitude of the vectorial sum
of the transverse momenta of all particles reconstructed by
an online particle-flow algorithm~\cite{Chatrchyan:2011ds} at the HLT.

The L1 muon trigger allows for late arriving particles (such
as slow moving HSCPs) by accepting tracks that produce signals
in the RPCs within either the 25\unit{ns} time window corresponding
to the interaction bunch crossing or the following 25\unit{ns} time
window.  For the data used in this analysis, the second 25\unit{ns} time
window is empty of proton-proton collisions because of the 50\unit{ns} LHC bunch spacing during the 2011 and 2012 operation.

Triggering on \MET\ allows for some recovery of events with hadron-like HSCPs in which
none of the $R$-hadrons in the event are charged in both the inner tracker and the muon system.
The \MET\ in the event arises because the particle-flow algorithm rejects
tracks not consistent with a SM particle. This rejection includes tracks reconstructed only in the inner tracker with a track \pt
much greater than the matched energy deposited in the calorimeter~\cite{ref:PAS-PFT-09-001}
as would be the case for $R$-hadrons becoming neutral in the calorimeter,
and tracks reconstructed only in the muon system as would be the case for $R$-hadrons that are initially neutral.
Thus, in both cases, only the energy these HSCPs deposit in the calorimeter, roughly 10--20\GeV, will be included in the \MET\ calculation.
In events in which no HSCPs are reconstructed as muon candidates, significant \MET\
will result if the vector sum of the HSCPs' momenta is large.  The \MET\ trigger will collect these events, allowing for sensitivity
to HSCP without a muon-like signature.

For all the analyses, the muon trigger
requires $\pt > 40$\GeVc measured in the inner tracker
and the \MET\ trigger requires
$\MET > 150$\GeV at the HLT.
The \muononly\ analysis uses the same two triggers, and additionally
a third trigger that requires both a reconstructed muon segment
with $\pt > 70$\GeVc (measured using only the muon system) and $\MET > 55$\GeV.
For the first part of the 2012 data (corresponding to an integrated luminosity of 700\pbinv), the
requirement was $\MET > 65$\GeV.
Using multiple triggers in all of the analyses allows for increased sensitivity
to HSCP candidates that arrive late in the muon system and to
hadron-like HSCPs that are sometimes charged in only one of the inner tracker
and muon system and sometimes charged in both.
The \muononly\ analysis uses only
$\sqrt{s} = 8$\TeV data as the necessary triggers were not available in 2011.

For the \tkonly\ analysis, all events are required
to have a track candidate in the region $|\eta| < 2.1$ with
$\pt > 45$\GeVc (as measured in the inner tracker).  In addition, a relative
uncertainty in \pt  ($\sigma_{\PT}/\PT$) less than 0.25 and
a track fit $\chi^2$ per number of degrees of freedom ($n_{d}$) less than 5 is required.
Furthermore, the magnitudes of the longitudinal ($d_z$) and transverse ($d_{xy}$) impact
parameters are both required to be less than 0.5\unit{cm}.
The impact parameters $d_z$ and $d_{xy}$ are both defined
with respect to the primary vertex that yields the smallest $|d_z|$ for the candidate track.
The requirements on the impact parameters are very loose compared with the resolutions for tracks ($\sigma(d_{xy,z})<0.1\unit{cm}$) in the inner tracker.
Candidates must pass isolation requirements in the tracker and
calorimeter.  The tracker isolation requirement is
$\Sigma \pt < 50$\GeVc where the sum is over all tracks (except the candidate's track)
within a cone about the candidate track
$\Delta R = \sqrt{(\Delta \eta)^2 + (\Delta \phi)^2} < 0.3$ radians.
The calorimeter isolation requirement is $E/p < 0.3$ where $E$
is the sum of energy deposited in the calorimeter towers
within $\Delta R < 0.3$ (including the candidate's energy deposit) and $p$ is the candidate track momentum reconstructed
from the inner tracker.
Candidates must have at least two measurements in the silicon
pixel detector and
at least eight measurements in the combination of the silicon strip and pixel detectors.
In addition, there must be measurements in at least 80\% of
the silicon layers between the first and last measurements on
the track.  To reduce the rate of contamination from clusters
with large energy deposition due to overlapping tracks, a
"cleaning" procedure is applied to remove clusters in the
silicon strip tracker that are not
consistent with passage of only one charged particle (\eg,
a narrow cluster with most of the energy deposited in one or
two strips).
After cluster cleaning, there must be at least six measurements
in the silicon strip detector that are used for the \dedx\
calculation.
Finally, $I_h > 3\MeVcm$ is required.

The \tktof\ analysis applies the same criteria, but
additionally requires a reconstructed muon matched to a track
in the inner detectors.  At least eight independent time measurements
are needed for the TOF computation.
Finally, $1/\beta > 1$ and $\sigma_{1/\beta} < 0.07$ are required.

The \muononly\ analysis uses separate criteria that include
requiring a reconstructed track in the muon system
with $\pt > 80\GeVc$ within
$|\eta| < 2.1$. The relative resolution in \pt  is approximately 10\% in the barrel
region and approaches 30\% for $|\eta| > 1.8$~\cite{MUO-10-004}.
However, charge flipping by $R$-hadrons can lead to an overestimate of \pt.
The effect is more pronounced for gluinos, where all of the electric charge
comes from SM quarks. The measured curvature in the muon system for gluinos is 60--70\% smaller than would be expected for a muon with the same transverse momentum.
The candidate track must have measurements in two or more DT or CSC stations,
and $|d_z|$ and $|d_{xy}| < 15$\unit{cm} (calculated using tracks from the
muon system and measured relative to the nominal beam spot position rather than to the
reconstructed vertex). The requirements on $|d_z|$ and $|d_{xy}|$ are approximately 90\% and 95\%
efficient for prompt tracks, respectively.
HSCPs are pair produced and often back-to-back in $\phi$ but not in $\eta$ because the collision is in general boosted along the $z$-axis. On the other hand, cosmic
ray muons passing close to the interaction point would pass through the top and bottom halves of CMS, potentially giving the appearance of two tracks back-to-back in both $\phi$ and $\eta$.
Often only one of these legs will be reconstructed as a track while the other will leave only a muon segment (an incomplete track) in the detector.
To reject cosmic ray muons, candidates are removed if there is a muon
segment both with $\eta$ within $\pm$0.1 of $-\eta_{\textnormal{cand}}$, where
$\eta_{\textnormal{cand}}$ is the pseudorapidity of the HSCP candidate, and with
$|\delta\phi| > 0.3$ radians, where $\delta\phi$ is the difference in $\phi$ between the candidate and the muon segment.
The $|\delta\phi|$ requirement prevents candidates with small $|\eta|$ from being rejected by their proximity to their own muon segments.
Additionally, candidates compatible with vertically downward cosmic ray muons, $1.2 <  |\phi| < 1.9$\unit{radians}, are rejected.
To reject muons from adjacent beam crossings, tracks are removed if their time leaving the interaction point
as measured by the muon system is within ${\pm}5\unit{ns}$ of a different LHC beam crossing. This veto makes the background
from muons from such crossings negligible while removing very little signal.
Finally, the same quality requirements used in the \tktof\ analysis are applied in the muon-only $1/\beta$ measurement.

The \fractionalcharge\ search uses the same preselection
criteria as the \tkonly\ analysis except
that \ih\ is required to be $<$2.8\MeVcm.
An additional veto on cosmic ray muons rejects candidates if a track with $\pt >
40\GeVc$ is found on the opposite side of the detector ($\Delta R > \pi - 0.3$).

The \multicharge\ particle search uses the same
preselection as the \tktof\ analysis except
that the $E/p$ selection is removed.
Furthermore, given that a multiply charged particle might have a cluster shape different from that of a $|Q|=1e$ particle,
the cluster cleaning procedure is not applied for the \multicharge\ analysis.

The preselection criteria applied on the inner tracker track for the analyses are summarized in Table~\ref{tab:preselectionTk}
while the criteria on the muon system track are summarized in Table~\ref{tab:preselectionSA}.

\begin{table}
 \begin{center}
  \topcaption{Preselection criteria on the inner tracker track used in the various analyses as defined in the text.
     \label{tab:preselectionTk}}
  \begin{tabular}{|l|c|c|c|c|} \hline
                                            & $|Q|<1e$ & \tktof\ & \tkonly\  &  $|Q|>1e$    \\ \hline
   $|\eta|$                                 & \multicolumn{4}{c|}{${<}2.1$}                            \\ \hline
   $\pt$ (\GeVcns{})                            & \multicolumn{4}{c|}{${>}45$}               \\ \hline
   $d_z$ and $d_{xy}$ (cm)                  & \multicolumn{4}{c|}{${<}0.5$}              \\ \hline
   $\sigma_{\pt}/\pt$                       & \multicolumn{4}{c|}{${<}0.25$}             \\ \hline
   Track $\chi^2/{n_d}$           & \multicolumn{4}{c|}{${<}5$}                    \\ \hline
   \# Pixel hits                            & \multicolumn{4}{c|}{${>}1$}                \\ \hline
   \# Tracker hits                          & \multicolumn{4}{c|}{${>}7$}                \\ \hline
   Frac. Valid hits                         & \multicolumn{4}{c|}{${>}0.8$}              \\ \hline
   $\Sigma \pt^\text{trk} (\Delta R < 0.3)$ (\GeVcns{})& \multicolumn{4}{c|}{${<}50$}             \\ \hline
   \# \dedx\ measurements                   & \multicolumn{4}{c|}{${>}5$}                \\ \hline
   \dedx\ strip shape test                  & \multicolumn{3}{c|}{yes}       & no       \\ \hline
   $E_\text{cal}(\Delta R < 0.3)/p$              & \multicolumn{3}{c|}{${<}0.3$}   & $-$      \\ \hline
   \ih\ (\!\MeV/cm)                            & ${<}2.8$ & \multicolumn{3}{c|}{ ${>}3.0$}    \\ \hline
   $\Delta R$ to another track              & ${<}\pi - 0.3$           & \multicolumn{3}{c|}{$-$}       \\ \hline
  \end{tabular}
 \end{center}
\end{table}

\begin{table}
 \begin{center}
  \topcaption{Preselection criteria on the muon system track used in the various analyses as defined in the text.
     \label{tab:preselectionSA}}
  \begin{tabular}{|l|c|c|c|} \hline
                                            & \tktof\  &  $|Q|>1e$  & \muononly\     \\ \hline
   \# TOF measurements                      & \multicolumn{3}{c|}{${>}7$}   \\ \hline
   $\sigma_{1/\beta}$                       & \multicolumn{3}{c|}{${<}0.07$}\\ \hline
   $1/\beta$                                & \multicolumn{3}{c|}{${>}1$}   \\ \hline
   $|\eta|$                                 & \multicolumn{2}{c|}{$-$}              & ${<}2.1$              \\ \hline
   $\pt$ (\GeVc)                            & \multicolumn{2}{c|}{$-$}              &  ${>}80$     \\ \hline
   $d_z$ and $d_{xy}$ (cm)                  & \multicolumn{2}{c|}{$-$}              & ${<}15$      \\ \hline
   \# DT or CSC stations                    & \multicolumn{2}{c|}{$-$}                   & ${>}1$      \\ \hline
   Opp. segment $|\eta|$ difference         & \multicolumn{2}{c|}{$-$}                   & ${>}0.1$    \\ \hline
   $|\phi|$                                 & \multicolumn{2}{c|}{$-$}                   & ${<}1.2$ OR ${>}1.9$    \\ \hline
   $|\delta t|$ to another beam crossing (ns) & \multicolumn{2}{c|}{$-$}                   & ${>}5$    \\ \hline
  \end{tabular}
 \end{center}
\end{table}

\section{Background prediction \label{sec:bkgpred}}

Candidates passing the preselection criteria (Section~\ref{sec:presel})
are subject to two (or three) additional selection criteria to further improve
the signal-to-background ratio.
For all of the analyses, results are based upon a comparison of the number of candidates passing the final section criteria with the number of background events estimated from the numbers of events that fail combinations of the criteria.

\begin{figure}%[tbhp]
 \begin{center}
  \includegraphics[clip=true, trim=0.0cm 0cm 2.8cm 0cm, width=0.32\textwidth]{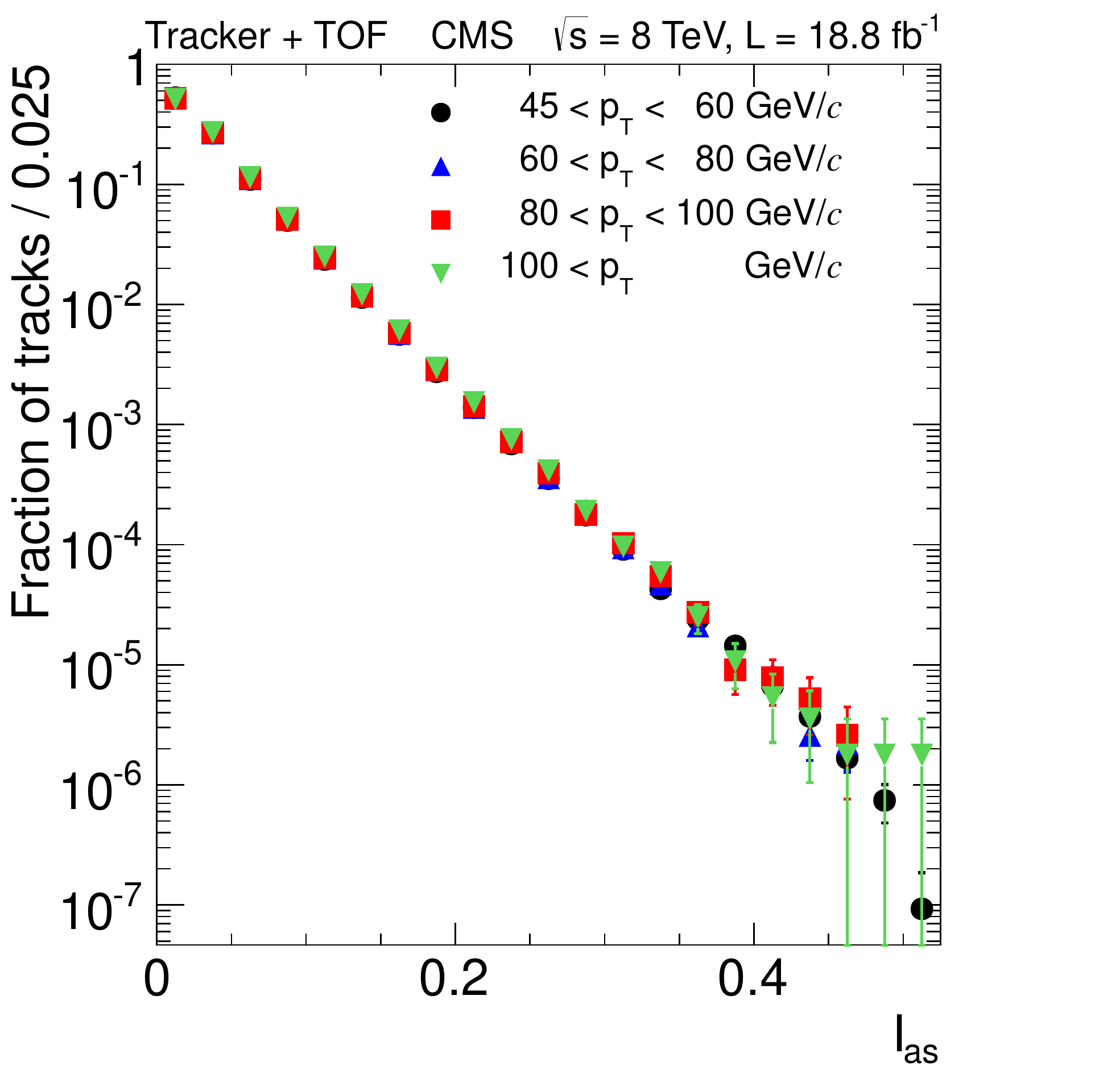}
  \includegraphics[clip=true, trim=0.0cm 0cm 2.8cm 0cm, width=0.32\textwidth]{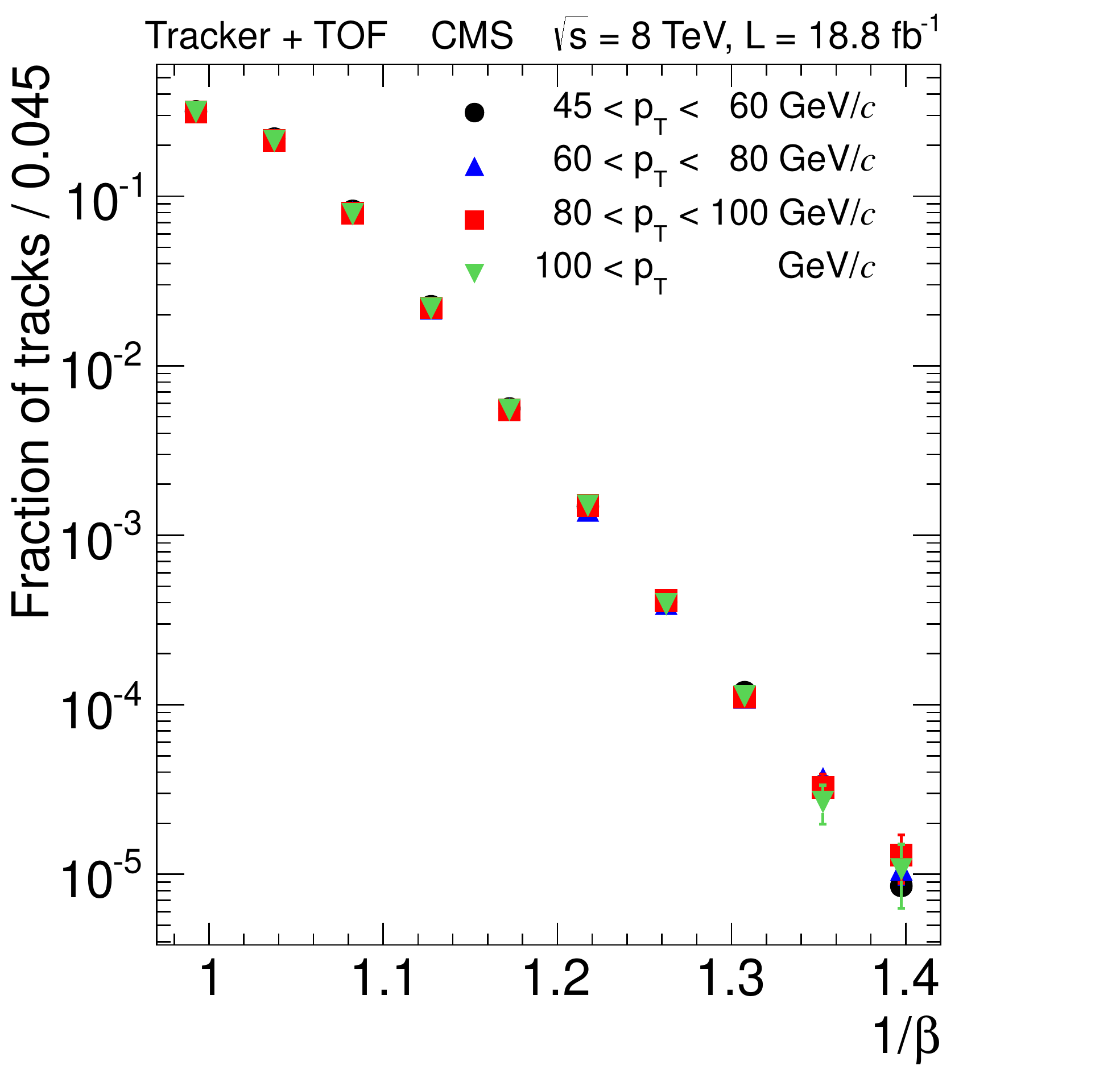}
  \includegraphics[clip=true, trim=0.0cm 0cm 2.8cm 0cm, width=0.32\textwidth]{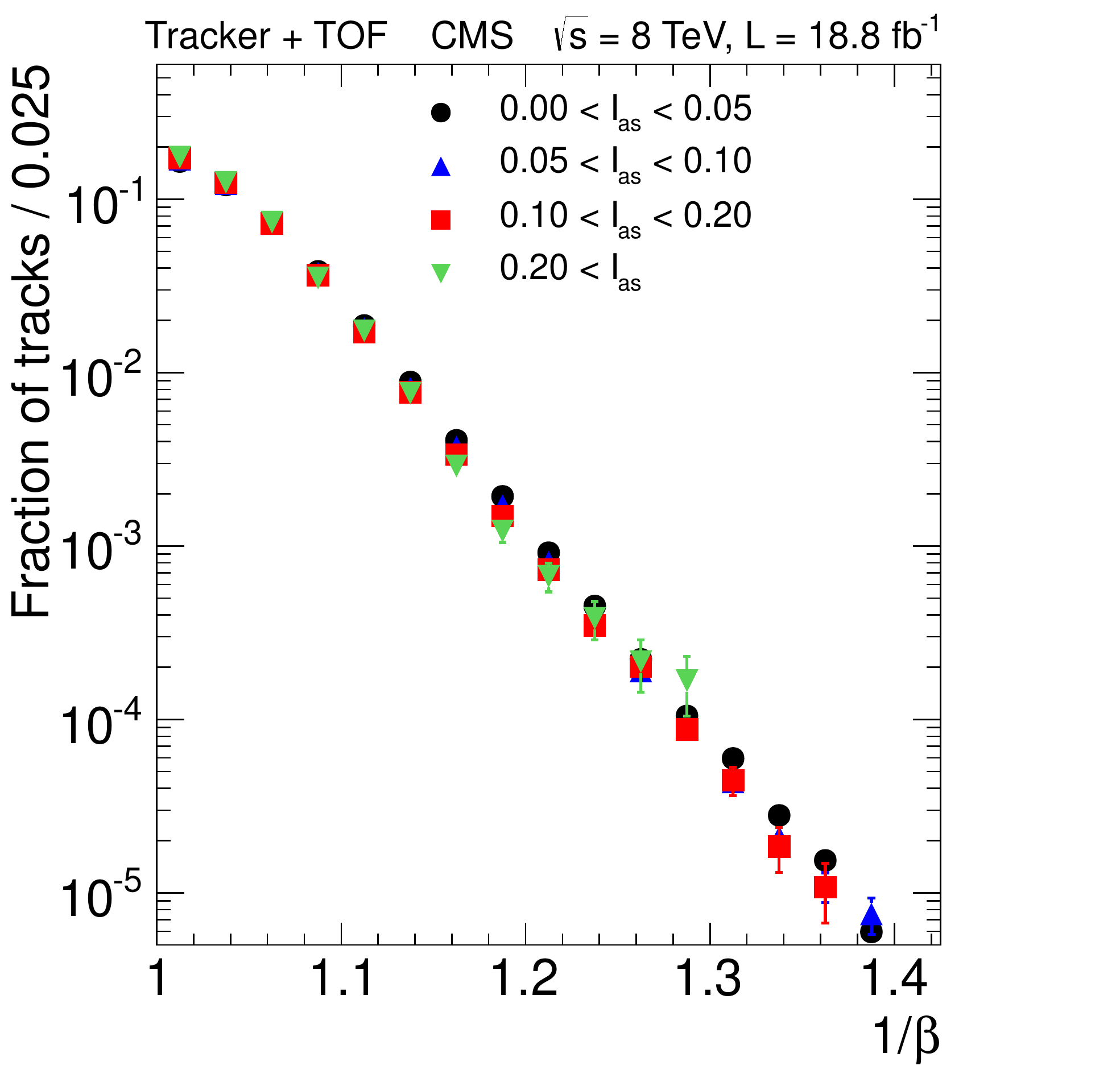}
 \end{center}
 \caption{Measured $I_{as}$ (left) and $1/\beta$ (middle) distributions for several \pt ranges and measured $1/\beta$ distributions for several $I_{as}$  ranges (right).
Results are for the \tktof\ selection at $\sqrt{s} = 8$\TeV.  The lack of variation of the distributions for different ranges of the other variables
demonstrates the lack of strong correlation between \invbeta, \ias, and \pt.
    \label{fig:Uncorrelation}}
\end{figure}

The background expectation in the signal region, $D$, is estimated as $D = BC/A$, where $B$ ($C$) is the number of candidates that fail
the first (second) criteria but pass the other one and $A$ is the number of candidates that fail both criteria. The method works if the probability for a background
candidate to pass one of the criteria is not correlated with whether it passes the other criteria. The lack of strong correlation between
the selection criteria is evident in Fig.~\ref{fig:Uncorrelation}.
Tests of the background prediction (described below) are used to
quantify any residual effect and to calculate the systematic error in the background estimate.
All tracks passing the preselection enter either the signal region $D$ or one of the control regions that is used for the background prediction.

For the \tkonly\ analysis, the two chosen criteria are
\pt  and \ias.  Threshold values ($\pt > 70$\GeVc and
$\ias >0.4$) are fixed such that failing
candidates passing only \pt(\ias) fall into the $B$ ($C$) regions.
The $B$ ($C$) candidates are then used to form a binned probability
density function in \ih($p$) such that, using the mass
determination (Eq.~(\ref{eq:MassFromHarmonicEstimator})),
the full mass spectrum of the background in the
signal region $D$ can be predicted.  The $\eta$
distribution of candidates at low \dedx\ differs from the
distribution of the candidates at high \dedx.
To correct for this effect, events in the $C$ region are weighted
such that the $\eta$ distribution matches that in the $B$ region.

For the \tktof\ analysis, three criteria are used,
\pt, \ias, and \invbeta, creating eight regions labeled $A$ through $H$.
The final threshold values are selected to be $\pt > 70$\GeVc,
$\ias> 0.125$, and $\invbeta > 1.225$.
 Region $D$ represents the signal region, with events passing all three criteria.
The candidates in the $A$, $F$, and $G$ regions pass only the \invbeta, \ias, and \pt criteria, respectively,
while the candidates in the $B$, $C$, and $H$ regions fail only the \pt, \ias, and \invbeta
criteria, respectively.  The $E$ region fails all three criteria.  The
background estimate can be made from several different combinations
of these regions.  The combination $D = AGF/E^2$ is used because it yields
the smallest statistical uncertainty.  Similar to the \tkonly\
analysis, events in the $G$ region are reweighted to match the
$\eta$ distribution in the $B$ region.
From a consideration of the observed spread in
background estimates from the other combinations, a 20\% systematic
uncertainty is assigned to the background estimate.
The 20\% systematic uncertainty is also assigned to the background estimate for the tracker-only analysis.

\begin{figure}%[tbhp]
 \begin{center}
  \includegraphics[clip=true, trim=0.0cm 0cm 2.8cm 0cm, width=0.44\textwidth]{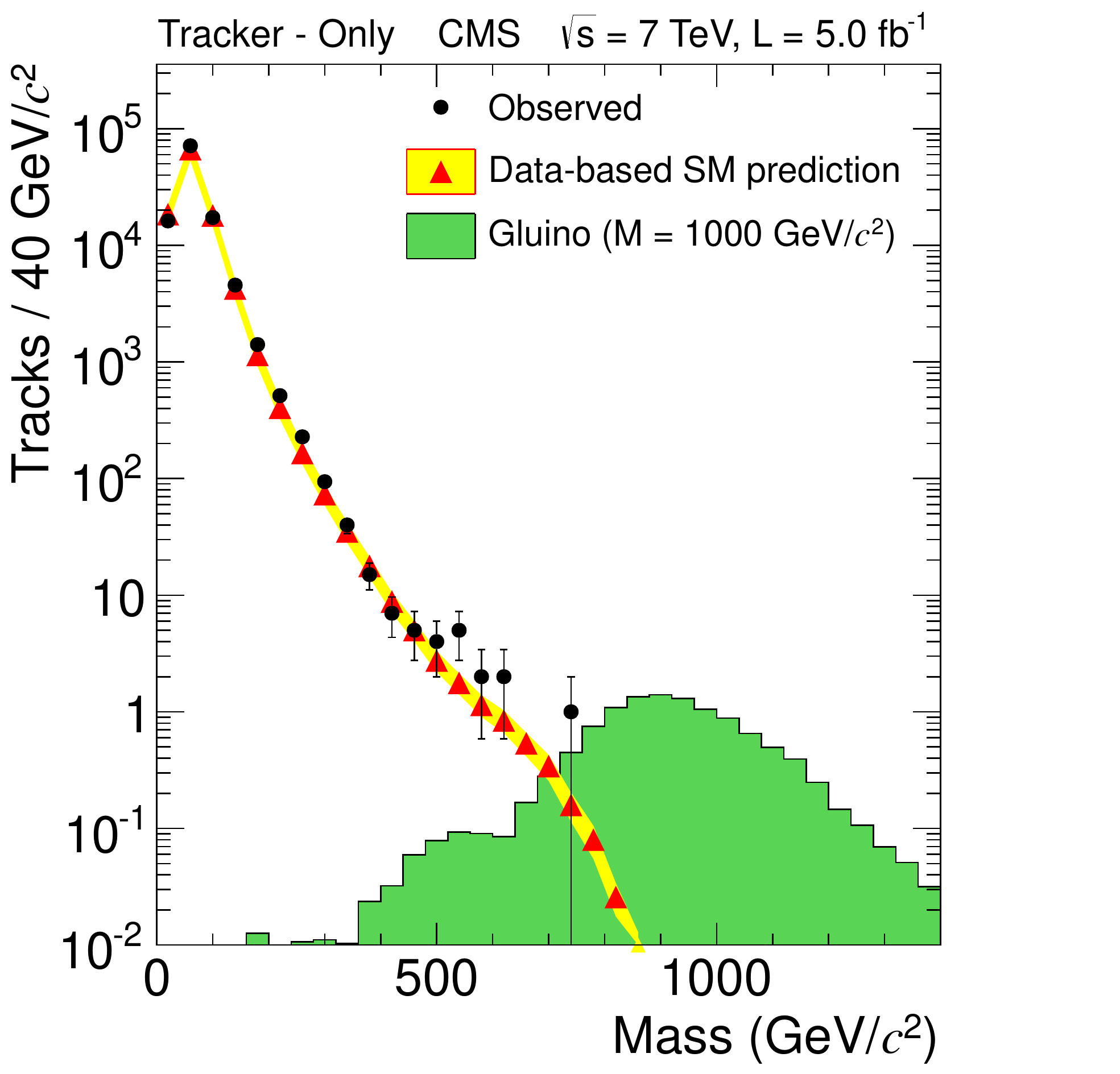}
  \includegraphics[clip=true, trim=0.0cm 0cm 2.8cm 0cm, width=0.44\textwidth]{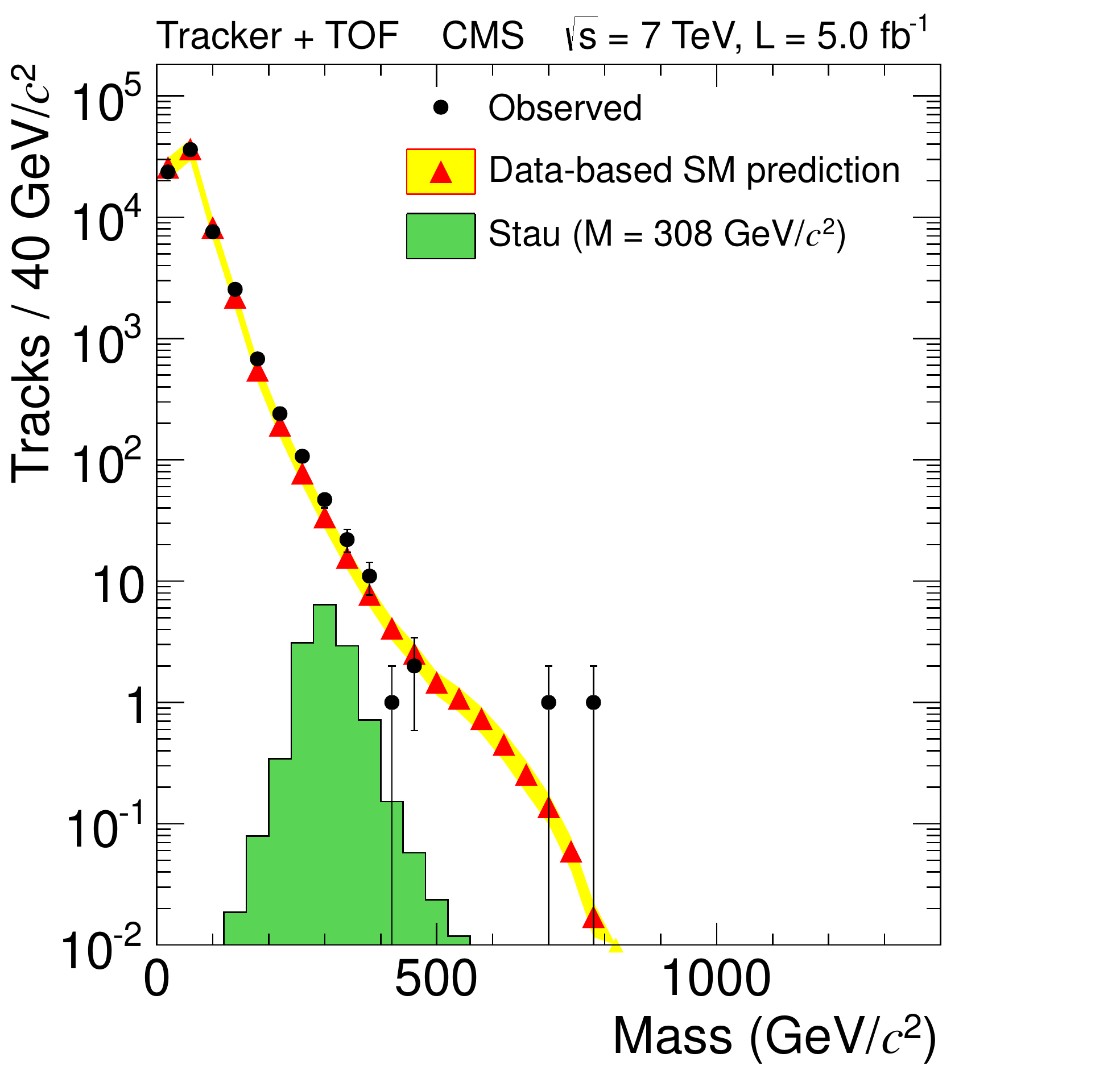} \\
  \includegraphics[clip=true, trim=0.0cm 0cm 2.8cm 0cm,width=0.44\textwidth]{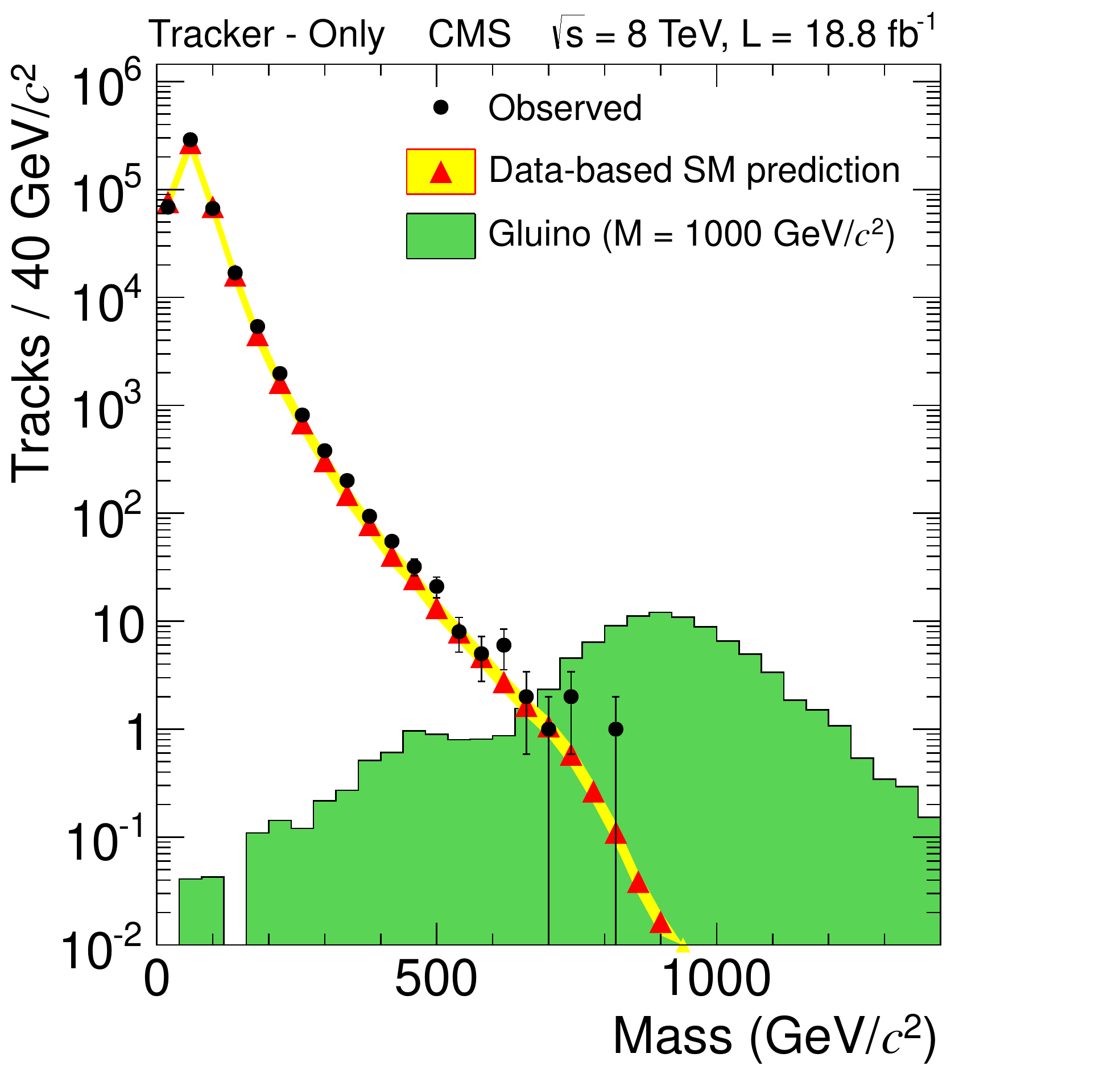}
  \includegraphics[clip=true, trim=0.0cm 0cm 2.8cm 0cm,width=0.44\textwidth]{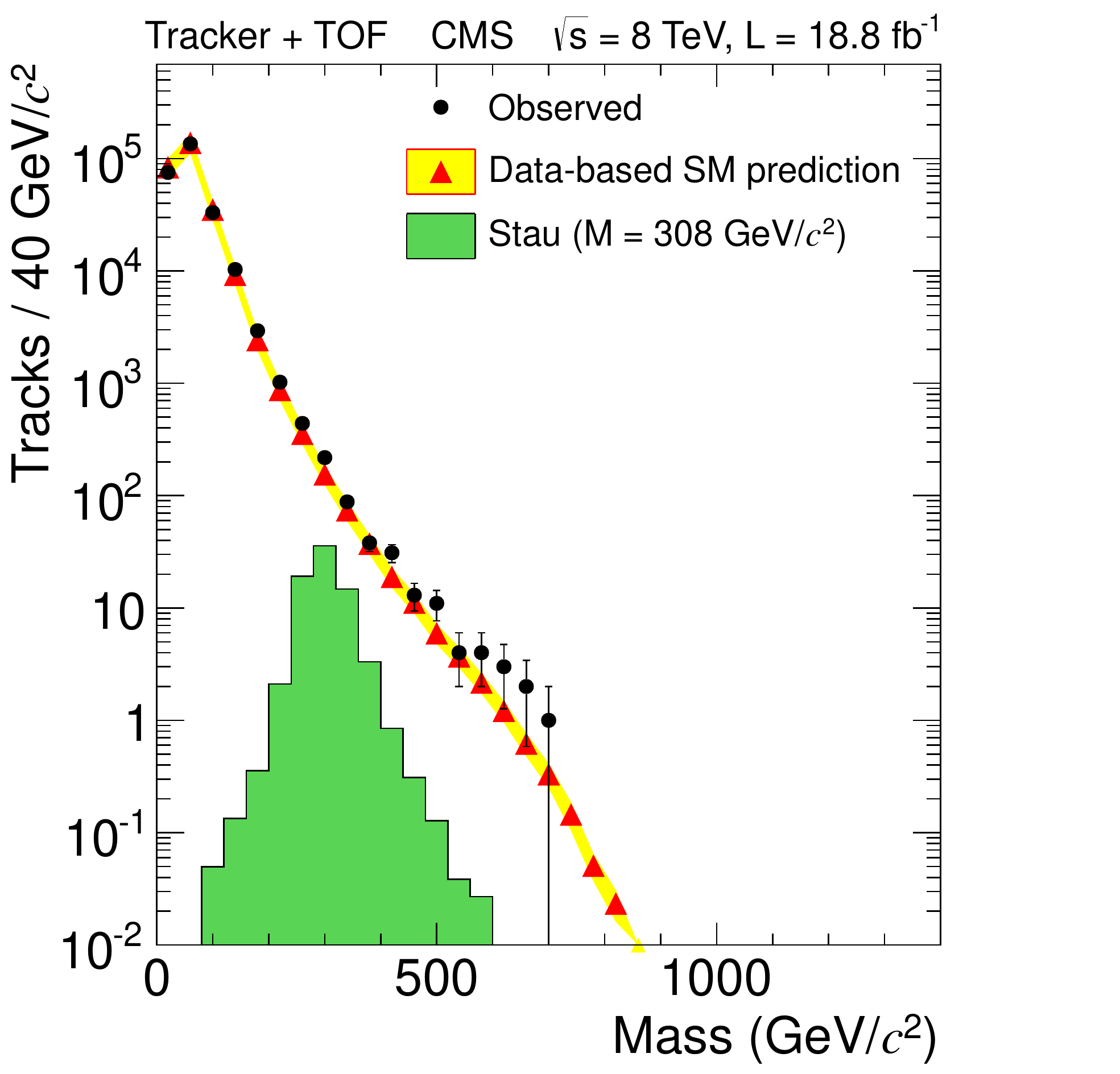}
 \end{center}
 \caption{Observed and predicted mass spectra for candidates
    entering the \tkonly\ (left column) or \tktof\ (right column) signal
    region for the loose selection.
    The expected distribution for a representative signal scaled to the integrated luminosity is shown as the shaded
    histogram.
    The top (bottom) row is for $\sqrt{s} = 7$ (8)\TeV.
    \label{fig:LooseMassDistribution}}
\end{figure}

In order to check the background prediction, loose selection samples,
which would be dominated by background tracks, are used for the \tkonly\ and \tktof\
analyses.
The loose selection sample for the \tkonly\ analysis is defined
as $\pt>50$\GeVc and $I_{as}>0.10$.  The loose selection
sample for the \tktof\ analysis is defined as
$\pt>50$\GeVc, $I_{as}>0.05$, and $1/\beta>1.05$.
Figure~\ref{fig:LooseMassDistribution} shows the observed and predicted
mass spectra for these samples.

The \muononly\ analysis uses the \pt  and \invbeta
criteria for the $ABCD$ method.  The final selections are
$\pt  > 230$\GeVc and $\invbeta> 1.4$.  It has been found that
these variables are correlated with $|\eta|$
and with the number of muon stations used to fit the candidate.
Therefore, the background prediction is performed in
six separate bins (2/3/4 muon stations in central ($|\eta| < 0.9$) and forward ($0.9 < |\eta| < 2.1$) regions).  The
final result is computed from a sum of these six bins.
The systematic uncertainty in this background estimate is
determined by defining four additional regions
$A^\prime, B^\prime, C^\prime,$ and $D^\prime$.  Events in
$B^\prime$ ($A^\prime$) pass (fail) the \pt  requirement, but
with $0.8 < 1/\beta < 1.0$, while events in $D^\prime$ ($C^\prime$)
pass (fail) the \pt  requirement with $1/\beta < 0.8$.
Two complementary predictions now become possible,
$D = CB^\prime/A^\prime$ and $D = CD^\prime/C^\prime$.
From a consideration of the spread of the three estimates, a systematic uncertainty of
20\% is assigned to the background estimate for the \muononly\
analysis using this method.

\begin{figure}
 \begin{center}
  \includegraphics[clip=true, trim=0.0cm 0cm 2.7cm 0cm, width=0.44\textwidth]{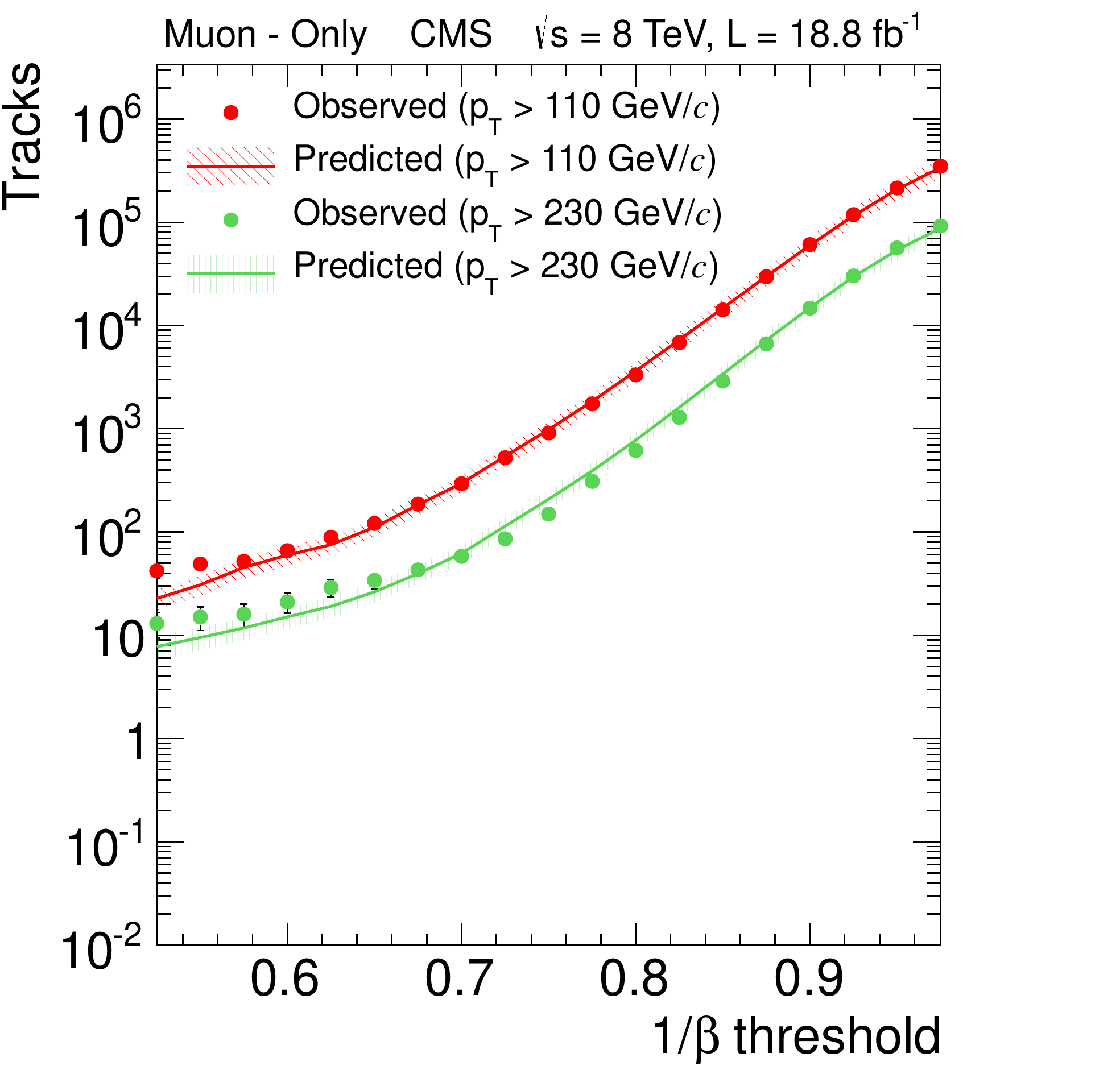}
  \includegraphics[clip=true, trim=0.0cm 0cm 2.7cm 0cm, width=0.44\textwidth]{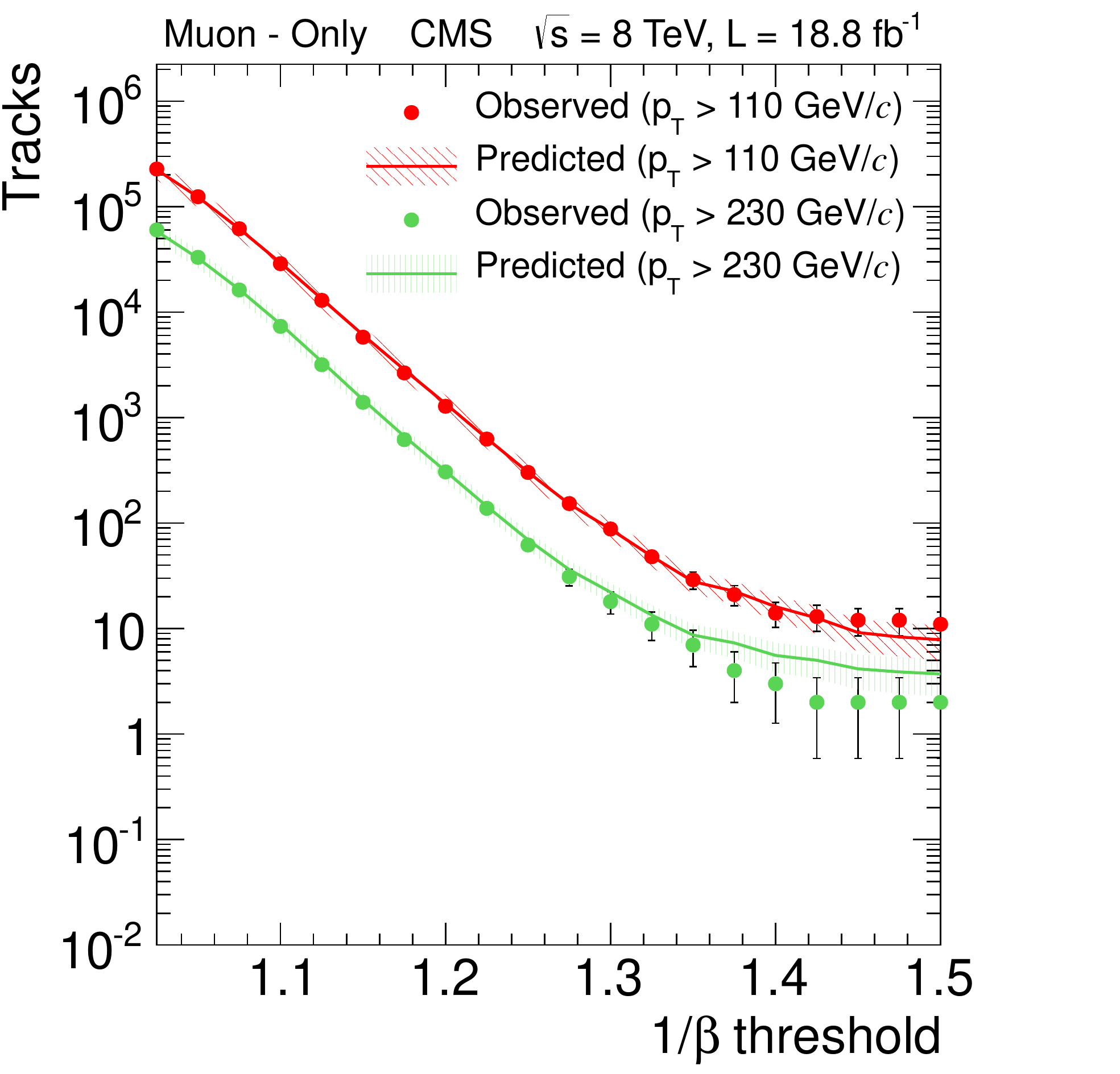}
 \end{center}
 \caption{Observed and predicted numbers of tracks in both the control region with $1/\beta < 1$ (left) and the signal region (right)
as functions of the \invbeta threshold and for two different \pt  thresholds for the \muononly\ analysis at $\sqrt{s} = 8\TeV$. Only statistical uncertainties are shown.
    \label{fig:PredVsObsMuOnly}}
\end{figure}

The \muononly\ analysis also has background contributions from
cosmic ray muons even after the previously mentioned cosmic ray muon veto requirements are applied.
The number of cosmic ray muons expected to pass the selection criteria is determined
by using the sideband region of $70 < |d_{z}| < 120$\unit{cm}. To increase the number of cosmic ray muons
in the sideband region, the veto requirements are not applied here.
To reduce the contamination in the sideband region due to muons from collisions, the tracks are
required to not be reconstructed in the inner tracker.
The number of tracks ($N$) in the sideband with $1/\beta$ greater than the threshold is counted.
To determine the ratio ($R_{\mu}$) of candidates in the signal region with respect to
the sideband region, a pure cosmic ray sample is used.
The sample is collected using a
trigger requiring a track from the muon system with $\pt > 20$\GeVc, rejecting
events within ${\pm}50$\unit{ns} of a beam crossing and events triggered as beam halo.
The cosmic ray muon contribution to the \muononly\ analysis signal region is determined as
$N \times R_{\mu}$. A similar procedure is used to subtract the estimated cosmic ray muon contribution to the
$A$, $B$, and $C$ regions prior to estimating the collision muon background
in the $D$ region.
The cosmic ray muon contribution to the
signal region constitutes approximately 60\% of the total expected background.
The systematic uncertainty in the
cosmic ray muon contribution is determined by comparing
estimates using $|d_{z}|$ ranges of 30--50\unit{cm}, 50--70\unit{cm},
70--120\unit{cm}, and $>$120\unit{cm}.  It is found to be 80\%.
Figure~\ref{fig:PredVsObsMuOnly} shows the numbers of predicted and observed candidates in both the control region with $1/\beta<1$ and the signal region for various
\pt and $1/\beta$ thresholds for the $\sqrt{s} = 8\TeV$ data.
The number of predicted events includes both the cosmic ray muon and collision muon contributions. Only statistical uncertainties are shown.

\begin{figure}
 \begin{center}
 \includegraphics[clip=true, trim=0.0cm 0cm 2.7cm 0cm, width=0.44\textwidth]{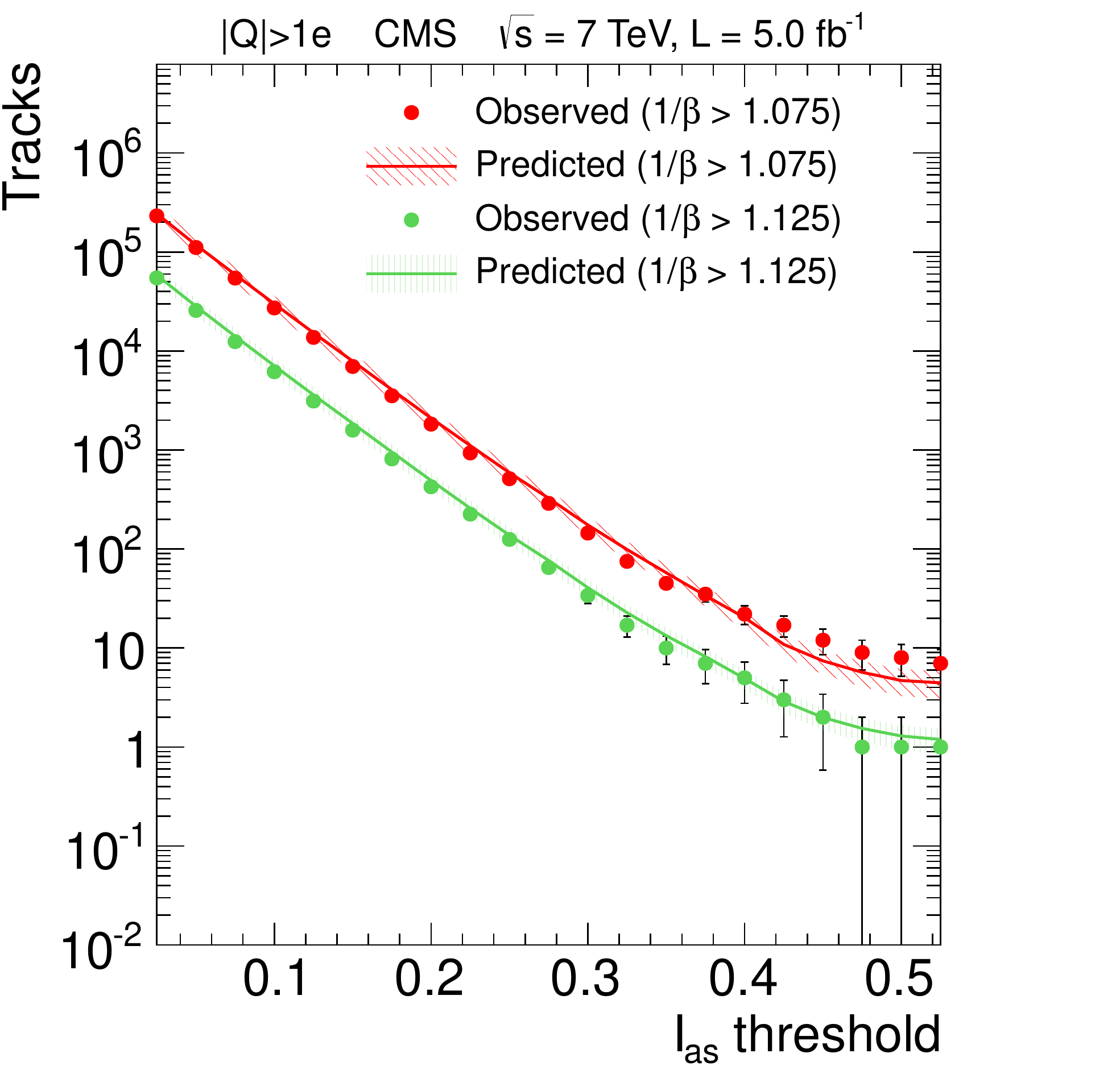}
  \includegraphics[clip=true, trim=0.0cm 0cm 2.7cm 0cm, width=0.44\textwidth]{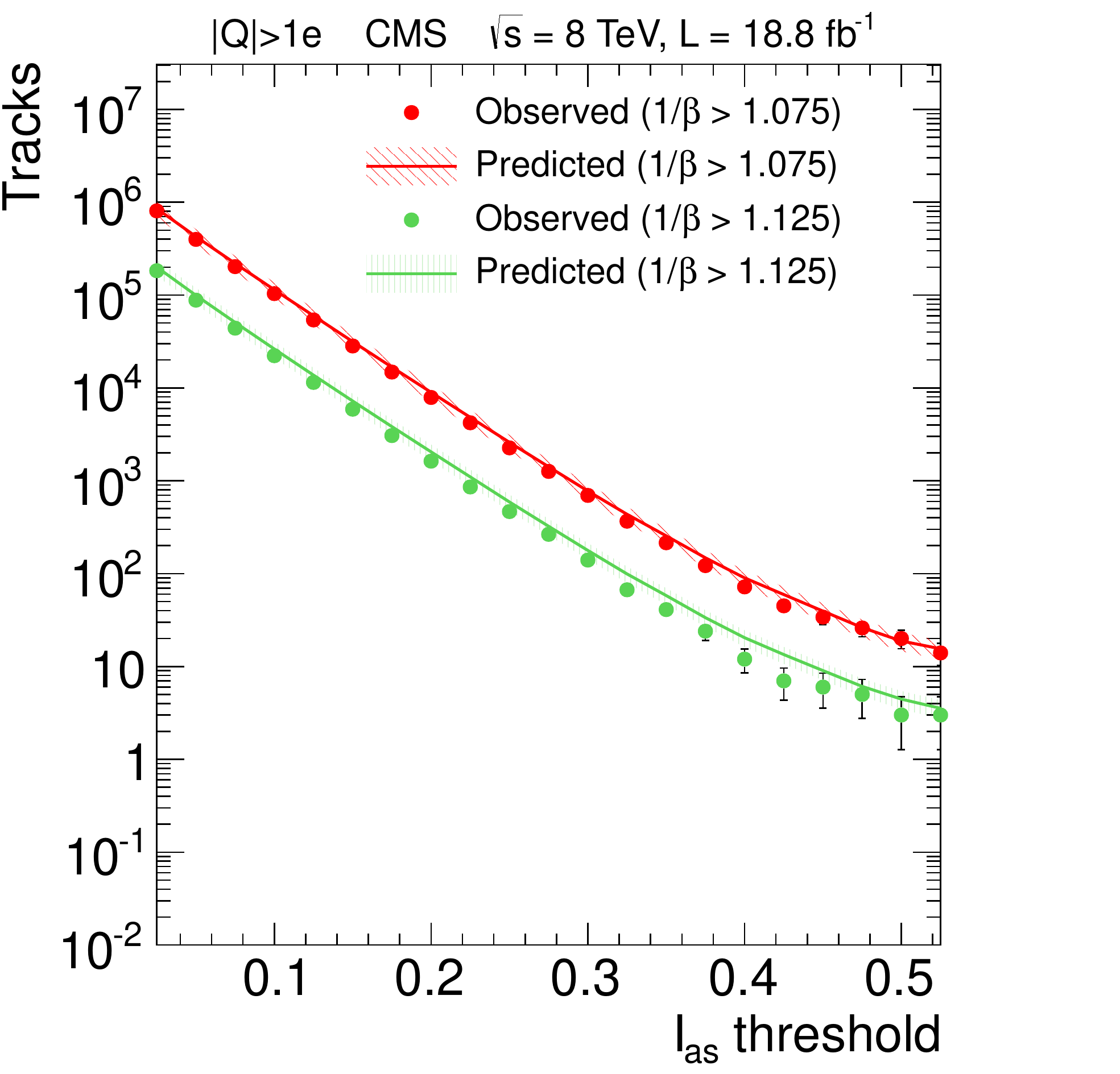}
 \end{center}
 \caption{Observed and predicted numbers of tracks as a function of the \ias\ threshold for two different \invbeta thresholds at $\sqrt{s} = 7\TeV$ (left) and $8\TeV$ (right) for the \multicharge\ analysis.
Only statistical uncertainties are shown.
    \label{fig:PredVsObsmCHAMP}}
\end{figure}

The \multicharge\ analysis uses the \ias\ and \invbeta
criteria.  Since the default track reconstruction code assumes
$|Q|=1e$ for \pt  determination, the transverse momentum for
$|Q|>1e$ particles is underestimated by a factor of $1e/|Q|$.  Therefore
\pt  is not used in the final selection.  In addition, while
\dedx\ scales as $Q^2$, the dynamic range of the silicon readout of individual strips
saturates for energy losses $\approx$3 times that of a $\beta \approx 1$, $|Q|=1e$ particle.
Since both the \pt scaling and the \dedx\ saturation effects can bias the reconstructed
mass to lower values (less separation from background),
the reconstructed mass is not used for this analysis.
Despite the saturation effect, $|Q|>1e$ particles have a larger
incompatibility of their \dedx\ measurements with the MIP
hypothesis, increasing the separation power of the \dedx\
discriminator for multiply charged particles, relative to that for $|Q|=1e$ HSCPs.
The systematic uncertainty in the background estimate for
the \multicharge\ analysis is
determined by the same method that is used for the collision muon background
in the \muononly\ analysis except with \pt  changed to be \ias.  The two complementary estimates from
the $1/\beta < 1.0$ region lead to a 20\% uncertainty.
Figure~\ref{fig:PredVsObsmCHAMP} shows the numbers of predicted and observed candidates for various \ias\ and \invbeta thresholds.
Only the statistical uncertainties are shown.

\begin{figure}
 \begin{center}
 \includegraphics[clip=true, trim=0.0cm 0cm 2.7cm 0cm, width=0.44\textwidth]{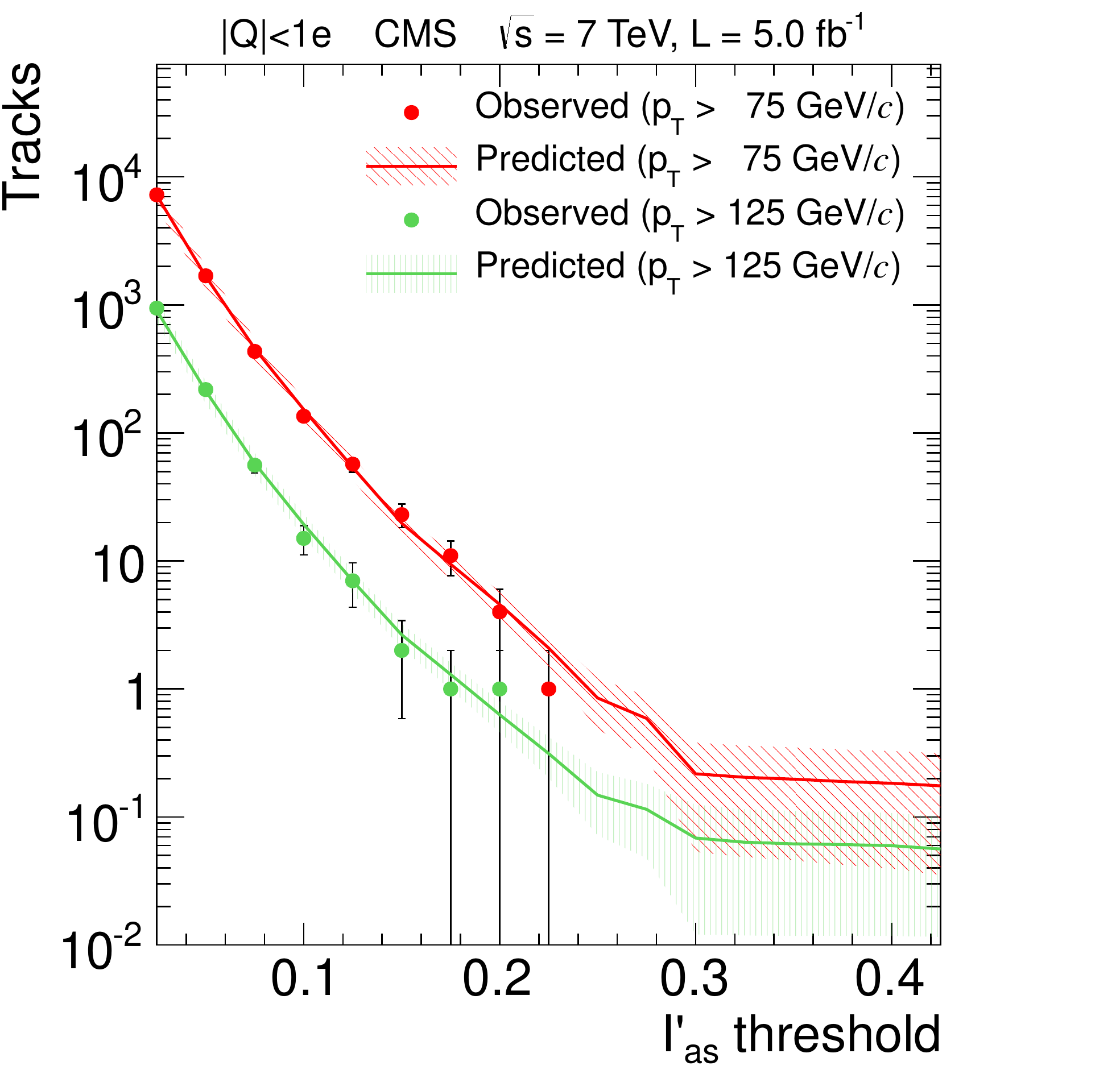}
 \includegraphics[clip=true, trim=0.0cm 0cm 2.7cm 0cm, width=0.44\textwidth]{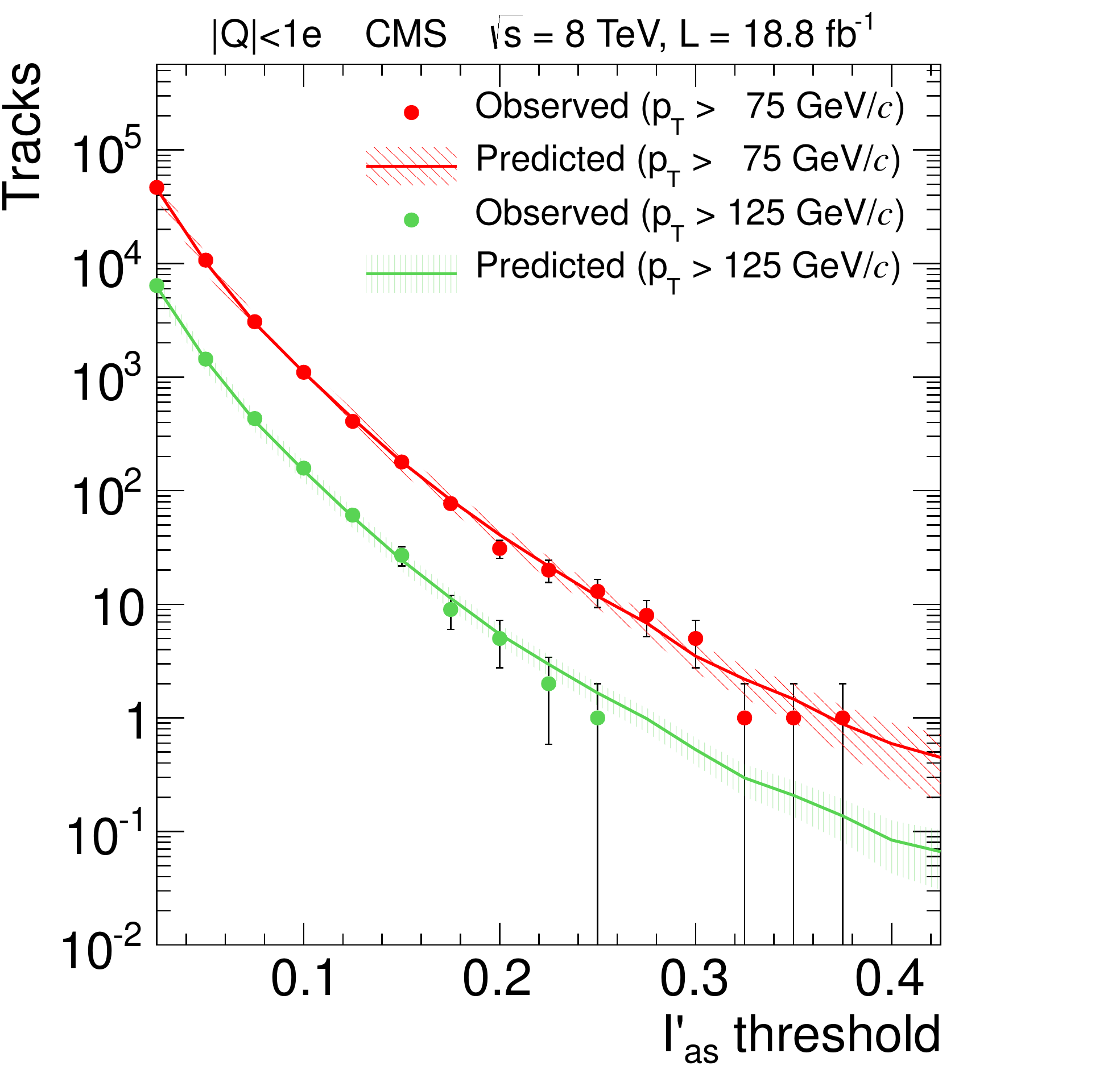}
 \end{center}
 \caption{Observed and predicted numbers of tracks as a function of the \iasp\ threshold for two different  \pt  thresholds at $\sqrt{s} = 7\TeV$ (left) and 8\TeV (right) for the \fractionalcharge\ analysis.
Only statistical uncertainties are shown.
    \label{fig:PredVsObsfrac}}
\end{figure}

The \fractionalcharge\ analysis uses the same
method to estimate the background as the \tkonly\
analysis, replacing the \ias\ variable with \iasp\ and not applying a mass requirement.
The systematic uncertainty in the prediction is taken from the \tktof\ analysis.
In addition, the cosmic ray muon background is studied, since
particles passing through the tracker not synchronized with the
LHC clock often produce tracker hits with low energy readout.
The cosmic ray muon background is found to be small and a 50\%
uncertainty is assigned to this prediction.
The numbers of predicted and observed candidates for various \pt  and \ias\ thresholds can be seen in Fig.~\ref{fig:PredVsObsfrac}.
Only the statistical uncertainties are shown.

For each analysis, fixed selections on the appropriate set of \ias,
\iasp, \pt, and \invbeta are used to define the final signal
region (and the regions for the background prediction).
These values are chosen to give discovery potential over
the signal mass regions of interest.  For the \tkonly\ and
\tktof\ analyses, an additional requirement on the reconstructed
mass is applied.  The mass requirement depends upon the HSCP
signal.  For a given model and HSCP mass, the range is $M_\text{reco} - 2\sigma$ to 2\TeVcc where $M_\text{reco}$ is
the average reconstructed mass for the given HSCP mass and
$\sigma$ is the expected resolution.
Both $M_\text{reco}$ and $\sigma$ are determined from simulation.

\begin{figure}%[tbhp]
\begin{center}
 \includegraphics[clip=true, trim=0.0cm 0cm 2.8cm 0cm, width=0.44\textwidth]{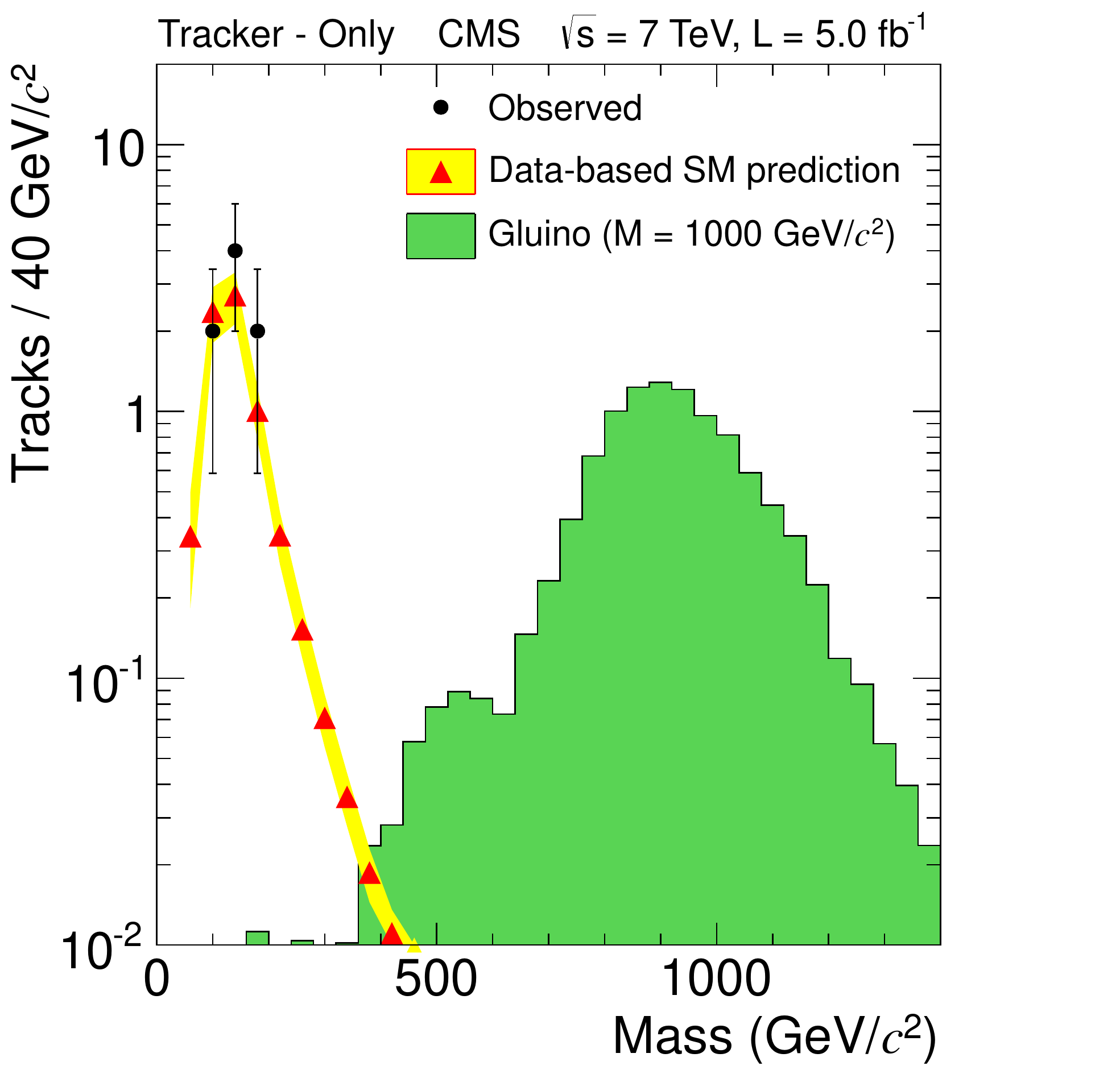}
 \includegraphics[clip=true, trim=0.0cm 0cm 2.8cm 0cm, width=0.44\textwidth]{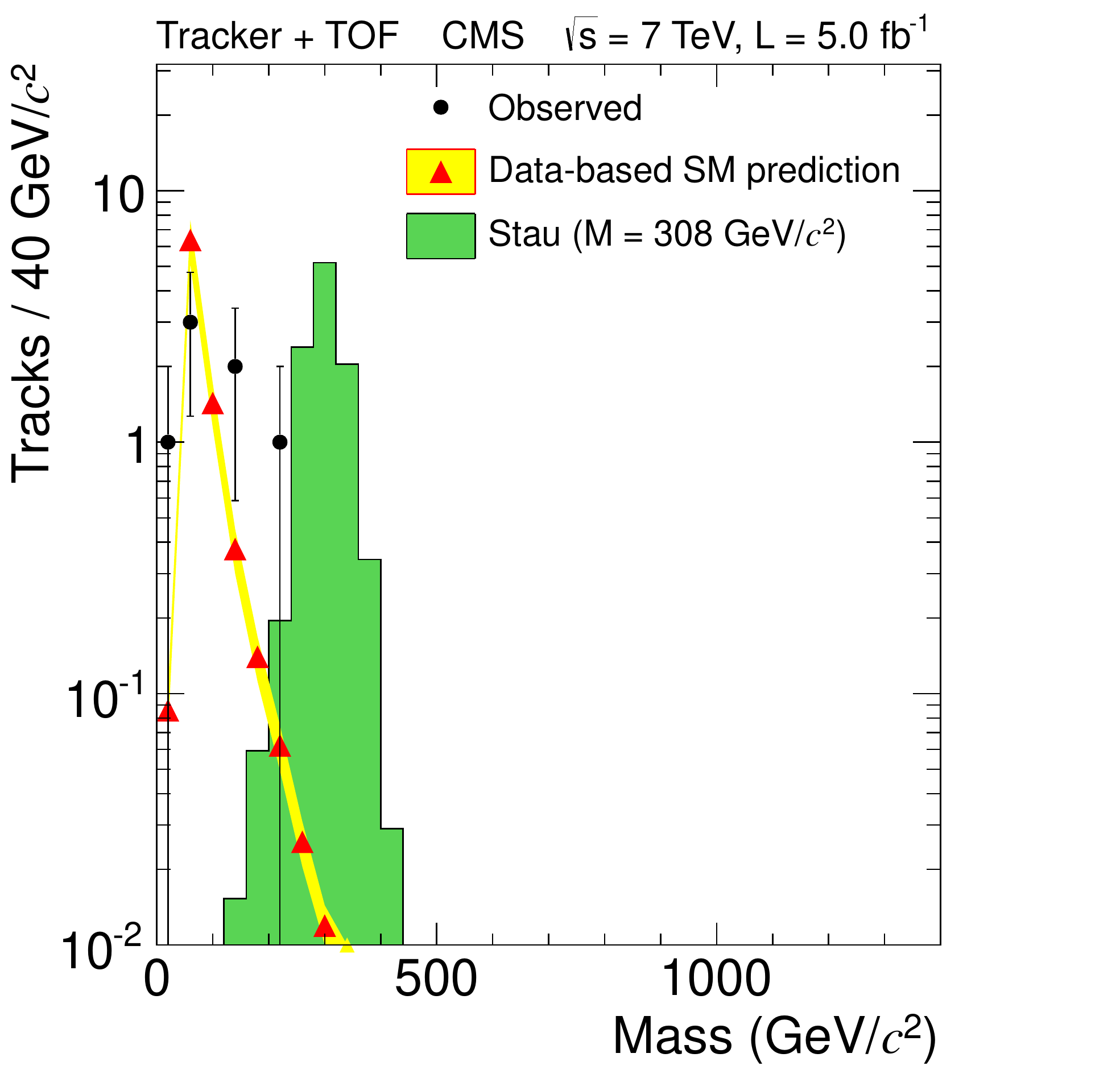} \\
 \includegraphics[clip=true, trim=0.0cm 0cm 2.8cm 0cm,width=0.44\textwidth]{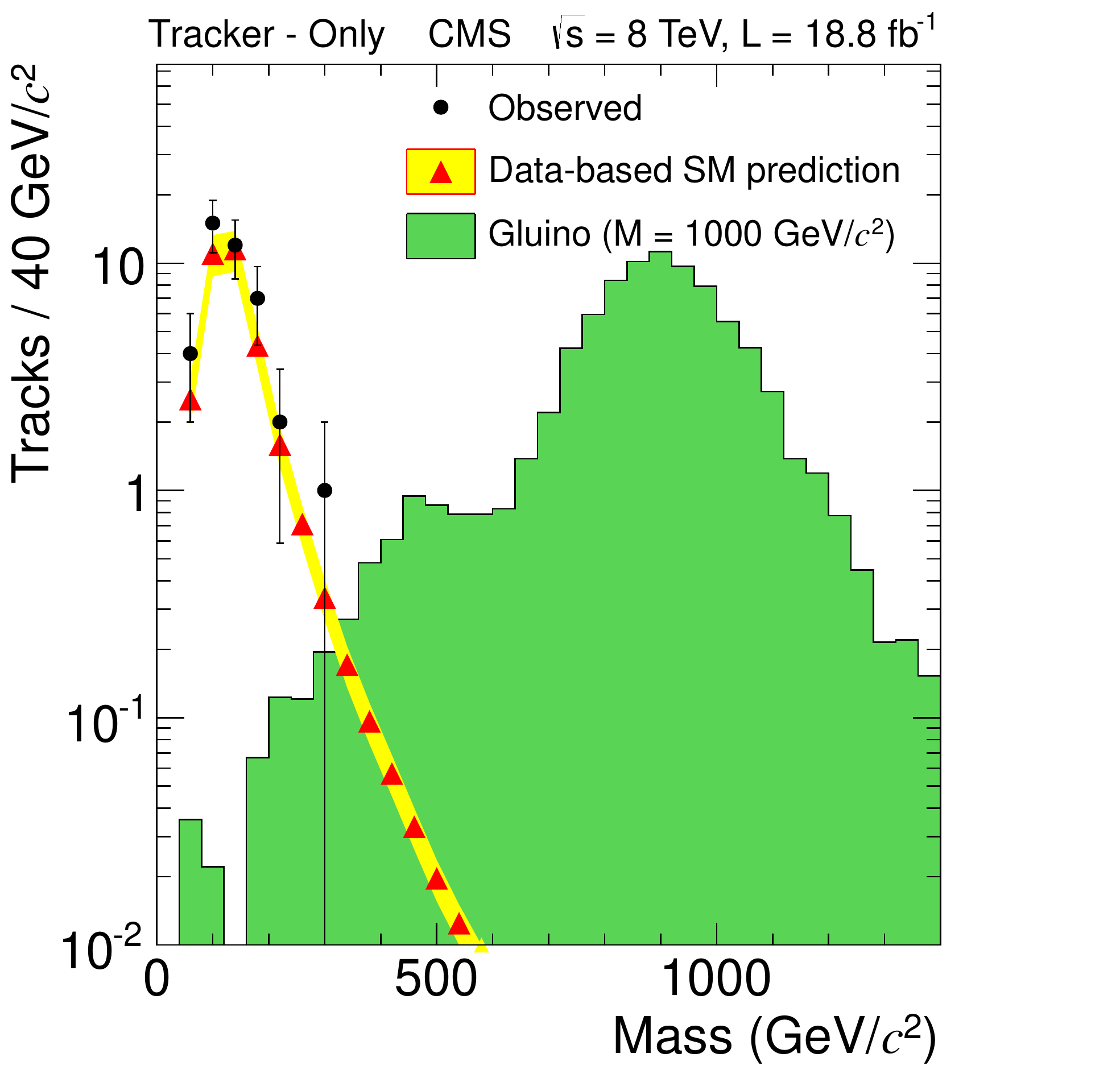}
 \includegraphics[clip=true, trim=0.0cm 0cm 2.8cm 0cm,width=0.44\textwidth]{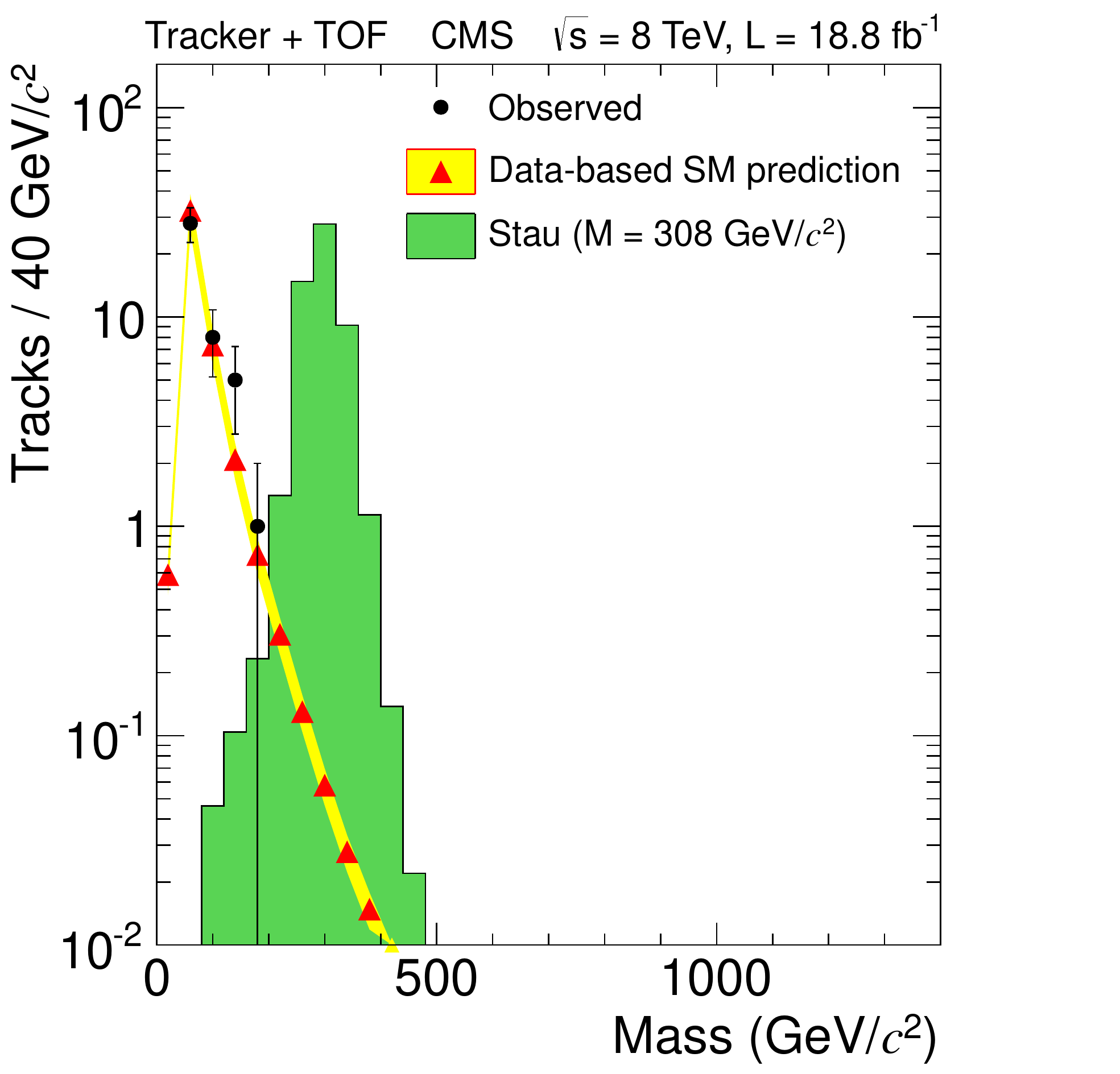}
 \end{center}
 \caption{Observed and predicted mass spectra for candidates
    entering the \tkonly\ (left column) or \tktof\ (right column) signal
    region for the tight selection.
    The expected distribution for a representative signal scaled to
    the integrated luminosity is shown as the shaded histogram.
    The top (bottom) row is for $\sqrt{s} = 7$ (8)\TeV.
    \label{fig:TightMassDistribution}}
\end{figure}

Table~\ref{tab:finalsel} lists the final selection criteria, the predicted
numbers of background events, and the numbers of events observed in
the signal region.  Agreement between prediction and observation
is seen over the full range of analyses. Figure~\ref{fig:TightMassDistribution} shows the observed and predicted mass
distributions for the \tkonly\ and \tktof\ analyses with the tight selection.
The bump at lower mass values expected from the signal MC is due to the saturation of the strip electronic readout.

\begin{table}
 \topcaption{Results of the final selections for the predicted background and
   the observed numbers of events. The uncertainties include both statistical and systematic contributions.
   \label{tab:finalsel}}
 \begin{center}
  \small
 \begin{tabular}{|p{1.8cm}|c|c|c|c|cc|cc|} \hline
                             & \multicolumn{4}{c|}{~}                               & \multicolumn{4}{c|}{Number of events} \\
                             & \multicolumn{4}{c|}{Selection criteria}              & \multicolumn{2}{c|}{$\sqrt{s}=7$\TeV}&\multicolumn{2}{c|}{$\sqrt{s}=8$\TeV} \\ \hline
                             & \pt                   & \multirow{2}{*}{$I_{as}^{(\prime)}$} & \multirow{2}{*}{$1/\beta$} & Mass    & \multirow{2}{*}{Pred.} & \multirow{2}{*}{Obs.}             & \multirow{2}{*}{Pred.}          & \multirow{2}{*}{Obs.}      \\
                             & (\GeVc)                 &                            &                            & (\GeVcc)&                &     &                &       \\ \hline
   \multirow{4}{*}{\hspace{-0.1cm}Tracker-only  }& \multirow{4}{*}{${>}70$} & \multirow{4}{*}{${>}0.4$}   & \multirow{4}{*}{$-$}       & ${>}0$ & $7.1\pm1.5$    & $8$ & $33\pm7$   & $41$ \\
                             &                         &                            &                            & ${>}100$ & $6.0\pm1.3$    & $7$ & $26\pm5$   & $29$ \\
                             &                         &                            &                            & ${>}200$ & $0.65\pm0.14$  & $0$ & $3.1\pm0.6$    & $3$  \\
                             &                         &                            &                            & ${>}300$ & $0.11\pm0.02$  & $0$ & $0.55\pm0.11$  & $1$  \\
                             &                         &                            &                            & ${>}400$ & $0.030\pm0.006$& $0$ & $0.15\pm0.03$  & $0$ \\ \hline
  \multirow{4}{*}{\hspace{-0.1cm}Tracker+\tof\ }& \multirow{4}{*}{${>}70$} & \multirow{4}{*}{${>}0.125$} & \multirow{4}{*}{${>}1.225$} & ${>}0$ & $8.5\pm1.7$    & $7$ & $44\pm9$   & $42$ \\
                             &                         &                            &                            & ${>}100$ & $1.0\pm0.2$    & $3$ & $5.6\pm1.1$    & $7$  \\
                             &                         &                            &                            & ${>}200$ & $0.11\pm0.02$  & $1$ & $0.56\pm0.11$  & $0$  \\
                             &                         &                            &                            & ${>}300$ & $0.020\pm0.004$& $0$ & $0.090\pm0.02$ & $0$ \\ \hline
   \hspace{-0.1cm}Muon-only  & ${>}230$                 &  $-$                       & ${>}1.40$                   &    $-$  & $-$            & $-$ & $6\pm3$    & $3$ \\ \hline
   \hspace{-0.1cm}$|Q|>1e$   &           $-$           &       ${>}0.500$            &        ${>}1.200 $          &    $-$  & $0.15\pm0.04$  & $0$ & $0.52\pm 0.11$ & $1$ \\ \hline
   \hspace{-0.1cm}$|Q|<1e$   & ${>}125$                 & ${>}0.275$                   &            $-$             &    $-$  & $0.12\pm0.07$  & $0$ & $1.0\pm0.2$  & $0$ \\ \hline
 \end{tabular}
 \normalsize
 \end{center}
\end{table}

\section{Systematic uncertainties}

The sources of systematic uncertainty include
those related to the integrated luminosity, the background prediction, and the
signal acceptance.  The uncertainty in the integrated
luminosity is 2.2\% (4.4\%) at
$\sqrt{s} = 7$ (8)\TeV~\cite{SMP-12-008, LUM-12-001}.
The uncertainties in the background predictions are
described in Section~\ref{sec:bkgpred}.

The signal acceptance is obtained from MC
simulations of the various signals processed through
the full detector simulation (Section~\ref{sec:signals}).
Systematic uncertainties in the final results are
dominated by uncertainties in the differences between
the simulation and data evaluated in control samples.  The relevant differences
are discussed below.
A summary of the systematic uncertainties is given in Table~\ref{tab:systuncertainties}.

The trigger acceptance is dominated by the muon triggers
for all the models except for the charge-suppressed scenarios.  The uncertainty in the muon
trigger acceptance arises from several effects.  A difference
of up to 5\% between data and MC simulation events has been
observed~\cite{MUO-10-004}.  For slow moving particles, the effect
of the timing synchronization of the muon system is tested by shifting
the arrival times in simulation to match the synchronization offset and width observed in data, resulting
in an acceptance change of 2\% (4\%) for $\sqrt{s}=7$ (8)\TeV.
For the $|Q|<1e$ samples, an additional uncertainty arises from
the possibility of losing hits because their ionization in the muon
system is closer to the hit threshold.  The uncertainty in the gains in
the muon system is evaluated by shifting the gain by 25\%, yielding an acceptance change of 15\% (3\%) for $|Q|=e/3 (2e/3)$
samples.  The uncertainty in the \MET\ trigger acceptance is found
by varying, at HLT level, the energy of simulated jets by the scale uncertainties.
The \MET\ uncertainty for $\sqrt{s}=7\TeV$ samples is estimated
to be less than 2\% for all scenarios except for the charge-suppressed ones,
where it is estimated to be ${<}5\%$.  For $\sqrt{s}=8$\TeV samples
it is less than 1\% for all samples.

The energy loss in the silicon tracker is important for
all the analyses except for the \muononly\ one.
Low-momentum protons are used to quantify the
agreement between the observed and simulated distributions
for \ih\ and \ias.  The \dedx\ distributions of signal samples
are varied in the simulation by the observed differences, in order to determine the systematic uncertainty.
Because the \fractionalcharge\
analysis is also sensitive to changes to the number of
hits on the track, track reconstruction is also performed
after shifting \dedx.
The uncertainty in the signal acceptance varies by
less than 24\% for the $|Q|=1e$ samples, being less than 10\% for all
masses above 200\GeVcc.  For the $|Q|<1e$
samples, the effect of the \dedx\ shift and the track
reconstruction combined is 25\% ($<$10\%) for
$|Q|=e/3 (2e/3)$.  The $|Q|>1e$ samples have sufficient
separation of the signal from the final \ias\ selection
that the effect of the \dedx\ shift is negligible.

The Z boson decays to muons are used to test the MC simulation of the \invbeta measurement.
At $\sqrt{s}=7\TeV$, the \invbeta measurement was observed to have
a disagreement of 0.02 in the CSC system and 0.003 in the DT system.
At $\sqrt{s}=8\TeV$ a disagreement of 0.005 is observed for both systems.
The uncertainty in the signal acceptance is
estimated to be between 0 and 15\% by shifting \invbeta by these amounts.  The
uncertainty is generally less than 7\% except for the
high-charge/low-mass samples in the \multicharge\
analysis.

The uncertainties in the efficiencies for muon reconstruction~\cite{MUO-10-004}
and track reconstruction~\cite{CMS-PAS-TRK-10-002} are
less than 2\% each.  The track momentum uncertainty for the \muononly\ analysis is determined by shifting
$1/$\pt of muon system tracks by 10\%.  %~\cite{MUO-10-004}.
For all other analyses, the momentum from the inner tracker track is
varied as in Ref.~\cite{Chatrchyan:2012sp}.  The uncertainty is estimated to be ${<}$5\% for all but the $|Q|<1e$ samples, low-mass $|Q|>1e$ samples, and the \muononly\ scenarios,
where the uncertainty is less than 10\%.

The uncertainty in the number of pileup events
is evaluated by varying by 5-6\% the minimum bias cross section used to
calculate the weights applied to signal events in order to reproduce the pileup observed in data.
This results in uncertainties due to pileup of less than 4\%.

The uncertainty in the amount of material in the detector simulation
results in an uncertainty in the signal trigger and reconstruction
acceptance, particularly for the $|Q|>1e$ samples.  This is
evaluated by increasing the amount of material in the hadronic
calorimeter by a conservative 5\%~\cite{Chatrchyan:2009si}.
Since it was not practical to evaluate the effect in detail for each value of $Q$ considered, the largest change in signal acceptance observed ($\sim$20\%) was assigned to all $|Q|>1e$ scenarios.
The change in signal acceptance is $\leq1\%$ for all $|Q|\leq1e$ scenarios.

The total systematic uncertainty in the signal acceptance for the \tkonly\ analysis is less than 32\% and is less than 11\% for all of the gluino and scalar top cases.
For the \tktof\ analysis it is less than 15\% for all cases except for $|Q|=2e/3$, where the uncertainty ranges from 15\% to 31\%, being
larger at low masses. The \muononly\ analysis has an uncertainty in the signal acceptance in the range of 7--13\%. The \multicharge\ analysis has an uncertainty in the signal
acceptance in the range of 21--29\% for $|Q|>1e$ samples and 7--13\% for $|Q|=1e$ samples with both being larger at low masses.
The \fractionalcharge\ analysis has an uncertainty in the signal acceptance of 31\% and 12\% for $|Q|=e/3$ and $2e/3$ samples, respectively.

The statistical uncertainty in the signal acceptance is small compared to the total systematic uncertainty for all the cases except for the low-mass highly charged scenarios, where the low acceptance leads to a statistical uncertainty that is comparable with the systematic uncertainties.  For example, in the $|Q|=6e$, $M=100$\GeVcc signal, the statistical uncertainty is as high as 30\%.  In all cases the statistical uncertainty is taken into account when setting limits on signal cross sections.

\begin{table}
 \topcaption{Systematic uncertainties for the various HSCP searches.
   \label{tab:systuncertainties}}
 \begin{center}
 \begin{tabular}{|ll|c|c|c|c|c|} \hline
   \multicolumn{2}{|l|}{Signal acceptance} & $|Q|<1e$ & Tracker-only & Tracker+\tof\ & $|Q|>1e$ & Muon-only \\ \hline

\multicolumn{2}{|l|}{~---~~Trigger acceptance}             & ${<}16\%$ & ${<}7\%$ & ${<}7\%$ & ${<}6\%$ & $~~~~~7\%$ \\ \hline
\multicolumn{2}{|l|}{~---~~Track momentum scale}           & $<10\%$ & ${<}4\%$ & ${<3}\%$ & ${<}10\%$ & ${<}10\%$ \\ \hline
\multicolumn{2}{|l|}{~---~~Track reconstruction eff.}      & \multirow{2}{*}{${<}25\%$} & ${<}2\%$ & ${<}2\%$ & ${<}2\%$ & $-$ \\ \cline{1-2} \cline{4-7}
\multicolumn{2}{|l|}{~---~~Ionization energy loss}         &       & ${<}18\%$ & ${<}15\%$ & ${<}12\%$ & $-$ \\ \hline
\multicolumn{2}{|l|}{~---~~Time-of-flight}                 & $-$ & $-$ & ${<}2\%$ & ${<}15\%$ & ${<}3\%$ \\ \hline
\multicolumn{2}{|l|}{~---~~Muon reconstruction eff.}       & $-$ & $-$ & $~~~~~2\%$ & $~~~~~2\%$ & $~~~~~2\%$ \\ \hline
\multicolumn{2}{|l|}{~---~~Pile-up}                        & ${<}2\%$ & ${<}2\%$ & ${<}2\%$ & ${<}2\%$ & ${<}4\%$ \\ \hline
\multicolumn{2}{|l|}{~---~~Detector material}              & ${<}1\%$ & ${<}1\%$ & ${<}1\%$ & $~~~~~20\%$ & ${<}1\%$ \\ \hline \hline

  \multicolumn{2}{|l|}{Total signal acceptance} & ${<}31\%$ & ${<}32\%$ & ${<}31\%$ & ${<}29\%$ & ${<}13\%$ \\ \hline
   \multicolumn{2}{|l|}{Expected collision bckg.} & $~~~~20\%$ & $~~~~20\%$ & $~~~~20\%$ & $~~~~20\%$ & $~~~~20\%$ \\ \hline
   \multicolumn{2}{|l|}{Expected cosmic ray bckg.} & $~~~~50\%$ & $-$ & $-$ & $-$ & $~~~~80\%$ \\ \hline
   \multicolumn{2}{|l|}{Integrated luminosity} & \multicolumn{5}{c|}{2.2\% (4.4\%) for $\sqrt{s}=7$ (8)\TeV} \\ \hline
 \end{tabular}
 \end{center}
\end{table}

\section{Results \label{sec:results}}

\begin{figure}
 \begin{center}
  \includegraphics[clip=true, trim=0.0cm 0cm 2.5cm 0cm, width=0.4\linewidth]{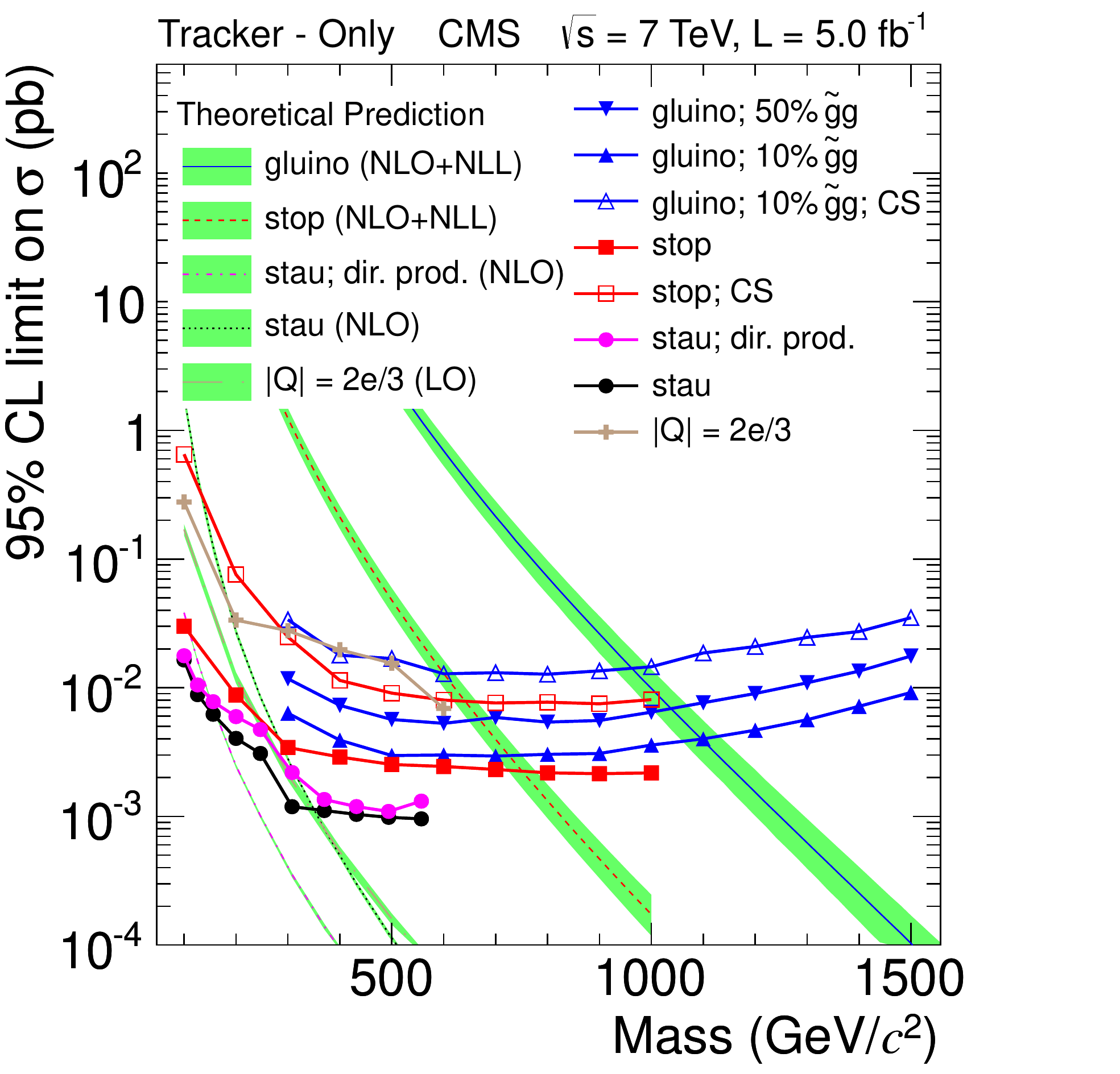}
  \includegraphics[clip=true, trim=0.0cm 0cm 2.5cm 0cm, width=0.4\linewidth]{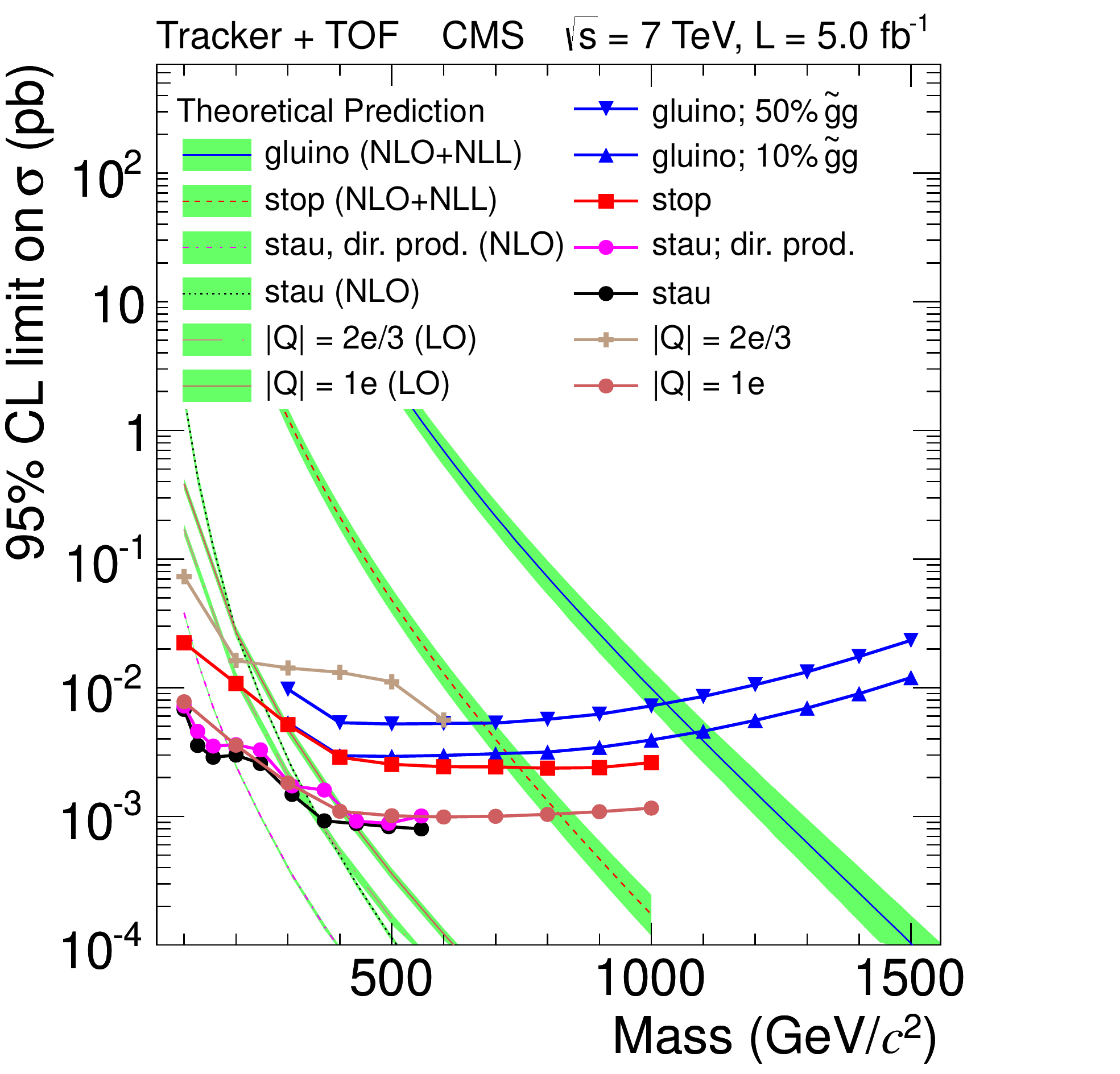}
  \includegraphics[clip=true, trim=0.0cm 0cm 2.5cm 0cm, width=0.4\linewidth]{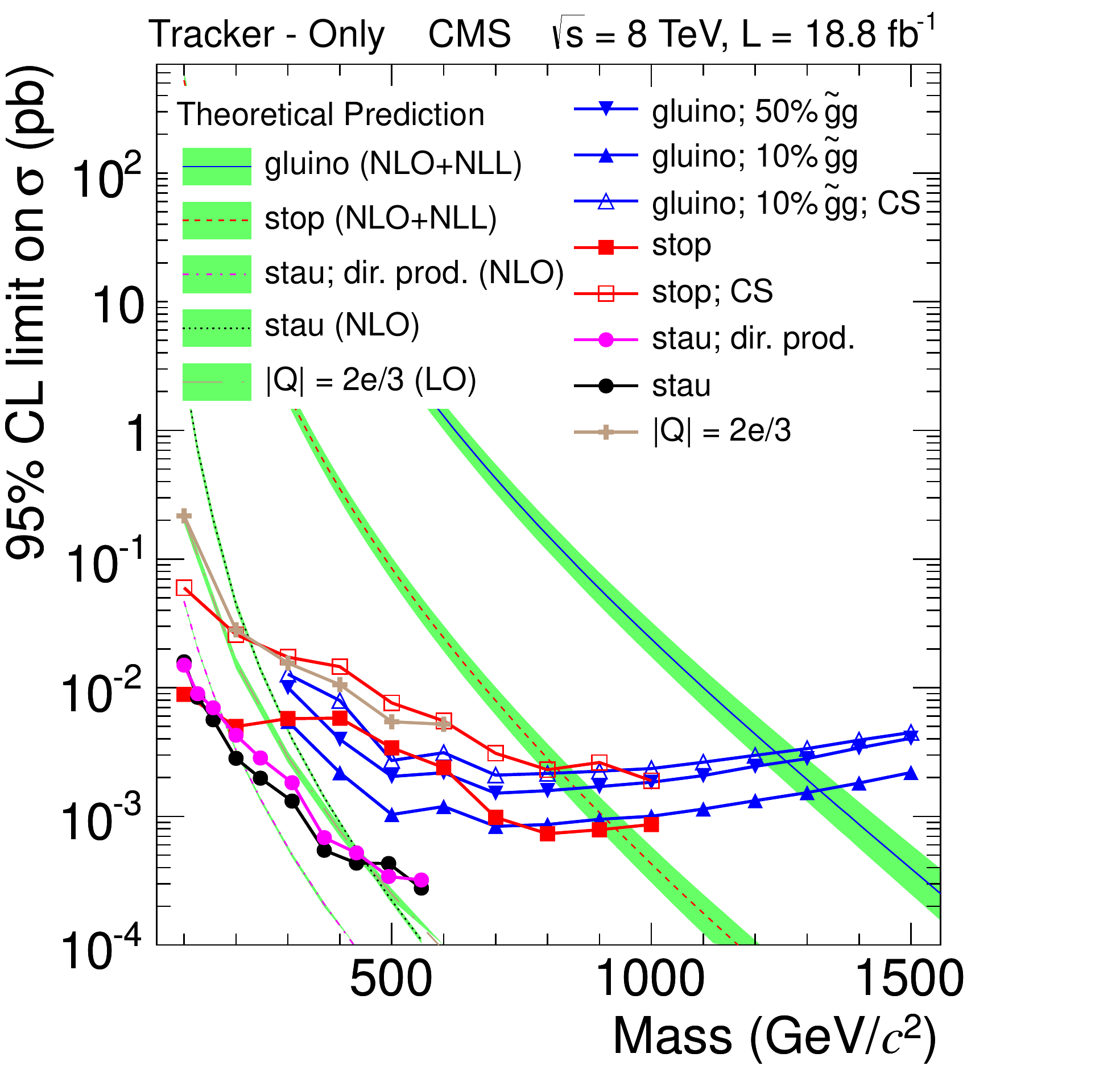}
  \includegraphics[clip=true, trim=0.0cm 0cm 2.5cm 0cm, width=0.4\linewidth]{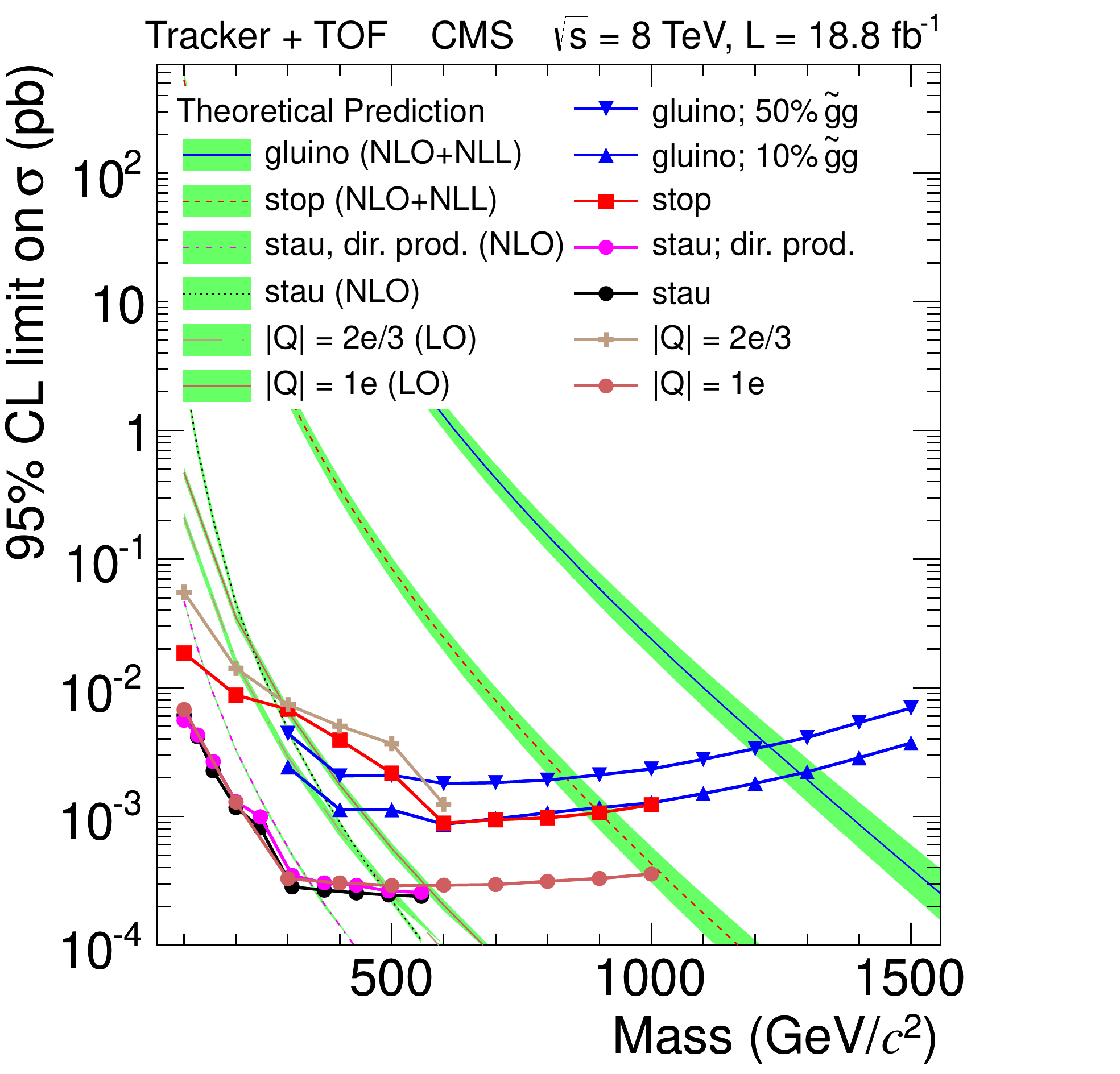}
  \includegraphics[clip=true, trim=0.0cm 0cm 2.5cm 0cm, width=0.4\linewidth]{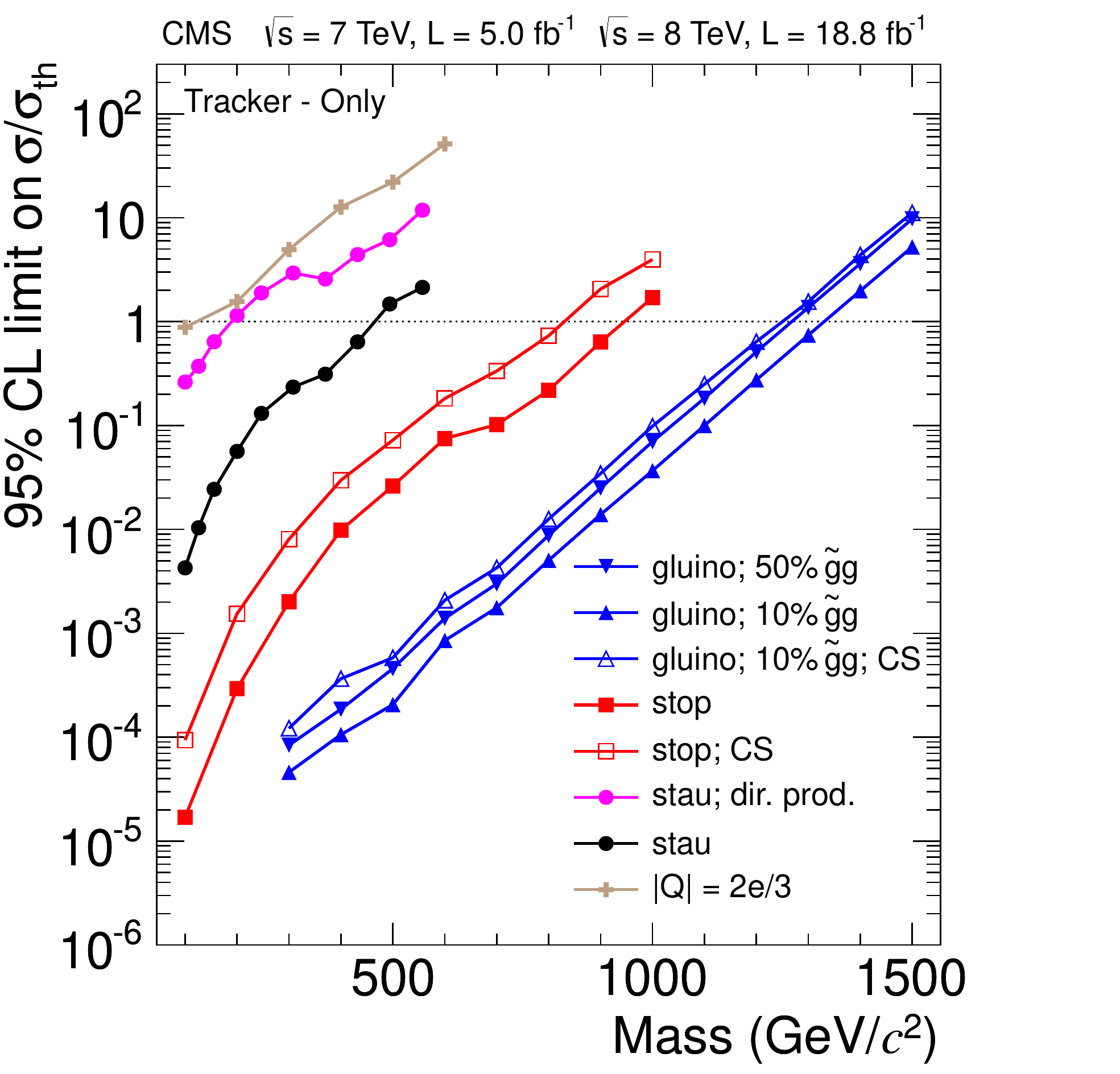}
  \includegraphics[clip=true, trim=0.0cm 0cm 2.5cm 0cm, width=0.4\linewidth]{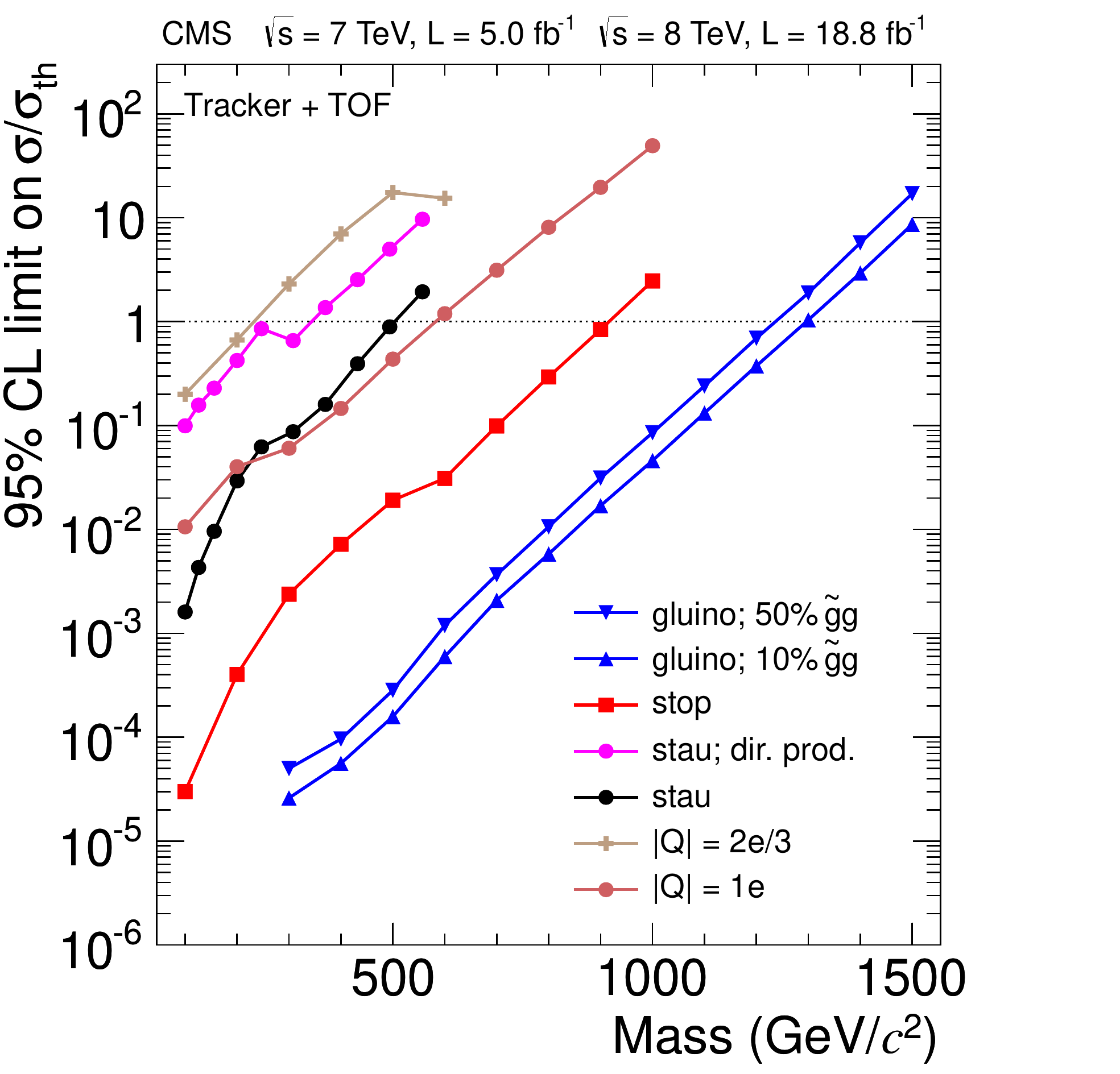}
 \end{center}
 \caption{Upper cross section limits at 95\% CL on various signal models for the
   \tkonly\ analysis (left column) and \tktof\
   analysis (right column).  The top row is for the data at $\sqrt{s} = 7$\TeV, the middle row is for the data at $\sqrt{s} = 8$\TeV,
   the bottom row shows the ratio of the limit to the theoretical value for the combined dataset.  In the legend, 'CS' stands for the charge-suppressed interaction model.
   \label{fig:limits1}}
\end{figure}

\begin{figure}
 \begin{center}
  \includegraphics[clip=true, trim=0.0cm 0cm 2.5cm 0cm, width=0.4\linewidth]{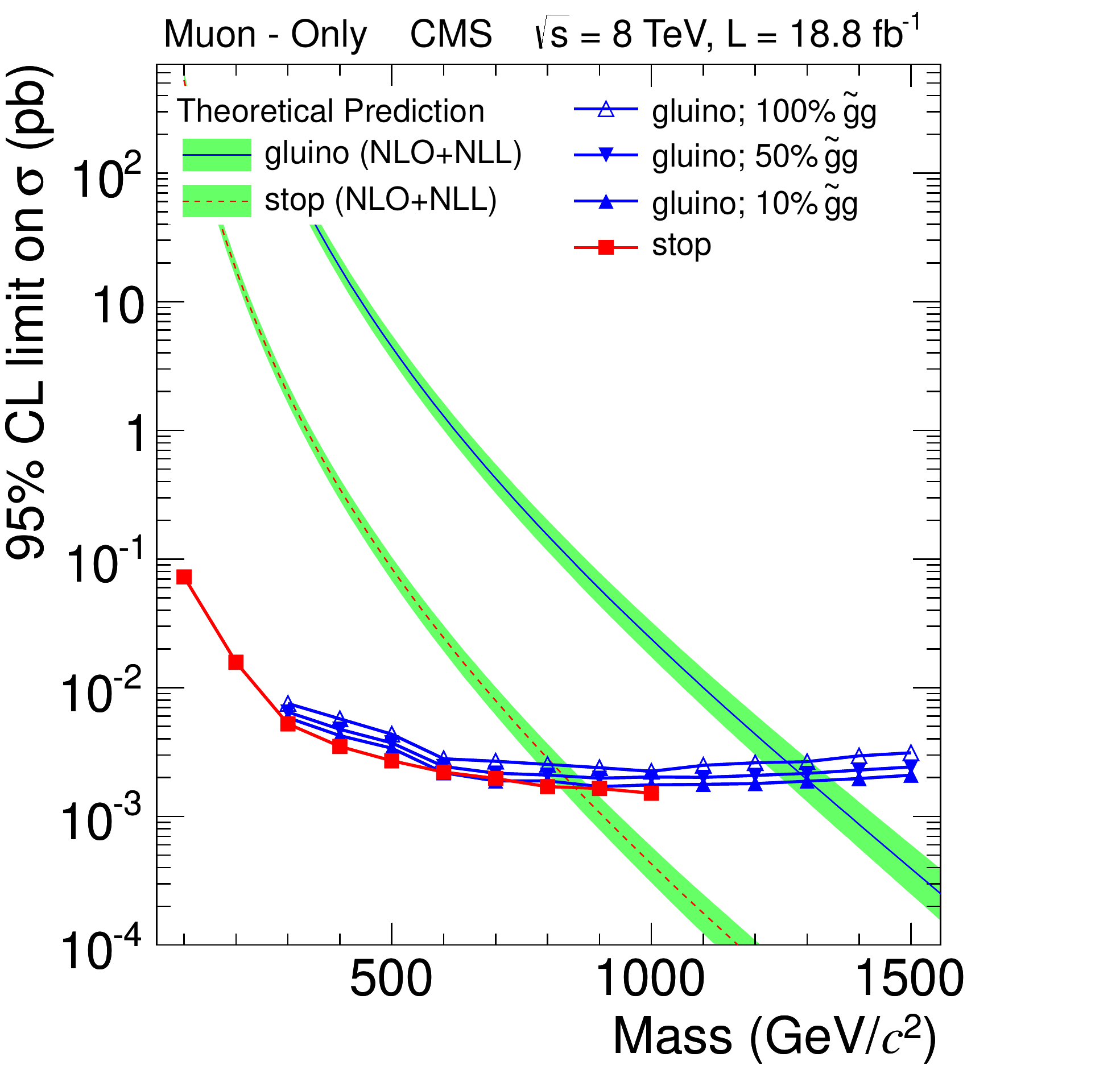}
  \includegraphics[clip=true, trim=0.0cm 0cm 2.5cm 0cm, width=0.4\linewidth]{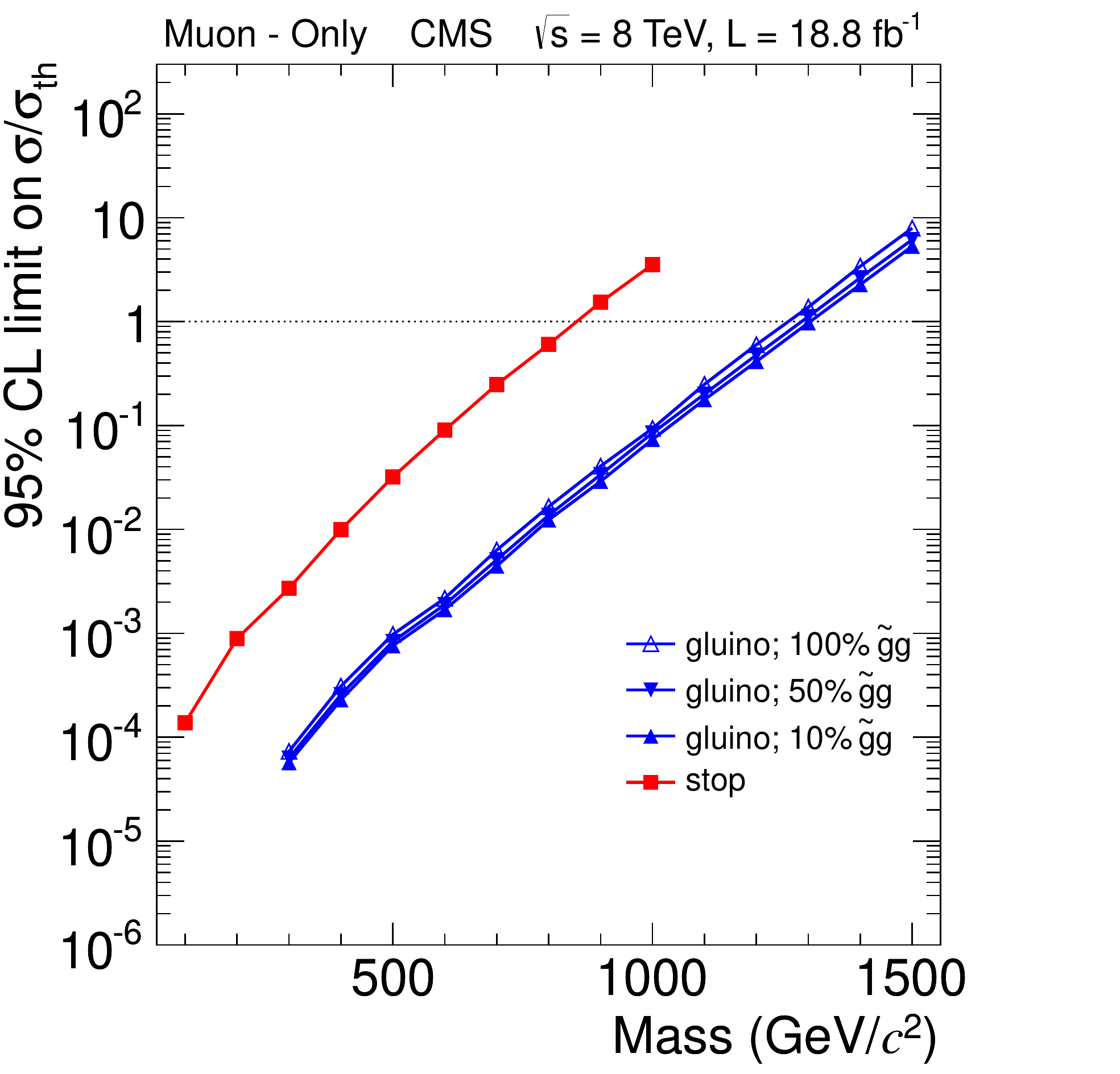}
 \end{center}
 \caption{Upper cross section limits at 95\% CL on various signal models for the \muononly\ analysis for the data at $\sqrt{s} = 8$\TeV (left).
   Limits on the signal strength ($\mu = \sigma/\sigma_{\textnormal{th}}$) for the same data (right).
   \label{fig:limits2}}
\end{figure}

\begin{figure}
 \begin{center}
  \includegraphics[clip=true, trim=0.0cm 0cm 2.5cm 0cm, width=0.4\linewidth]{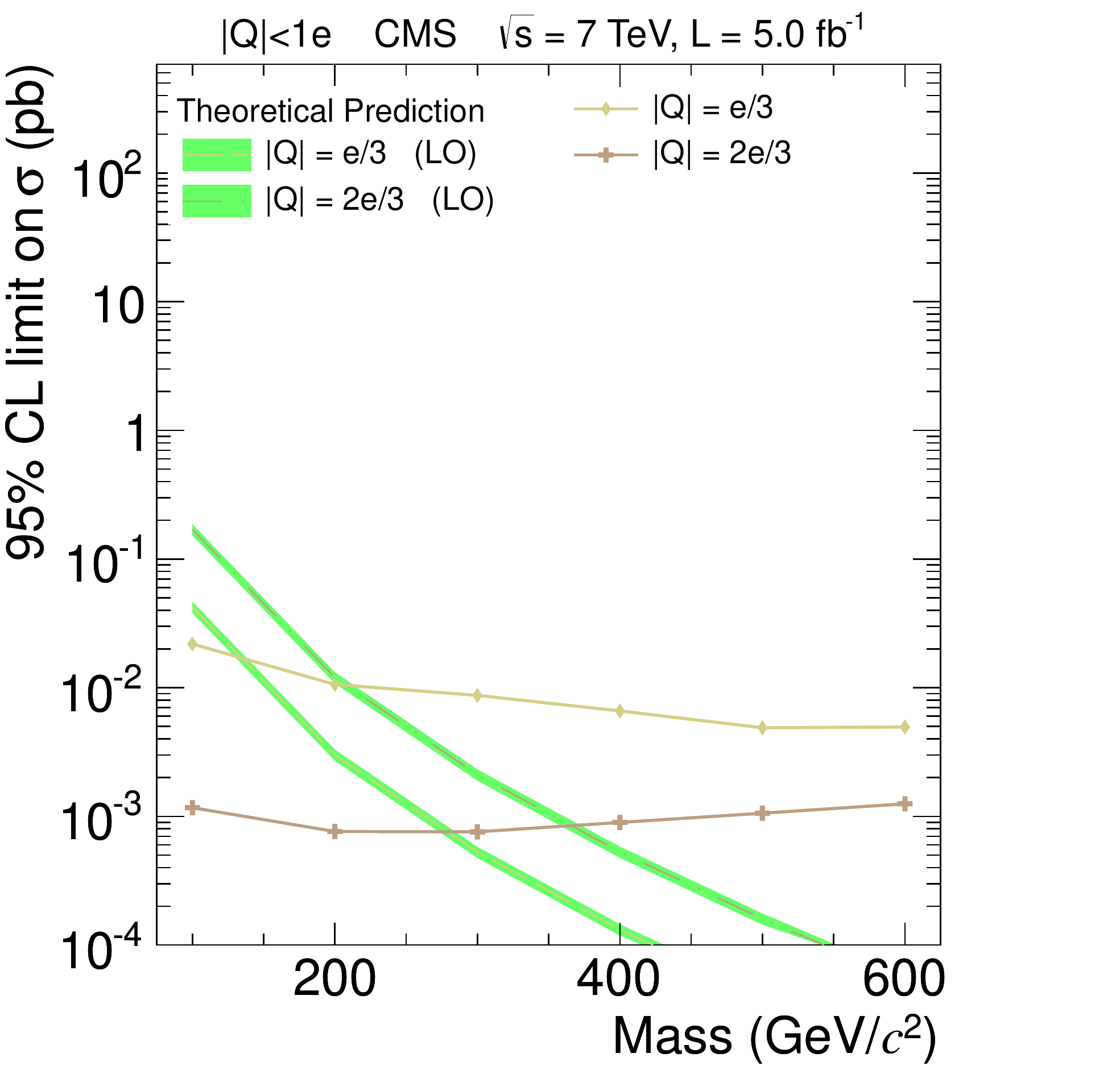}
  \includegraphics[clip=true, trim=0.0cm 0cm 2.5cm 0cm, width=0.4\linewidth]{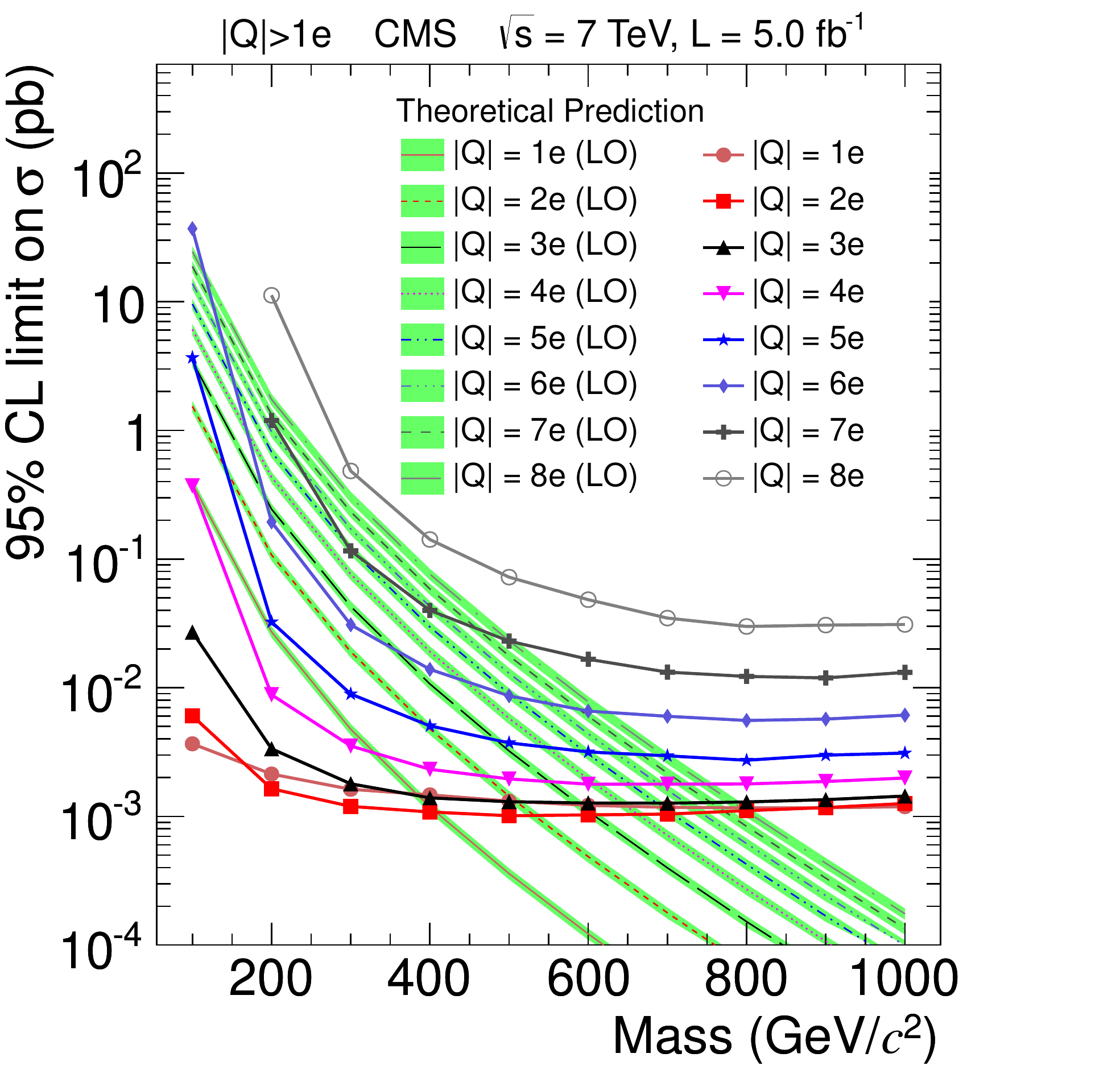}
  \includegraphics[clip=true, trim=0.0cm 0cm 2.5cm 0cm, width=0.4\linewidth]{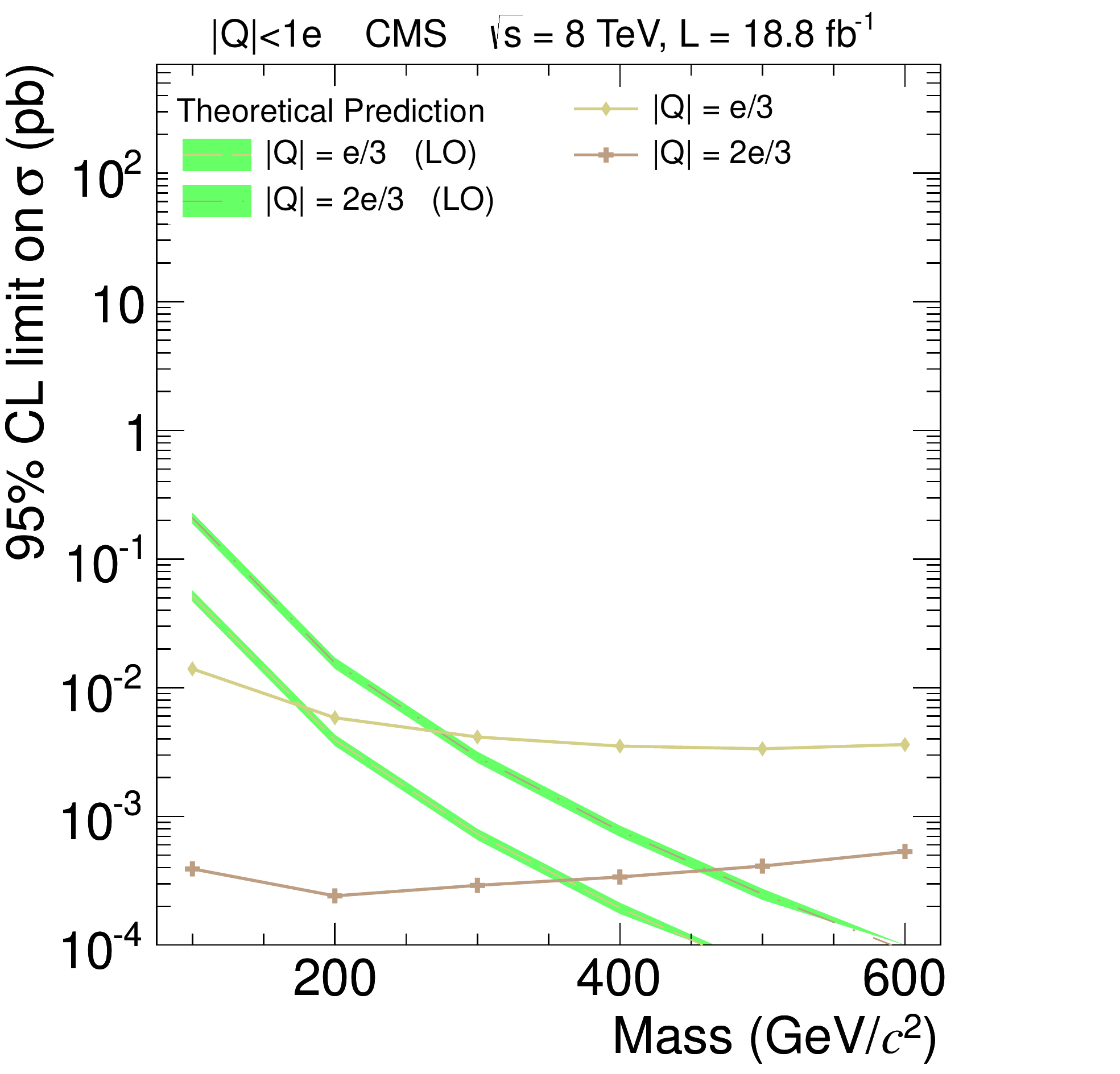}
  \includegraphics[clip=true, trim=0.0cm 0cm 2.5cm 0cm, width=0.4\linewidth]{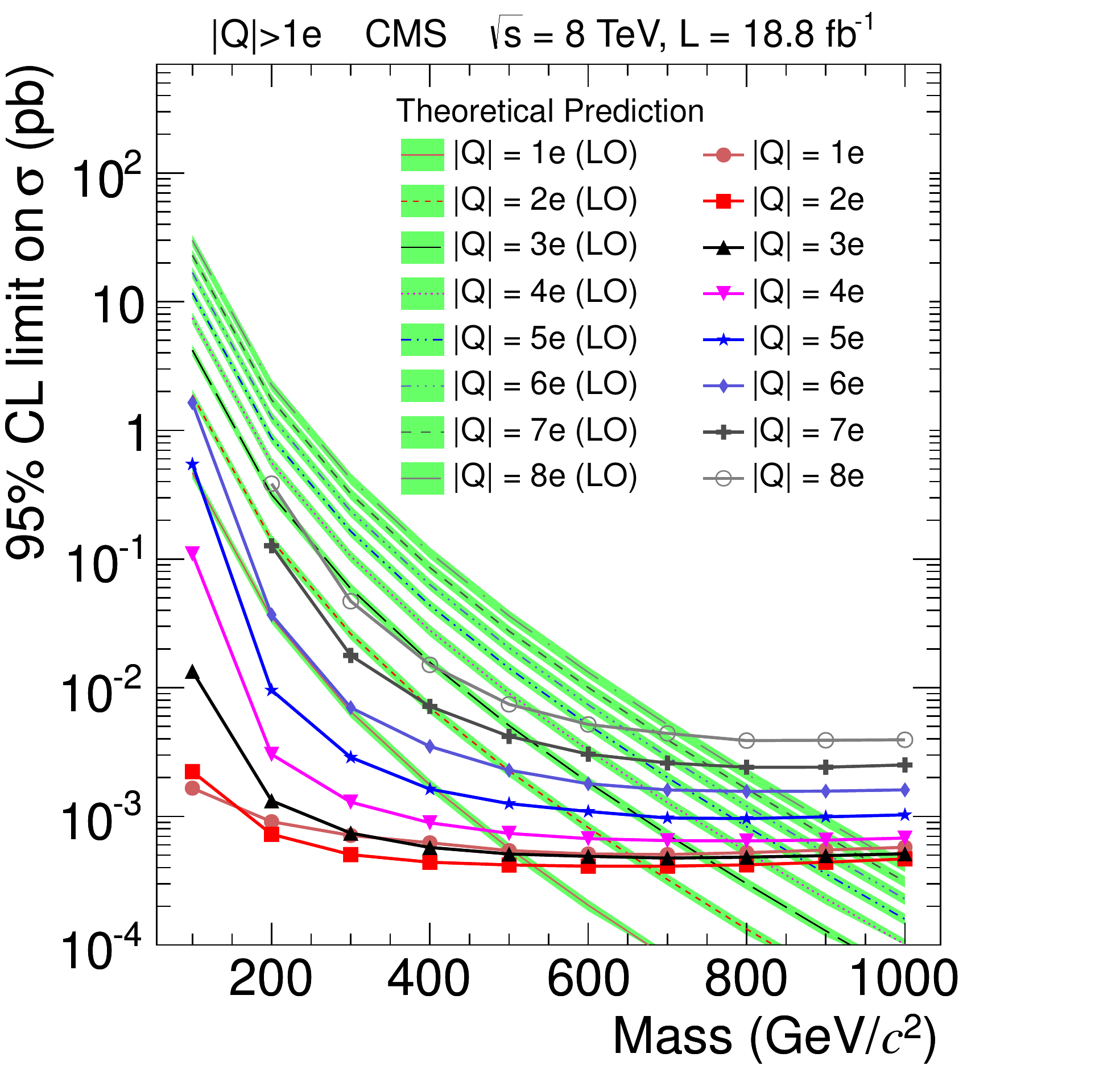}
  \includegraphics[clip=true, trim=0.0cm 0cm 2.5cm 0cm, width=0.4\linewidth]{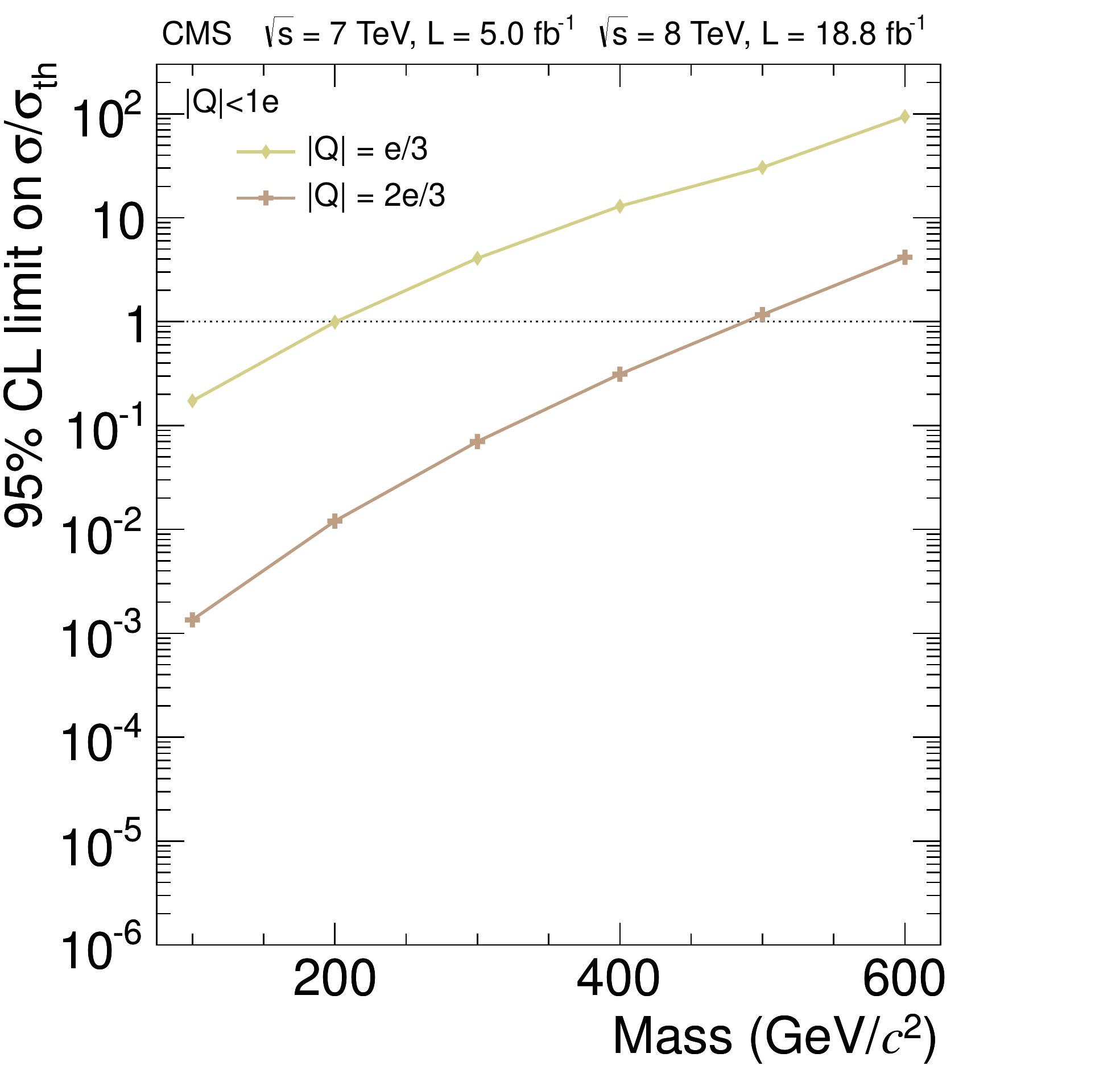}
  \includegraphics[clip=true, trim=0.0cm 0cm 2.5cm 0cm, width=0.4\linewidth]{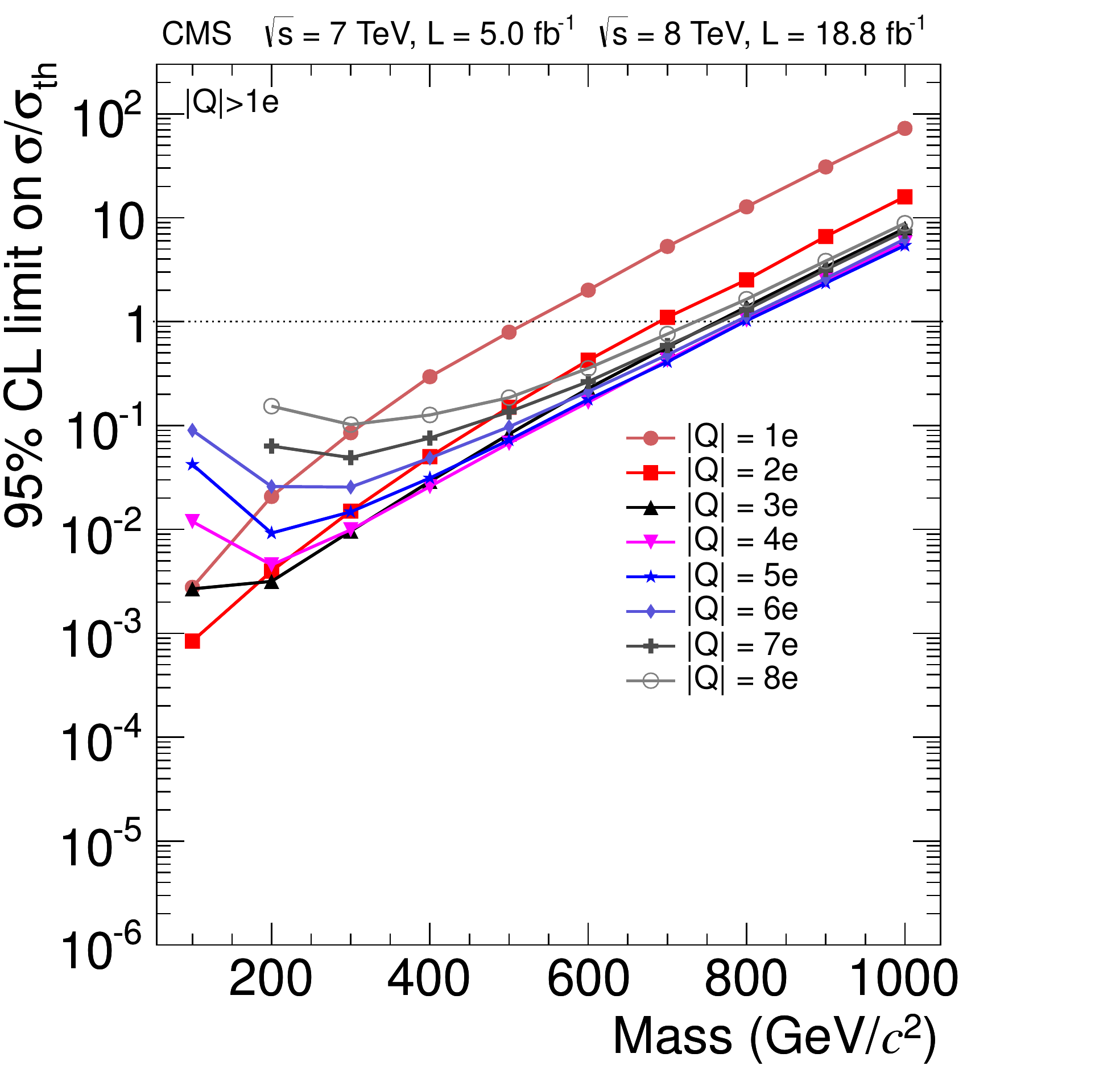}
 \end{center}
 \caption{Upper cross section limits at 95\% CL on various signal models for the
   \fractionalcharge\ analysis (left column) and \multicharge\
   analysis (right column).  The top row is for the data at $\sqrt{s} = 7$\TeV, the middle row is for the data at $\sqrt{s} = 8$\TeV,
   the bottom row shows the ratio of the limit to the theoretical value for the combined dataset.
   \label{fig:limits3}}
\end{figure}

No significant excess of events is observed over the predicted backgrounds.
The largest excess for any of the selections shown in Table~\ref{tab:finalsel} has a significance of 1.3
standard deviations.
Cross section limits are placed at 95\% confidence level (CL) for both $\sqrt{s} = 7$ and 8\TeV using the
CL$_s$ approach~\cite{Junk:1999kv, READ:JPG2002} where $p$-values are computed with a
hybrid bayesian-frequentist technique~\cite{ATLAS:1379837} that uses a lognormal
model~\cite{Eadie, James} for the nuisance parameters. The latter are the
integrated luminosity, the signal acceptance, and the
expected background in the signal region.
The uncertainty in the theoretical cross section is not considered as a nuisance parameter.
For the combined dataset,
the limits are instead placed on the signal strength ($\mu = \sigma/\sigma_{\text{th}}$).
Limits on the signal strength using only the 8\TeV dataset for the \muononly\ analysis are also presented.
The observed limits are shown in Figs.~\ref{fig:limits1}--\ref{fig:limits3} for all the analyses
along with the theoretical predictions.
For the gluino and scalar top pair production,
the theoretical cross sections are computed at NLO+NLL~\cite{Kulesza:2008jb,Kulesza:2009kq, Beenakker:2009ha, Beenakker:2010nq} using
\PROSPINO~\cite{Beenakker:1996ed} with CTEQ6.6M PDFs~\cite{Nadolsky:2008zw}.
The uncertainty bands on the theoretical cross sections include the PDF uncertainty, as well as the $\alpha_{s}$ and scale uncertainties.
Mass limits are obtained from the intersection of the observed
limit and the central value of the theoretical cross section.
For the combined result, the masses for which the signal strength is less than one are excluded.

From the final results, 95\% CL limits on the production cross section
are shown in Tables~\ref{tab:limitsGluino}, ~\ref{tab:limitsStop}, ~\ref{tab:limitsStau}, and ~\ref{tab:limitsDY}
for gluino, scalar top, stau, and for Drell--Yan like production of fractionally, singly, or multiply charged particles, respectively.
The limits are determined from the numbers of events passing
all final criteria (including the mass criteria for the \tkonly\
and \tktof\ analyses).
Figure~\ref{fig:limits1} shows the limits as a function of mass
for the \tkonly\ and \tktof\ analyses.
The \tkonly\ analysis excludes gluino masses below 1322 and 1233\GeVcc for $f=0.1$
in the cloud interaction model and charge-suppressed model, respectively.
Stop masses below 935 (818)\GeVcc are excluded for the
cloud (charge-suppressed) models.
In addition, the \tktof\ analysis excludes \stau masses below
$500\,(339)$\GeVcc for the direct+indirect (direct only) production. Drell--Yan signals with
$|Q| = 2e/3$ and $|Q|=1e$ are excluded below $220$ and $574$\GeVcc, respectively.

\begin{table}
 \topcaption{Expected and observed cross section limits and the signal acceptance for gluino signals at $\sqrt{s} = 7$ and
   8\TeV, as well as the ratio of the cross section limit to the
   theoretical value for the combined dataset.
The limit on the ratio for the \muononly\ analysis uses only $\sqrt{s} = 8$\TeV data.
The minimum reconstructed mass required ($M$ req.) for each sample in the \tkonly\ analysis is also given.
   \label{tab:limitsGluino}}
\centering
 \footnotesize
 \begin{tabular}{|c|c|ccc|ccc|cc|} \hline
 Mass & $M$ req.& \multicolumn{3}{c|}{$\sigma$ (pb) ($\sqrt{s} = 7$\TeV)} & \multicolumn{3}{c|}{$\sigma$ (pb) ($\sqrt{s} = 8$\TeV)} & \multicolumn{2}{c|}{$\sigma/\sigma_{\text{th}}$ (7+8\TeV)} \\
 (\GeVcc) & (\GeVcc) & Exp. & Obs. & Acc. & Exp. & Obs. & Acc. & Exp. & Obs. \\ \hline
 \multicolumn{10}{|c|}{Gluino ($f=0.1$) ~~~---~~~ \tkonly\ analysis}  \\ \hline
 300 & ${>}100$  & 0.0046 & 0.0063 &   0.17 & 0.0055 & 0.0055 &   0.15 & $      4.0 \times 10^{-5}$ & $      4.6 \times 10^{-5}$\\
 700 & ${>}370$  & 0.0028 & 0.0029 &   0.21 & 0.00081 & 0.00084 &   0.19 & 0.0017 & 0.0018\\
 1100 & ${>}540$  & 0.0039 & 0.0040 &   0.15 & 0.0010 & 0.0011 &   0.14 & 0.098 & 0.10\\
 1500 & ${>}530$  & 0.0088 & 0.0092 &  0.066 & 0.0021 & 0.0022 &  0.073 & 5.1 & 5.2\\

\hline
 \multicolumn{10}{|c|}{Gluino charge-suppressed ($f=0.1$) ~~~---~~~ \tkonly\ analysis} \\ \hline
  300 & ${>}130$  & 0.035 & 0.034 &  0.021 & 0.013 & 0.013 &  0.048 & 0.00011 & 0.00012\\
  700 & ${>}340$  & 0.012 & 0.013 &  0.046 & 0.0020 & 0.0021 &  0.077 & 0.0044 & 0.0043\\
 1100 & ${>}410$  & 0.018 & 0.019 &  0.033 & 0.0025 & 0.0026 &  0.061 & 0.24 & 0.25\\
 1500 & ${>}340$  & 0.034 & 0.035 &  0.017 & 0.0045 & 0.0045 &  0.035 & 11 & 11\\

\hline
 \multicolumn{10}{|c|}{Gluino ($f=0.5$) ~~~---~~~ \muononly\ analysis} \\ \hline
  300 &  --      &     --     &   --      &   --      & 0.0060 & 0.0065 &  0.058 & $      5.8 \times 10^{-5}$ & $      6.3 \times 10^{-5}$\\
  700 &  --      &     --     &   --      &   --      & 0.0026 & 0.0022 &   0.12 & 0.0062 & 0.0051\\
 1100 &  --      &     --     &   --      &   --      & 0.0024 & 0.0020 &   0.13 & 0.24 & 0.20\\
 1500 &  --      &     --     &   --      &   --      & 0.0030 & 0.0024 &   0.11 & 7.5 & 6.2\\

\hline
 \multicolumn{10}{|c|}{Gluino ($f=1.0$) ~~~---~~~ \muononly\ analysis} \\ \hline
 300 &  --      &     --     &   --      &   --      & 0.0070 & 0.0075 &   0.050 & $      6.8 \times 10^{-5}$ & $      7.3 \times 10^{-5}$\\
 700 &  --      &     --     &   --      &   --      & 0.0032 & 0.0027 &    0.10 & 0.0075 & 0.0063\\
 1100 &  --      &     --     &   --      &   --      & 0.0030 & 0.0025 &   0.11 & 0.30 & 0.25\\
 1500 &  --      &     --     &   --      &   --      & 0.0037 & 0.0031 &  0.087 & 9.5 & 7.9\\

\hline
 \end{tabular}
 \normalsize
\end{table}

\begin{table}
 \topcaption{Expected and observed cross section limits and the signal acceptance for scalar top signals at $\sqrt{s} = 7$ and
   8\TeV, as well as the ratio of the cross section limit to the
   theoretical value for the combined dataset.
The minimum reconstructed mass required ($M$ req.) for each sample in the \tkonly\ analysis is also given.
   \label{tab:limitsStop}}
 \centering
 \footnotesize
 \begin{tabular}{|c|c|ccc|ccc|cc|} \hline
 Mass & $M$ req.& \multicolumn{3}{c|}{$\sigma$ (pb) ($\sqrt{s} = 7$\TeV)} & \multicolumn{3}{c|}{$\sigma$ (pb) ($\sqrt{s} = 8$\TeV)} & \multicolumn{2}{c|}{$\sigma/\sigma_{\text{th}}$ (7+8\TeV)} \\
 (\GeVcc) & (\GeVcc) & Exp. & Obs. & Acc. & Exp. & Obs. & Acc. & Exp. & Obs. \\ \hline
 \multicolumn{10}{|c|}{Stop ~~~---~~~ \tkonly\ analysis} \\ \hline
  200 & ${>}0$    & 0.0080 & 0.0088 &   0.14 & 0.0051 & 0.0050 &   0.18 & 0.00026 & 0.00029\\
  500 & ${>}120$  & 0.0024 & 0.0025 &   0.24 & 0.0027 & 0.0034 &   0.23 & 0.022 & 0.026\\
  800 & ${>}330$  & 0.0021 & 0.0022 &   0.28 & 0.00072 & 0.00073 &   0.22 & 0.21 & 0.22\\

\hline
 \multicolumn{10}{|c|}{Stop charge-suppressed ~~~---~~~ \tkonly\ analysis } \\ \hline
  200 & ${>}0$    & 0.063 & 0.075 &   0.020 & 0.018 & 0.026 &   0.050 & 0.0011 & 0.0014\\
  500 & ${>}120$  & 0.0086 & 0.0089 &   0.066 & 0.0068 & 0.0081 &   0.10 & 0.062 & 0.070\\
  800 & ${>}270$  & 0.0071 & 0.0076 &   0.079 & 0.0019 & 0.0023 &   0.10 & 0.61 & 0.74\\
\hline
 \end{tabular}
 \normalsize
\end{table}

\begin{table}
 \topcaption{Expected and observed cross section limits and the signal acceptance for stau signals at $\sqrt{s} = 7$ and
   8\TeV, as well as the ratio of the cross section limit to the
   theoretical value for the combined dataset.
The minimum reconstructed mass required ($M$ req.) for each sample in the \tktof\ analysis is also given.
   \label{tab:limitsStau}}
  \centering
  \footnotesize
 \begin{tabular}{|c|c|ccc|ccc|cc|} \hline
 Mass & $M$ req.& \multicolumn{3}{c|}{$\sigma$ (pb) ($\sqrt{s} = 7$\TeV)} & \multicolumn{3}{c|}{$\sigma$ (pb) ($\sqrt{s} = 8$\TeV)} & \multicolumn{2}{c|}{$\sigma/\sigma_{\text{th}}$ (7+8\TeV)} \\
 (\GeVcc) & (\GeVcc) & Exp. & Obs. & Acc. & Exp. & Obs. & Acc. & Exp. & Obs. \\ \hline
 \multicolumn{10}{|c|}{Direct+indirect produced stau ~~~---~~~ \tktof\ analysis} \\ \hline
  126 & ${>}40$   & 0.0046 & 0.0035 &   0.29 & 0.0042 & 0.0042 &   0.25 & 0.0050 & 0.0043\\
  308 & ${>}190$  & 0.00094 & 0.0015 &   0.63 & 0.00029 & 0.00028 &   0.56 & 0.065 & 0.087\\
  494 & ${>}330$  & 0.00079 & 0.00084 &   0.74 & 0.00023 & 0.00024 &   0.66 & 0.66 & 0.84\\

\hline
 \multicolumn{10}{|c|}{Direct produced stau ~~~---~~~ \tktof\ analysis} \\ \hline
  126 & ${>}40$   & 0.0056 & 0.0046 &   0.26 & 0.0044 & 0.0043 &   0.24 & 0.18 & 0.16\\
  308 & ${>}190$  & 0.0011 & 0.0017 &   0.54 & 0.00035 & 0.00035 &   0.46 & 0.62 & 0.66\\
  494 & ${>}330$  & 0.00084 & 0.00088 &   0.69 & 0.00025 & 0.00026 &   0.61 & 4.7 & 5.0\\

\hline
 \end{tabular}
 \normalsize
\end{table}

\begin{table}
 \topcaption{Expected and observed cross section limits and the signal acceptance for the Drell--Yan like production of fractionally, singly, and multiply charged particles at $\sqrt{s} = 7$ and
   8\TeV, as well as the ratio of the cross section limit to the
   theoretical value for the combined dataset.
The minimum reconstructed mass required ($M$ req.) for each sample in the \tktof\ analysis is also given.
   \label{tab:limitsDY}}
  \centering
  \footnotesize
 \begin{tabular}{|c|c|ccc|ccc|cc|} \hline
 Mass & $M$ req.& \multicolumn{3}{c|}{$\sigma$ (pb) ($\sqrt{s} = 7$\TeV)} & \multicolumn{3}{c|}{$\sigma$ (pb) ($\sqrt{s} = 8$\TeV)} & \multicolumn{2}{c|}{$\sigma/\sigma_{\text{th}}$ (7+8\TeV)} \\
 (\GeVccns{}) & (\GeVccns{}) & Exp. & Obs. & Acc. & Exp. & Obs. & Acc. & Exp. & Obs. \\ \hline
\multicolumn{10}{|c|}{Drell--Yan like production $|Q|$ = $e/3$ ~~~---~~~ $|Q|<1e$ analysis} \\ \hline
  100 &  --      & 0.019 & 0.022 &  0.029 & 0.016 & 0.014 &  0.012 & 0.19 & 0.17\\
  200 &  --      & 0.0094 & 0.011 &   0.060 & 0.0066 & 0.0058 &   0.030 & 1.2 & 0.99\\
  400 &  --      & 0.0058 & 0.0066 &  0.098 & 0.0041 & 0.0035 &  0.048 & 15 & 13\\
\hline
\multicolumn{10}{|c|}{Drell--Yan like production $|Q|$ = $2e/3$ ~~~---~~~ $|Q|<1e$ analysis} \\ \hline
  100 &  --      & 0.0011 & 0.0012 &   0.53 & 0.00042 & 0.00039 &   0.45 & 0.0015 & 0.0013\\
  200 &  --      & 0.00071 & 0.00076 &   0.81 & 0.00027 & 0.00024 &   0.68 & 0.014 & 0.012\\
  400 &  --      & 0.00083 & 0.00090 &   0.68 & 0.00033 & 0.00034 &   0.56 & 0.35 & 0.31\\
\hline
\multicolumn{10}{|c|}{Drell--Yan like production $|Q|$ = $1e$ ~~~---~~~ \tktof\ analysis} \\ \hline
  200 & ${>}120$  & 0.0015 & 0.0036 &   0.41 & 0.00077 & 0.0013 &   0.36 & 0.019 & 0.040\\
  500 & ${>}340$  & 0.00098 & 0.0010 &    0.60 & 0.00028 & 0.00029 &   0.56 & 0.41 & 0.44\\
  800 & ${>}530$  & 0.0010 & 0.0010 &   0.58 & 0.00030 & 0.00031 &   0.52 & 7.5 & 8.1\\
\hline
  \multicolumn{10}{|c|}{Drell--Yan like production $|Q|$ = $2e$ ~~~---~~~ $|Q|>1e$ analysis} \\ \hline
  200 &  --      & 0.0016 & 0.0016 &   0.36 & 0.00050 & 0.00073 &   0.33 & 0.0028 & 0.0040\\
  500 &  --      & 0.00098 & 0.0010 &   0.59 & 0.00029 & 0.00042 &   0.56 & 0.11 & 0.15\\
  800 &  --      & 0.0011 & 0.0011 &   0.55 & 0.00029 & 0.00042 &   0.56 & 1.9 & 2.5\\
\hline
  \multicolumn{10}{|c|}{Drell--Yan like production $|Q|$ = $3e$ ~~~---~~~ $|Q|>1e$ analysis} \\ \hline
  200 &  --      & 0.0031 & 0.0034 &   0.18 & 0.00090 & 0.0013 &   0.18 & 0.0023 & 0.0032\\
  500 &  --      & 0.0012 & 0.0013 &   0.47 & 0.00035 & 0.00051 &   0.46 & 0.059 & 0.083\\
  800 &  --      & 0.0012 & 0.0013 &   0.47 & 0.00033 & 0.00048 &   0.49 & 0.99 & 1.4\\
\hline
  \multicolumn{10}{|c|}{Drell--Yan like production $|Q|$ = $4e$ ~~~---~~~ $|Q|>1e$ analysis} \\ \hline
  200 &  --      & 0.0082 & 0.0088 &  0.069 & 0.0021 & 0.0030 &  0.078 & 0.0031 & 0.0045\\
  500 &  --      & 0.0018 & 0.0020 &   0.31 & 0.00051 & 0.00074 &   0.32 & 0.048 & 0.068\\
  800 &  --      & 0.0017 & 0.0018 &   0.34 & 0.00045 & 0.00064 &   0.37 & 0.75 & 1.0\\
\hline
  \multicolumn{10}{|c|}{Drell--Yan like production $|Q|$ = $5e$ ~~~---~~~ $|Q|>1e$ analysis} \\ \hline
  200 &  --      & 0.030 & 0.032 &  0.019 & 0.0066 & 0.0096 &  0.025 & 0.0064 & 0.0092\\
  500 &  --      & 0.0035 & 0.0037 &   0.16 & 0.00086 & 0.0013 &   0.19 & 0.052 & 0.073\\
  800 &  --      & 0.0026 & 0.0027 &   0.22 & 0.00066 & 0.00096 &   0.24 & 0.71 & 1.0\\
\hline
  \multicolumn{10}{|c|}{Drell--Yan like production $|Q|$ = $6e$ ~~~---~~~ $|Q|>1e$ analysis} \\ \hline
  200 &  --      & 0.17 & 0.19 & 0.0032 & 0.026 & 0.037 & 0.0064 & 0.018 & 0.026\\
  500 &  --      & 0.0079 & 0.0086 &  0.072 & 0.0016 & 0.0023 &    0.10 & 0.055 & 0.077\\
  800 &  --      & 0.0054 & 0.0056 &   0.11 & 0.0011 & 0.0016 &   0.15 & 0.81 & 1.1\\
\hline
  \multicolumn{10}{|c|}{Drell--Yan like production $|Q|$ = $7e$ ~~~---~~~ $|Q|>1e$ analysis} \\ \hline
  200 &  --      & 1.1 & 1.2 & 0.00053 & 0.086 & 0.13 & 0.0019 & 0.047 & 0.063\\
  500 &  --      & 0.022 & 0.023 &  0.026 & 0.0028 & 0.0042 &  0.057 & 0.079 & 0.11\\
  800 &  --      & 0.012 & 0.012 &  0.049 & 0.0017 & 0.0024 &  0.099 & 0.96 & 1.3\\
\hline
  \multicolumn{10}{|c|}{Drell--Yan like production $|Q|$ = $8e$ ~~~---~~~ $|Q|>1e$ analysis} \\ \hline
  200 &  --      & 9.8 & 11 & $7.1 \times 10^{-5}$ & 0.26 & 0.38 & 0.00064 & 0.11 & 0.15\\
  500 &  --      & 0.068 & 0.072 & 0.0084 & 0.0051 & 0.0074 &  0.032 & 0.11 & 0.16\\
  800 &  --      & 0.028 & 0.030 &   0.020 & 0.0027 & 0.0039 &  0.062 & 1.0 & 1.4\\
\hline
 \end{tabular}
 \normalsize
\end{table}

The limits from the \muononly\ analysis for the scalar top and
the gluino with
various hadronization fractions $f$ are shown
in Fig.~\ref{fig:limits2}.  The \muononly\ analysis excludes gluino masses
below $1250 (1276)$\GeVcc for $f=1.0 (0.5)$.

Figure~\ref{fig:limits3} shows the limits applied to the
Drell--Yan production model for both the \fractionalcharge\ and
\multicharge\ analyses.  The \fractionalcharge\ analysis
excludes masses below 200 and 480\GeVcc for
$|Q| = e/3$ and $2e/3$, respectively.  The \multicharge\
analysis excludes masses below 685, 752, 793, 796, 781, 757, and 715\GeVcc for $|Q|$ = $2e$, $3e$, $4e$, $5e$, $6e$, $7e$, and $8e$, respectively.
The \multicharge\ analysis is not optimized for singly charged particles but can set a limit and is able to exclude masses below 517\GeVcc.
As expected, this limit is not as stringent as the one
set by the \tktof\ analysis but does allow results to be
interpolated to non-integer charge values (such as $|Q|$= $3e/2$, $4e/3$)
using results from the same analysis.

\begin{figure}%[tbhp]
 \begin{center}
  \includegraphics[clip=true, trim=0.0cm 0cm 1.0cm 0cm, width=0.44\linewidth]{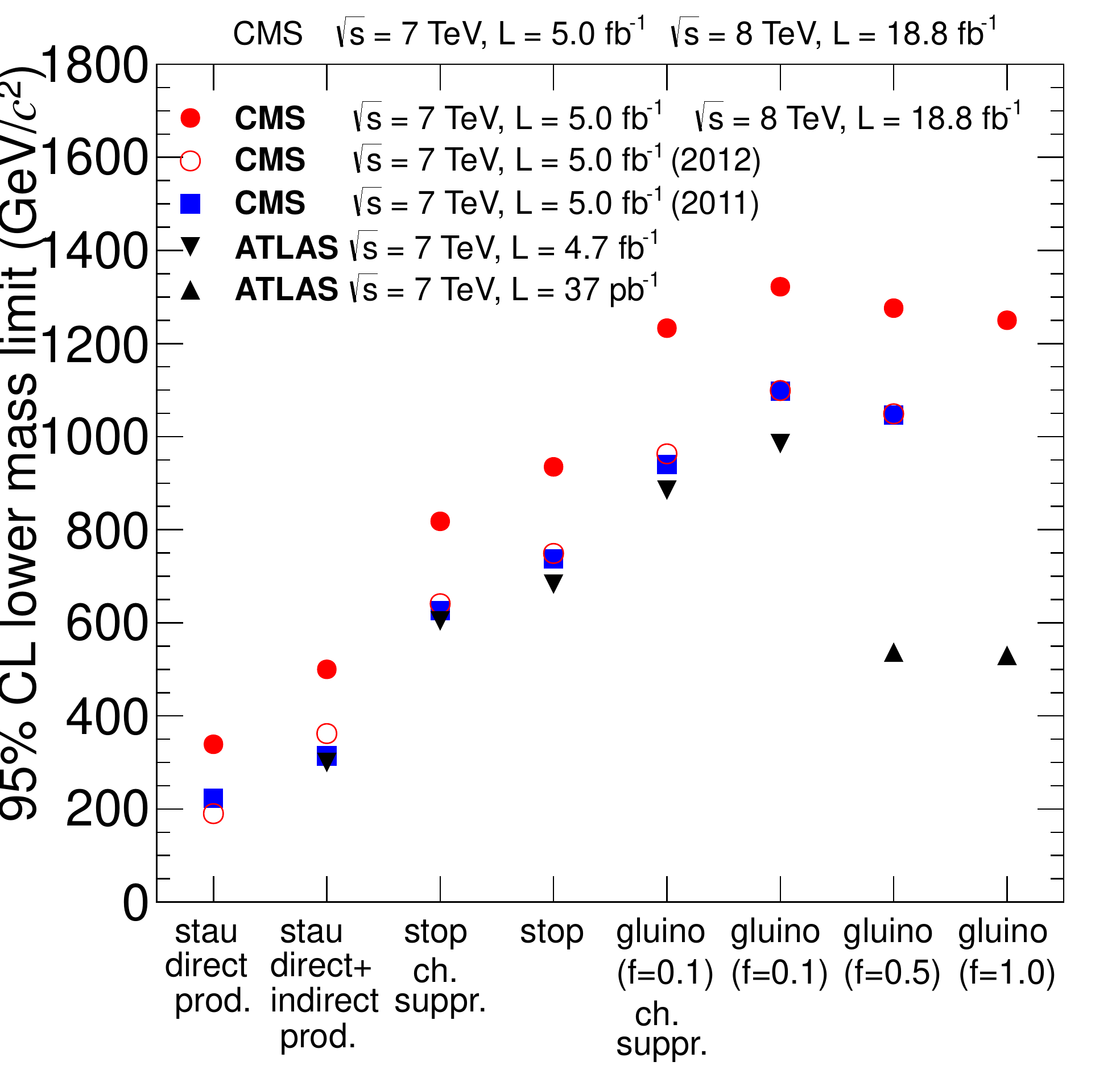} ~~
  \includegraphics[clip=true, trim=0.0cm 0cm 1.0cm 0cm, width=0.44\linewidth]{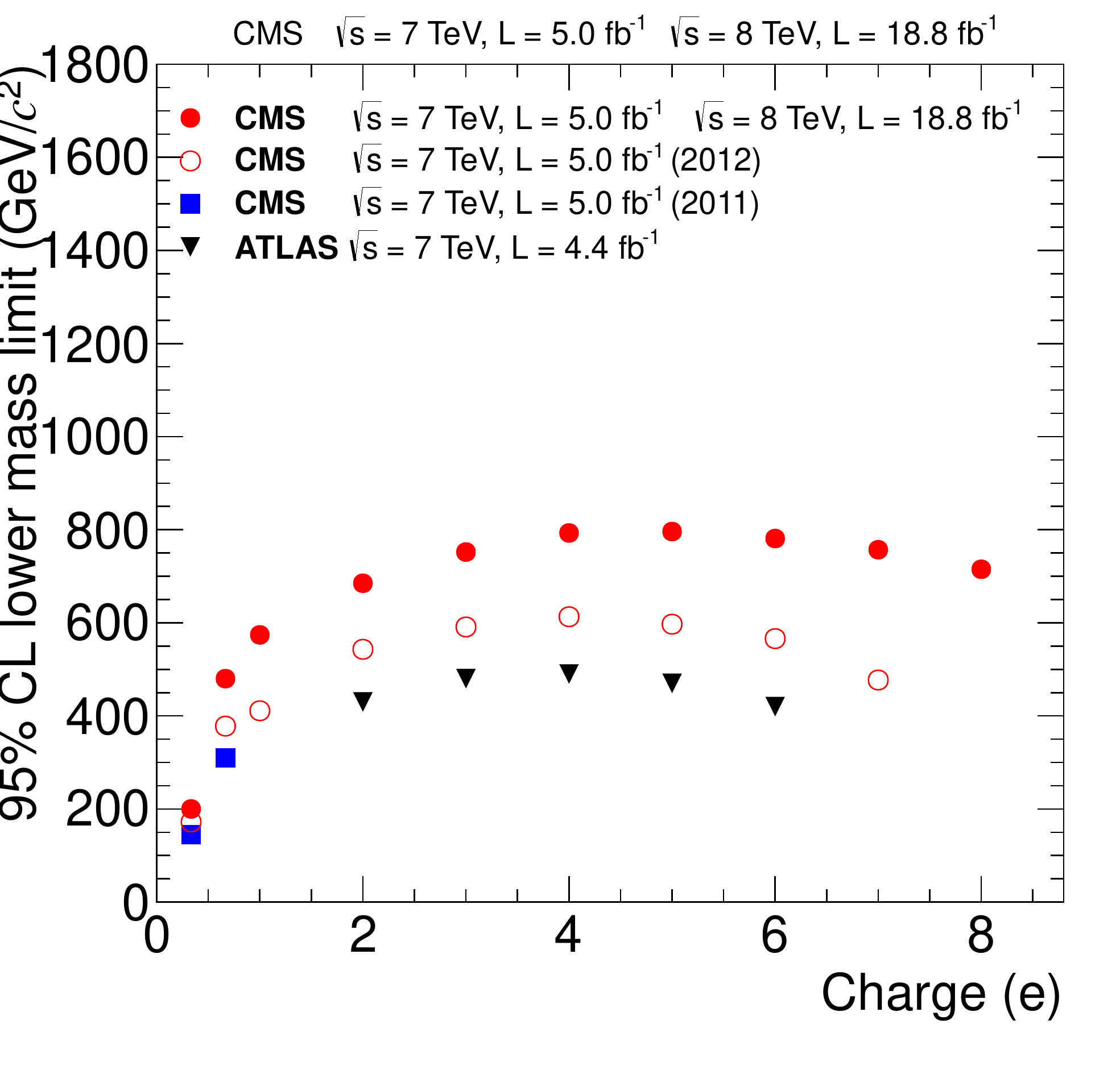}
 \end{center}
 \caption{Lower mass limits at 95\% CL for various models compared with previously published results~\cite{Aad:2011hz, Aad:2011yf, Aad:2011mb,
Aad:2012vd, ATLASmCHAMPs, Khachatryan:2011ts, Chatrchyan::2012dr, Chatrchyan:2012sp}. The model type
is given on the x-axis (left). Mass limits are shown for Drell--Yan like production of fractionally, singly, and multiply charged particles (right).
These particles were assumed to be neutral under $SU(3)_C$ and $SU(2)_L$.
   \label{fig:masslimits}}
\end{figure}

The mass limits for various signals and electric charges are shown in Fig.~\ref{fig:masslimits} and are compared with previously published results.

The limits obtained for the reanalyzed $\sqrt{s}=7$~TeV dataset are similar to the previously published CMS results except for the stau scenarios,
where the new cross section limits are slightly worse than the previously published ones.  This result is a consequence of having a common selection for all mass points
and models in contrast to what was done in Ref.~\cite{Chatrchyan:2012sp}, where the selection was optimized separately for each mass point and model.
However, the use of a higher NLO cross section for the indirect production of staus than in Ref.~\cite{Chatrchyan:2012sp} results in more stringent limits on the stau mass.

The mass limit for $|Q|<1e$ samples are significantly improved
with respect to Ref.~\cite{Chatrchyan::2012dr}, thanks to a different analysis approach and to the use of the \iasp\ likelihood discriminator that maximally exploits all
the \dedx information associated with a track.

\section{Summary}

A search for heavy stable charged particles has been presented,
based on several different signatures, using data recorded at
collision energies of 7 and 8\TeV.
Five complementary analyses have been performed:
a search with only the inner tracker, a search with
both the inner tracker and the muon system,
a search with only the muon system, a search for low
ionizing tracks, and a search for tracks with very large ionization energy loss.
No significant excess is
observed in any of the analyses.
Limits on cross sections are presented for models with the
production of gluinos, scalar tops, and staus, and for Drell--Yan like
production of fractionally, singly, and multiply charged particles.
The models for $R$-hadron-like HSCPs
include a varying fraction of $\PSg-$gluon production
and two different interaction schemes
leading to a variety of non-standard experimental signatures.
Lower mass limits for these models are given.
Gluino masses below 1322 and 1233\GeVcc are excluded for $f=0.1$
in the cloud interaction model and the charge-suppressed model, respectively.
For $f=0.5$ ($1.0$), gluino masses below $1276$ ($1250$)\GeVcc
are excluded.
For stop production, masses below 935 (818)\GeVcc are excluded for the
cloud (charge-suppressed) models.
In addition, these analyses exclude \stau masses below
$500$ ($339$)\GeVcc for the direct+indirect (direct only) production.
Drell--Yan like signals with $|Q| = e/3$, $2e/3$, $1e$, $2e$, $3e$, $4e$, $5e$,
$6e$, $7e$, and $8e$ are excluded with masses below 200, 480, 574,
685, 752, 793, 796, 781, 757, and 715\GeVcc, respectively.
These limits are the most stringent to date.

\section*{Acknowledgments}
We congratulate our colleagues in the CERN accelerator departments for the excellent performance of the LHC and thank the technical and administrative staffs at CERN and at other CMS institutes for their contributions to the success of the CMS effort. In addition, we gratefully acknowledge the computing centres and personnel of the Worldwide LHC Computing Grid for delivering so effectively the computing infrastructure essential to our analyses. Finally, we acknowledge the enduring support for the construction and operation of the LHC and the CMS detector provided by the following funding agencies: BMWF and FWF (Austria); FNRS and FWO (Belgium); CNPq, CAPES, FAPERJ, and FAPESP (Brazil); MEYS (Bulgaria); CERN; CAS, MoST, and NSFC (China); COLCIENCIAS (Colombia); MSES (Croatia); RPF (Cyprus); MoER, SF0690030s09 and ERDF (Estonia); Academy of Finland, MEC, and HIP (Finland); CEA and CNRS/IN2P3 (France); BMBF, DFG, and HGF (Germany); GSRT (Greece); OTKA and NKTH (Hungary); DAE and DST (India); IPM (Iran); SFI (Ireland); INFN (Italy); NRF and WCU (Republic of Korea); LAS (Lithuania); CINVESTAV, CONACYT, SEP, and UASLP-FAI (Mexico); MSI (New Zealand); PAEC (Pakistan); MSHE and NSC (Poland); FCT (Portugal); JINR (Armenia, Belarus, Georgia, Ukraine, Uzbekistan); MON, RosAtom, RAS and RFBR (Russia); MSTD (Serbia); SEIDI and CPAN (Spain); Swiss Funding Agencies (Switzerland); NSC (Taipei); ThEPCenter, IPST and NSTDA (Thailand); TUBITAK and TAEK (Turkey); NASU (Ukraine); STFC (United Kingdom); DOE and NSF (USA).

Individuals have received support from the Marie-Curie programme and the European Research Council and EPLANET (European Union); the Leventis Foundation; the A. P. Sloan Foundation; the Alexander von Humboldt Foundation; the Belgian Federal Science Policy Office; the Fonds pour la Formation \`a la Recherche dans l'Industrie et dans l'Agriculture (FRIA-Belgium); the Agentschap voor Innovatie door Wetenschap en Technologie (IWT-Belgium); the Ministry of Education, Youth and Sports (MEYS) of Czech Republic; the Council of Science and Industrial Research, India; the Compagnia di San Paolo (Torino); the HOMING PLUS programme of Foundation for Polish Science, cofinanced by EU, Regional Development Fund; and the Thalis and Aristeia programmes cofinanced by EU-ESF and the Greek NSRF.

\bibliography{auto_generated}   % will be created by the tdr script.

\cleardoublepage \appendix\section{The CMS Collaboration \label{app:collab}}\begin{sloppypar}\hyphenpenalty=5000\widowpenalty=500\clubpenalty=5000\textbf{Yerevan Physics Institute,  Yerevan,  Armenia}\\*[0pt]
S.~Chatrchyan, V.~Khachatryan, A.M.~Sirunyan, A.~Tumasyan
\vskip\cmsinstskip
\textbf{Institut f\"{u}r Hochenergiephysik der OeAW,  Wien,  Austria}\\*[0pt]
W.~Adam, T.~Bergauer, M.~Dragicevic, J.~Er\"{o}, C.~Fabjan\cmsAuthorMark{1}, M.~Friedl, R.~Fr\"{u}hwirth\cmsAuthorMark{1}, V.M.~Ghete, N.~H\"{o}rmann, J.~Hrubec, M.~Jeitler\cmsAuthorMark{1}, W.~Kiesenhofer, V.~Kn\"{u}nz, M.~Krammer\cmsAuthorMark{1}, I.~Kr\"{a}tschmer, D.~Liko, I.~Mikulec, D.~Rabady\cmsAuthorMark{2}, B.~Rahbaran, C.~Rohringer, H.~Rohringer, R.~Sch\"{o}fbeck, J.~Strauss, A.~Taurok, W.~Treberer-Treberspurg, W.~Waltenberger, C.-E.~Wulz\cmsAuthorMark{1}
% ...
\vskip\cmsinstskip
% ...
1:~~Also at Vienna University of Technology, Vienna, Austria\\
2:~~Also at CERN, European Organization for Nuclear Research, Geneva, Switzerland\\
% ...

\end{sloppypar}
\end{document}